%% file: main.tex
\documentclass{article}
\usepackage{graphicx} % Required for inserting images

\usepackage{amsxtra}
\usepackage{mathtools}
\usepackage{float}
\usepackage{xcolor}

\usepackage{macros}

\newcommand{\Figref}[1]{Figure~\ref{#1}}
\newcommand{\Quiver}[1]{$\mathcal Q_{\ref{#1}}$}
\newcommand{\q}[1]{``#1''}

\tikzset{gaugebl/.style={circle,draw=black,fill=black,inner sep=2.5pt}}
\tikzset{hasse/.style={circle, fill,inner sep=2pt}}

\allowdisplaybreaks

\preprint{Imperial/TP/23/AH/03}

\title{Actions on the quiver -- Discrete quotients on the Coulomb branch}

\author[a]{Amihay Hanany}
\author[a]{, Guhesh Kumaran}
\author[a]{, Chunhao Li}
\author[a]{, Deshuo Liu}
\author[b]{, Marcus Sperling}

\affiliation[a]{Theoretical Physics Group, The Blackett Laboratory, Imperial College London,\\ Prince Consort Road
London, SW7 2AZ, UK}
\affiliation[b]{Fakultät für Physik, Universität Wien,\\
Boltzmanngasse 5, 1090 Wien, Austria}
\emailAdd{a.hanany@imperial.ac.uk}
\emailAdd{guhesh.kumaran18@imperial.ac.uk}
\emailAdd{chunhao.li21@imperial.ac.uk}
\emailAdd{deshuo.liu21@imperial.ac.uk}
\emailAdd{marcus.sperling@univie.ac.at}

\date{May 2023}

\abstract{
This paper introduces two operations in quiver gauge theories. The first operation takes a quiver with a permutation symmetry $\Sfrak_n$ and gives a quiver with adjoint loops. The corresponding 3d ${\cal N}=4$ Coulomb branches are related by an orbifold of $\Sfrak_n$. The second operation takes a quiver with $n$ nodes connected by edges of multiplicity $k$ and replaces them by $n$ nodes of multiplicity $qk$. The corresponding Coulomb branch moduli spaces are related by an orbifold of type $\Z_q^{n-1}$. The first operation generalises known cases that appeared in the literature \cite{Hanany:2018vph,Hanany:2018cgo,Hanany:2018dvd}. These two operations can be combined to generate new relations between moduli spaces that are constructed using the magnetic construction.}
\begin{document}

\maketitle

\section{Introduction}
In recent years diagrammatic techniques on $3d$ $\mathcal N=4$ quiver gauge theories have been developed and provide insight and simple tools to study their moduli spaces of vacua.
Examples of diagrammatic techniques on $3d$ $\mathcal N=4$ quiver gauge theories include \textit{quiver subtraction} \cite{Cabrera:2018ann} which can be used to study the stratification of the Coulomb branch into a finite number of symplectic leaves \cite{2003math.....10186K,Bourget:2019aer,Grimminger:2020dmg} or to study hyper-Kähler quotients of Coulomb branches \cite{Hanany:2023tvn}. The former also has a realisation in string theory \cite{Cabrera:2016vvv,Cabrera:2017njm}.

More pertinent to this work is the study of discrete actions on the Coulomb branch via diagrammatic techniques. Examples include \textit{folding} \cite{Bourget:2020bxh,Bourget:2021xex} and \textit{discrete gauging} \cite{Hanany:2018vph,Hanany:2018cgo,Hanany:2018dvd,Bourget:2021xex}. These realise discrete actions on the Coulomb branch.

One result of this work is a generalisation of the discrete gauging of a bouquet of $\urm(1)$ in $3d$ $\mathcal N=4$ unitary quiver gauge theories introduced in \cite{Hanany:2018vph,Hanany:2018cgo,Hanany:2018dvd}. The discrete gauging of such a bouquet has a realisation in M-theory as a stack of M5 branes probing an $A_k$ singularity and making these M5 branes coincident. There is a natural $\Sfrak_n$ action permuting the $n$ M5. The reduction of the M-theory set-up to Type IIA corresponds to the same $\Sfrak_n$ action on NS5 branes. The magnetic quiver \cite{Cabrera:2019izd} for the theory on the worldvolume of the D6 branes has a bouquet of the $\urm(1)$ gauge nodes when the NS5 are not coincident and becomes a $\urm(n)$ gauge node with an adjoint hypermultiplet when the NS5 are coincident. This is illustrated below with the $\urm(1)$ in the bouquet shown in \textcolor{orange}{orange}:
\begin{equation}
\Ccal\left(
    \vcenter{\hbox{
        \begin{tikzpicture}
            \node[gauge,label=below:{$1$}] (6) at (-2,0) {};
            \node[gauge,label=below:{$2$}] (3) at (-1.2,0) {};
            \node[] (2) at (-0.6,0) {$\cdots$};
            \node[gauge,label=below:{$1$}] (7) at (2,0) {};
            \node[gauge,label=below:{$2$}] (4) at (1.2,0) {};
            \node[] (5) at (0.6,0) {$\cdots$};
            \node[gauge,label=below:{$k$}] (1) at (0,0) {};
            \draw[] (6)--(3) (7)--(4);
            \draw[-] (3)--(-0.9,0) (0.9,0)--(4)(-0.3,0)--(1)--(0.3,0);
            \node[gauge,label=left:{$1$},color=orange!100] (31) at (-0.5,1) {};
           
            \node at (0,1) {$\cdots$};
            \node[gauge,label=right:{$1$},color=orange!100] (33) at (0.5,1) {};
            
            \draw (31)--(1) (33)--(1);
		    \draw [decorate,decoration={brace,amplitude=5pt}] (-0.7,1.2)--(0.7,1.2);
		    \node at (0,1.6) {$n$};
        \end{tikzpicture}
    }}\right) \quad \overset{\Sfrak_n}{\longrightarrow} \quad
    \Ccal\left(
        \vcenter{\hbox{
        \begin{tikzpicture}
        \node[gauge,label=below:{$1$}] (6) at (-2,0) {};
            \node[gauge,label=below:{$2$}] (3) at (-1.2,0) {};
            \node[] (2) at (-0.6,0) {$\cdots$};
            \node[gauge,label=below:{$1$}] (7) at (2,0) {};
            \node[gauge,label=below:{$2$}] (4) at (1.2,0) {};
            \node[] (5) at (0.6,0) {$\cdots$};
            \node[gauge,label=below:{$k$}] (1) at (0,0) {};
            \draw[] (6)--(3) (7)--(4);
            \draw[-] (3)--(-0.9,0) (0.9,0)--(4)(-0.3,0)--(1)--(0.3,0);
            \node[gauge,label=left:{$n$},color=orange!100] (3) at (0,1) {};
            \draw (3) to [out=45,in=135,looseness=8] (3);
            \draw (3)--(1);
        \end{tikzpicture}
        }}\right)
\end{equation}

An application of the discrete quotients of Coulomb branches was to physically realise the results of Kostant-Brylinski \cite{1992math......4227B}, which relate certain nilpotent orbit closures under discrete quotients. 

The notion of a bouquet of $\urm(1)$ gauge nodes in a unitary $3d$ $\mathcal N=4$ quiver gauge theory is encompassed and extended to a \textit{complete graph} of $\urm(1)$ gauge nodes. There is a natural $\Sfrak_n$ permutation of these $\urm(1)$ gauge nodes. This outer automorphism symmetry may be gauged and consequently there is a discrete $\Sfrak_n$ action on the Coulomb branch.

A second result of this work is to realise orbifold actions on the Coulomb branch by turning simply laced edges into non-simply laced edges \cite{Cremonesi:2014xha}. The relationships between quivers with simply laced and non-simply laced edges were studied in \cite{Hanany:2020jzl} and are studied further here.

These two new diagrammatic techniques are applied to various quivers and examples are presented for each technique applied on its own and also in combination.

Mathematical results can be realised as a Coulomb branch using these techniques. For example, the $\Sfrak_2$ quotient of the Kleinian A-type singularity to the D-type singularity is constructed as a Coulomb branch. In addition many new discrete quotients of known moduli spaces such as $\overline{min. A_2}=a_2$, $\overline{min. A_3}=a_3$, and $\overline{min D_4}=d_4$ are constructed. Another outcome of this work is the Coulomb branch construction of $a_2/\Sfrak_4$ and $d_4/\Sfrak_4$, which appear respectively as a Slodowy slice in the nilpotent cone of $\mathfrak{e}_8$ between the nilpotent orbits with Bala-Carta labels $E_8(a_6)$ and $E_8(b_6)$ and as a Slodowy slice in the nilpotent cone of $\mathfrak{f}_4$ between the nilpotent orbits with Bala-Carter labels $F_4(a_3)$ and $A_2$ \cite{2023arXiv230807398F,2015arXiv150205770F}. One way of constructing $d_4/\Sfrak_4$ as a Coulomb branch is through the discrete gauging of a bouquet. The discrete gauging of a bouquet is not amenable for the case of $a_2/\Sfrak_4$. Instead, the isomorphism $\Sfrak_4\cong \Z_2^2\!\rtimes\!\Sfrak_3$ is employed. On the quiver, a combination of the two new techniques, orbifold and complete graph discrete quotients, are used to achieve the desired result. Specifically, we have the first construction of $a_2/\Sfrak_4$ as a Coulomb branch and now two ways of constructing $d_4/\Sfrak_4$.

Another outcome of this work is the discrete quotient on the Coulomb branch by a semi-direct product of discrete groups rather than a direct product of discrete groups.

The quivers for the $3d$ mirrors of certain Argyres-Douglas theories contain a complete graph \cite{Xie:2012hs,giacomelli:2021new}. The discrete quotient on the Coulomb branch of the $3d$ mirrors are computed.

\paragraph{Outline of the paper.}
The paper is organised as follows. In Section \ref{sec:loops_complete_graphs} the complete graph quiver is introduced and the notation that is used throughout this paper is defined. The claim that there is an $\Sfrak_n$ quotient between the Coulomb branches of quivers with complete graphs and a $\urm(n)$ gauge node with some number of adjoint hypermultiplets is stated. In Section \ref{sec:GeneralisedMolienSum} the monopole formula is introduced as well as the Molien sum on the monopole formula. The claim made in Section \ref{sec:loops_complete_graphs} is proved. The Molien sum on the monopole formula realising the orbifold relation between simply and non-simply laced quivers is also presented. In Section \ref{sec:abel} the Abelianisation procedure is introduced. In Section \ref{sec:SnExamples} many examples of the $\Sfrak_n$ quotient are presented. These include the $\Sfrak_2$ gauging of the A-type singularity to the D-type singularity. In addition the $\Sfrak_2$ gauging of the $A_3$ affine quiver. In Section \ref{sec:ZqQuot} examples of $\mathbb Z_q$ quotient are presented. In Section \ref{sec:CombinedQuot} a combination of the $\Sfrak_n$ and $\mathbb Z_q$ quotients on the Coulomb branch are combined. In Section \ref{sec:general} the $\Sfrak_n$ quotient is generalised to other types of gauge group. In Section \ref{sec:conclusions} conclusion are drawn and future work is discussed.

\section{Discrete quotient on the quiver}

\paragraph{Loops and complete graphs.}
\label{sec:loops_complete_graphs}

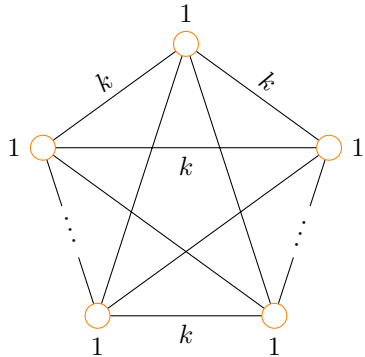
\begin{figure}[H]
    \centering
    \begin{tikzpicture}[main/.style={draw,circle}]
    \node[main, label=above:$1$,color=orange!100] (top) at (0,2) {};
    \node[main, label=right:$1$,color=orange!100] (topright) at ({2*cos(18)},{2*sin(18)}){};
    \node[main, label=left:$1$,color=orange!100] (topleft) at ({-2*cos(18)},{2*sin(18)}){};
    \node[main, label=below:$1$,color=orange!100] (bottomright) at ({2*cos(54)},{-2*sin(54)}){};
    \node[main, label=below:$1$,color=orange!100] (bottomleft) at ({-2*cos(54)},{-2*sin(54)}){};
 
    \path (topright) -- (bottomright) node[midway,sloped] (dotsr) {$\cdots$};
    \path (topleft) -- (bottomleft) node[midway,sloped] (dotsl) {$\cdots$};

     \draw[-] (top)--(topright) node[pos=0.5,above,sloped]{$k$}--(dotsr)--(bottomright)--(bottomleft)node[pos=0.5,below,sloped]{$k$}--(dotsl)--(topleft)--(topright)node[pos=0.5,below,sloped]{$k$}--(bottomleft)--(top)--(bottomright)--(topleft)--(top)node[pos=0.5,above,sloped]{$k$};

    \end{tikzpicture}
    \caption{The complete graph quiver $\CG_{n,k}$ with $n$ nodes of rank 1 and edge multiplicity $k$. Henceforth $k=2g-2$.}
    \label{fig:mysticrose}
\end{figure}

The first family of quivers studied in this paper is $\CG_{n,q}$, the family of complete graphs shown in \Figref{fig:mysticrose}.

This quiver consists of $n$ nodes of $\urm(1)$ with each pair of $\urm(1)$s connected by $k$ hypermultiplets transforming in the bi-fundamental representation. An overall $\urm(1)$ acts freely and the actual gauge group is $ \urm(1)^n/\urm(1)$. This family received some attention as the $3d$ mirror of the Argyres Douglas theories of type $(A_{n-1},A_{nk-1})$ \cite{Xie:2012hs}, i.e. the Coulomb/Higgs branch of the $\CG_{n,k}$ theory is the Higgs/Coulomb branch of the 3d $(A_{n-1},A_{nk-1})$ theory, respectively. For most quivers studied in this paper only even $k$ is needed, and furthermore a convenient parametrisation is $k=2g-2$, where $g\in\Z^+$ is suggestive of the genus of a Riemann surface, as becomes clearer in Section \ref{sec:dpsun}. Note that there is an $\Sfrak_n$ permutation symmetry of the $\urm(1)$ gauge nodes which comes from the outer automorphism symmetry of $\CG_{n,2g-2}$. Correspondingly, the Coulomb branch has this $\Sfrak_n$ symmetry, and one can construct many orbifolds by gauging a subgroup of this symmetry.

\begin{figure}[H]
    \centering
    \begin{tikzpicture}[main/.style={draw,circle}]
    \node[main, label=below:$n$,color=orange!100] (n) at (0,0) {};
    \draw (n) to [out=135, in=45,looseness=8] node[pos=0.5,above]{$g$} (n);
    \end{tikzpicture}
    \caption{Multi-loop quiver $\mathcal{ML}_{n,g}$ of $\urm(n)$ gauge group with $g$ adjoints.}
    \label{fig:UnGadj}
\end{figure}
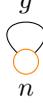

The second family of quivers studied here is the multi-loop quiver $\mathcal{ML}_{n,g}$ shown in \Figref{fig:UnGadj}. This quiver consists of a $\urm(n)$ gauge group with $g$ hypermultiplets in the adjoint representation. As the centre $\urm(1)$ is acting trivially on the matter content, the actual gauge group is $\psurm(n)$.

The two main results of this paper are
\begin{enumerate}
\item The Coulomb branches of $\ML_{n,g}$ and $\CG_{n,2g-2}$ are related by a quotient of the $\Sfrak_n$ symmetry:
\begin{equation}
\Ccal(\ML_{n,g})=\Ccal(\CG_{n,2g-2})/\Sfrak_n \,.
    \label{eq1}
\end{equation}
Here the edge multiplicity $k$ is required to be even. Odd multiplicities gives non-integer $g$ which makes the quiver of \Figref{fig:UnGadj} ill defined.

\item Given a complete graph quiver $\CG_{n,k}$, if the edge multiplicity is increased to be $q$-fold, the quiver becomes $\CG_{n,qk}$, and the Coulomb branches are related by a quotient of $\Z_q^{n-1}$:
\begin{equation}
\Ccal(\CG_{n,qk})=\Ccal(\CG_{n,k})/\Z_q^{n-1} \,.
    \label{eqzq}
\end{equation}

\end{enumerate}
There are some corollaries of these two results.
\paragraph{Combined action.}
Notice that the $\Sfrak_n$ symmetry is preserved after the $\Z_q^{n-1}$ quotient: $\CG_{n,1}$ and $\CG_{n,q}$ have the same outer-automorphism symmetry, so one can combine the two relations \eqref{eq1} and \eqref{eqzq} above, and result in:
\begin{equation}
\Ccal(\ML_{n,g})=\Ccal(\CG_{n,1})/\Z_{2g-2}^{n-1}\!\rtimes\!\Sfrak_n.
\label{zqsnrelation}
\end{equation}
The composition is a semi-direct product, since the $\Sfrak_n$ symmetry also permutes these $n-1$ $\Z_q$s.

\paragraph{Connecting to a background quiver.}

The two families of quivers $\CG_{n,2g-2}$ and $\ML_{n,g}$ can be connected to a \textit{pivot} node of a quiver $\mathcal{Q}_B$.

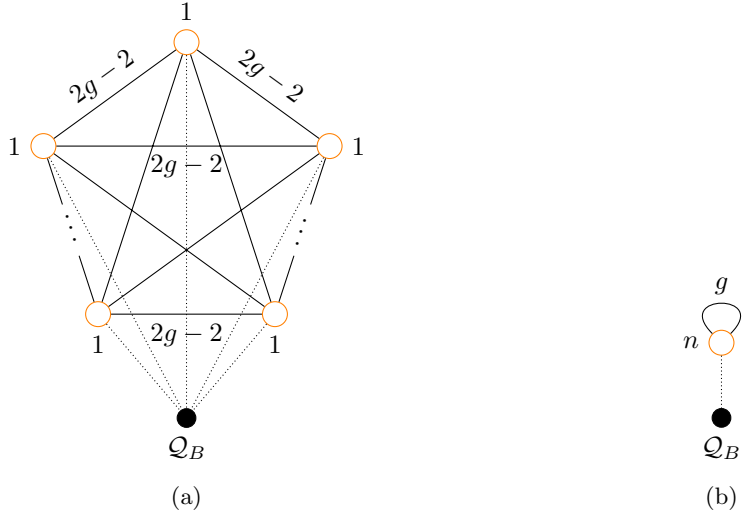
\begin{figure}[H]
    \centering
    
    \begin{subfigure}[t]{0.45\textwidth}
 \centering
    \begin{tikzpicture}[main/.style={draw,circle}]
    \node[main, label=above:$1$,color=orange!100] (top) at (0,2) {};
    \node[main, label=right:$1$,color=orange!100] (topright) at ({2*cos(18)},{2*sin(18)}){};
    \node[main, label=left:$1$,color=orange!100] (topleft) at ({-2*cos(18)},{2*sin(18)}){};
    \node[main, label=below:$1$,color=orange!100] (bottomright) at ({2*cos(54)},{-2*sin(54)}){};
    \node[main, label=below:$1$,color=orange!100] (bottomleft) at ({-2*cos(54)},{-2*sin(54)}){};

    \node (QB) at (0,-3) [gaugebl,label=below:{$\mathcal{Q}_B$}](QB) {};
 
    \path (topright) -- (bottomright) node[midway,sloped] (dotsr) {$\cdots$};
    \path (topleft) -- (bottomleft) node[midway,sloped] (dotsl) {$\cdots$};
 
     \draw[-] (top)--(topright) node[pos=0.5,above,sloped]{$2g-2$}--(dotsr)--(bottomright)--(bottomleft)node[pos=0.5,below,sloped]{$2g-2$}--(dotsl)--(topleft)--(topright)node[pos=0.5,below,sloped]{$2g-2$}--(bottomleft)--(top)--(bottomright)--(topleft)--(top)node[pos=0.5,above,sloped]{$2g-2$};
 
    \draw[densely dotted] (top)--(QB) (topright)--(QB) (topleft)--(QB) (bottomleft)--(QB) (bottomright)--(QB);
 
    \end{tikzpicture}
    \caption{}
    \label{fig:q4}
   \end{subfigure}
  \begin{subfigure}[t]{0.45\textwidth}
      \centering
    \begin{tikzpicture}[main/.style={draw,circle}]
    \node[main, label=left:$n$,color=orange!100] (n) at (0,0) {};
 
    \node (QB) at (0,-1) [gaugebl,label=below:$\mathcal{Q}_B$](QB){};
     
    \draw (n) to [out=135, in=45,looseness=8] node[pos=0.5,above]{$g$} (n);
 
    \draw[densely dotted] (n)--(QB);
    \end{tikzpicture}
    \caption{}
    \label{fig:q3}
   \end{subfigure}
   
    \caption{\subref{fig:q4}: Quiver \Quiver{fig:q4} which has the complete graph subquiver $\CG_{n,2g-2}$, connecting with background $\mathcal{Q}_B$. \subref{fig:q3}: Quiver \Quiver{fig:q3} for $\urm(n)$ gauge group with $g$ adjoints, connecting with background $\mathcal{Q}_B$.}
    \label{fig:Un+adj_result_bckgrd}
\end{figure}

Then there are two families of quivers \Quiver{fig:q3} and \Quiver{fig:q4}, shown in \Figref{fig:q3} and \Figref{fig:q4} respectively. The dotted lines in the quivers \Quiver{fig:q3} and \Quiver{fig:q4} refer to a link of any type to the same pivot node in $\mathcal Q_B$. For example, the dotted links could be a single bi-fundamental hypermultiplet, multiple bi-fundamental hypermultiplets, or directed non-simply laced edges. They can also be connected to multiple pivot nodes in $\mathcal{Q}_B$ as long as the links to that pivot are all the same for the $\urm(1)$ nodes in the complete graph.

By adding this pivot node, the quiver outer-automorphism still contains the same $\Sfrak_n$, hence the same quotient relation on the Coulomb branch holds:
\begin{equation}
    \Ccal(\text{\Quiver{fig:q3}})=\Ccal(\text{\Quiver{fig:q4}
    })/\Sfrak_n \,.
    \label{eq2}
\end{equation}

For the special case of $g=1$ and each connection to some gauge node in $\mathcal{Q}_B$ is a single bi-fundamental, claim \eqref{eq2} reproduces the $\Sfrak_n$ quotient result for a bouquet of $\urm(1)$ in  \cite{Hanany:2018vph,Hanany:2018cgo,Hanany:2018dvd}.

Moreover, the complete graph part of \Quiver{fig:q4} contains subgraphs which themselves are complete graphs of $m$ nodes ($m\leq n$). So Claim \eqref{eq2} implies that any $\Sfrak_m$ with $m\leq n$ can be quotiented from \Quiver{fig:q4}. An important consequence of this relation is that every unitary quiver with a loop can be replaced by a quiver that has no loops, but instead admits a discrete permutation symmetry.

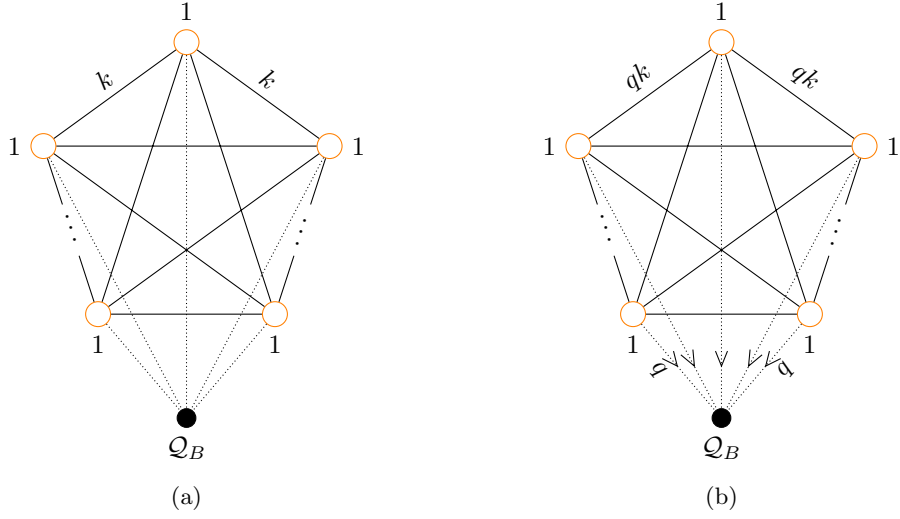
\begin{figure}[H]
    \centering
    
   \begin{subfigure}[H]{0.45\textwidth}
      \centering
    \begin{tikzpicture}[main/.style={draw,circle}]
    \node[main, label=above:$1$,color=orange!100] (top) at (0,2) {};
    \node[main, label=right:$1$,color=orange!100] (topright) at ({2*cos(18)},{2*sin(18)}){};
    \node[main, label=left:$1$,color=orange!100] (topleft) at ({-2*cos(18)},{2*sin(18)}){};
    \node[main, label=below:$1$,color=orange!100] (bottomright) at ({2*cos(54)},{-2*sin(54)}){};
    \node[main, label=below:$1$,color=orange!100] (bottomleft) at ({-2*cos(54)},{-2*sin(54)}){};
 
    \node (QB) at (0,-3) [gaugebl,label=below:{$\mathcal{Q}_B$}](QB) {};
 
    \path (topright) -- (bottomright) node[midway,sloped] (dotsr) {$\cdots$};
    \path (topleft) -- (bottomleft) node[midway,sloped] (dotsl) {$\cdots$};
 
     \draw[-] (top)--(topright) node[pos=0.5,above,sloped]{$k$}--(dotsr)--(bottomright)--(bottomleft)--(dotsl)--(topleft)--(topright)--(bottomleft)--(top)--(bottomright)--(topleft)--(top)node[pos=0.5,above,sloped]{$k$};
 
    \draw[densely dotted] (top)--(QB) (topright)--(QB) (topleft)--(QB) (bottomleft)--(QB) (bottomright)--(QB);
 
    \end{tikzpicture}
    \caption{}
    \label{fig:zqbefore}
   \end{subfigure}
    \begin{subfigure}[H]{0.45\textwidth}
 \centering
    \begin{tikzpicture}[main/.style={draw,circle}]
    \node[main, label=above:$1$,color=orange!100] (top) at (0,2) {};
    \node[main, label=right:$1$,color=orange!100] (topright) at ({2*cos(18)},{2*sin(18)}){};
    \node[main, label=left:$1$,color=orange!100] (topleft) at ({-2*cos(18)},{2*sin(18)}){};
    \node[main, label=below:$1$,color=orange!100] (bottomright) at ({2*cos(54)},{-2*sin(54)}){};
    \node[main, label=below:$1$,color=orange!100] (bottomleft) at ({-2*cos(54)},{-2*sin(54)}){};
 
    \node (QB) at (0,-3) [gaugebl,label=below:{$\mathcal{Q}_B$}](QB) {};
 
    \path (topright) -- (bottomright) node[midway,sloped] (dotsr) {$\cdots$};
    \path (topleft) -- (bottomleft) node[midway,sloped] (dotsl) {$\cdots$};
 
     \draw[-] (top)--(topright) node[pos=0.5,above,sloped]{$qk$}--(dotsr)--(bottomright)--(bottomleft)--(dotsl)--(topleft)--(topright)--(bottomleft)--(top)--(bottomright)--(topleft)--(top)node[pos=0.5,above,sloped]{$qk$};
 
    \draw[densely dotted] (top)--(QB) (topright)--(QB) (topleft)--(QB) (bottomleft)--(QB)node[pos=0.4,below,sloped]{$q$} (bottomright)--(QB)node[pos=0.4,below,sloped]{$q$};

    \draw ({-cos(54)-0.2},{-sin(54)-1.5+0.1})--({-cos(54)},{-sin(54)-1.5})--({-cos(54)-0.05},{-sin(54)-1.5+0.2})  ({cos(54)+0.2},{-sin(54)-1.5+0.1})--({cos(54)},{-sin(54)-1.5})--({cos(54)+0.05},{-sin(54)-1.5+0.2}) (-0.08,{-sin(54)-1.5+0.2})--(0,{-sin(54)-1.5})--(0.08,{-sin(54)-1.5+0.2})  (-0.36-0.18,{-sin(54)-1.5+0.15})--(-0.36,{-sin(54)-1.5})--(-0.36-0.02,{-sin(54)-1.5+0.22}) (0.36+0.18,{-sin(54)-1.5+0.15})--(0.36,{-sin(54)-1.5})--(0.36+0.02,{-sin(54)-1.5+0.22});

    \end{tikzpicture}
    \caption{}
    \label{fig:zqafter}
   \end{subfigure}
   
    \caption{\subref{fig:zqbefore}: $\CG_{n,k}$ with simply laced edges connecting a simply laced $\mathcal{Q}_B$. \subref{fig:zqafter}: $\CG_{n,qk}$ with $qk$ laced edges connecting a simply laced $\mathcal{Q}_B$.}
    \label{fig:zqba}
\end{figure}

The $\Z_q$ quotient relation also holds when connecting to a background quiver. The following claims are made\footnote{This claim does not rely on the complete graph structure. In fact, one can replace the complete graph with an arbitrary simply-laced Abelian quiver, and the quotient relation on Coulomb branch  still holds.}:
\begin{equation}
\Ccal\left(\text{\Quiver{fig:zqafter}}\right) = \begin{cases}
\Ccal\left(\text{\Quiver{fig:zqbefore}}/\Z_{q}^{n-1}\right) &\text{when $\mathcal{Q}_B$ is empty or unframed}\\
\Ccal\left(\text{\Quiver{fig:zqbefore}}/\Z_{q}^n\right) &\text{when $\mathcal{Q}_B$ is framed}
\end{cases} \,,
\label{eqzqb}
\end{equation}
the reason for the different action for the framed and unframed $\mathcal{Q}_B$ is discussed in Section \ref{sec:monopolezqn}.

Notice that the $\Sfrak_n$ symmetry is preserved after $\Z_q^{n}$ quotient, and the $\Sfrak_n$ symmetry permutes these $n$ $\Z_q$s.
The following claims are made:
\begin{equation}
\Ccal\left(\text{\Quiver{fig:zqafter}}/\Sfrak_n\right) = \begin{cases}
\Ccal\left(\text{\Quiver{fig:zqbefore}}/\Z_{q}^{n-1}\!\rtimes\!\Sfrak_n\right) &\text{when $\mathcal{Q}_B$ is empty or unframed}\\
\Ccal\left(\text{\Quiver{fig:zqbefore}}/\Z_{q}^n\!\rtimes\!\Sfrak_n\right) &\text{when $\mathcal{Q}_B$ is framed}
\end{cases} \,.
\label{eqzqsn}
\end{equation}

\section{Generalised Molien sum and Coulomb branch Hilbert series}
\label{sec:GeneralisedMolienSum}

The Molien sum is a method to calculate the Hilbert series of an orbifold from a known Hilbert series, more details are in Appendix \ref{app:molien}. In this section the generalised Molien sum on the monopole formula is introduced and \eqref{eq1} and \eqref{eq2} are proved using the Hilbert series.

\subsection{Conformal dimension and monopole formula}
The monopole formula, introduced in \cite{Cremonesi:2013lqa}, is a method to calculate the Coulomb branch Hilbert series of a $3\text{d} \,\,\mathcal{N}=4$ gauge theory. To evaluate the monopole formula, firstly the conformal dimension \cite{Gaiotto:2008ak} is determined, which gives the dimension of the bare monopole operator. Further details about the monopole formula are in Appendix \ref{app:monopole}.

The conformal dimension for the $\CG_{n,2g-2}$ theory is computed as \begin{equation}
    \Delta(m_1,\cdots,m_n)=(g-1)\sum_{1\leq i<j\leq n}|m_i-m_j|.
\end{equation} The key point is that the $\ML_{n,g}$ theory has exactly the same conformal dimension.

Given this fact for the standalone quivers $\CG_{n,2g-2}$ and $\ML_{n,g}$ it is a simple extension to show that the conformal dimension when these quivers are attached to a background quiver $\mathcal Q_B$, as in quivers \Quiver{fig:q3} and \Quiver{fig:q4}, is also the same,
\begin{equation}
    \Delta(m_1,\dots,m_n;\vec{B})=(g-1)\sum_{1\leq i<j\leq n}|m_i-m_j|+\sum_{1\leq j\leq l}\sum_{1\leq i\leq n}\Delta_j(m_i;\vec{B})+\Delta[\mathcal{Q}_B],.
    \label{cdq3}
\end{equation}
Here, $\Delta[\mathcal{Q}_B]$ is the contribution from $\mathcal{Q}_B$. There could be links to each of the $l$ gauge nodes in $\mathcal Q_B$, the contribution from these links to the $j^\text{th}$ node in $\mathcal Q_B$ is denoted $\Delta_j(m_i;\vec{B})$ for $1\leq j\leq l$. The contribution $\Delta_j(m_i;\vec{B})$ takes the same functional form for all $m_i$. This is the only requirement for the links to $\mathcal{Q}_B$ in the quivers. So those links can be a single bi-fundamental hypermultiplet, multiple bi-fundamental hypermultiplets, and also directed non-simply laced edges, as long as they are all the same. Note that if a non-simply laced edge is used then these must all be directed in the same way for all connections to $\mathcal Q_B$.

The conformal dimensions for the $\CG_{n,2g-2}$ and the $\ML_{n,g}$ are the same, but the difference in Hilbert series for the Coulomb branch of \Quiver{fig:q3} and \Quiver{fig:q4} only comes from the magnetic lattice and the dressing factors.

\subsection{\texorpdfstring{Generalised Molien sum of $\Sfrak_n$ on the monopole formula}{Sn action on monopole formula}}
\label{sec:Sn}

We take the specific action of $\Sfrak_n$ on the monopole formula and explain how the generalised Molien sum works.

Let us start with the quiver \Quiver{fig:q4}, which contains $n$ identical $\urm(1)$ nodes and a background $\mathcal{Q}_B$. Quiver \Quiver{fig:q4} is unchanged under permutation of those $n$ $\urm(1)$ nodes, i.e.\ there is a $\Sfrak_n \subset \text{Out}$(\Quiver{fig:q4}).

We label the magnetic flux associated to the $i$-th node $\urm(1)_i$ as $m_i$, and the collective magnetic fluxes associated to $\mathcal{Q}_B$ are denoted by $\vec{B}$. We label the dressing factor as $P_{\urm(1)^n}=\frac{1}{(1-t^2)^n}$ and $P_B(\vec{B})$. The monopole formula of \Quiver{fig:q4} can be written as:

\begin{equation}
\HS\left(\Ccal(\text{\Quiver{fig:q4}})\right)=\sum_{m_1,\dots,m_n,\vec{B}} P_{\urm(1)^n}P_B(\vec{B}) \cdot z_1^{m_1}\cdots z_n^{m_n} Z(\vec B)\cdot t^{\Delta(m_1,\dots, m_n;\vec{B})},
\end{equation}
where the $z_i$ are topological fugacities for the $\urm(1)_i$ and $Z(\vec B)$ is a monomial of topological fugacities associated to gauge nodes in $\mathcal Q_B$. The action of $\Sfrak_n$ permutes magnetic fluxes $m_i$, topological fugacities $z_i$ and Casimir invariants, however leaves $P_B(\vec B)$ and $Z(\vec B)$ invariant by construction. This can be implemented in the monopole formula by restricting it to the fixed loci of $\sigma \in \Sfrak_n$, which is discussed below in detail.

Elements of $\Sfrak_n$ can be written as cycles: $\sigma=(i_1\ \dots\ i_{\sigma_1})(i_{\sigma_1+1}\ \dots\ i_{\sigma_1+\sigma_2})\cdots(i_{n-\sigma_k+1}\ \dots\ i_{n})$, where $\sum_{l=1}^k\sigma_l=n$. For example, $(1\ 3)(2\ 5\ 4) \in \Sfrak_5$ acts on the magnetic fluxes as:
\begin{equation}
(m_1,m_2,m_3,m_4,m_5)\to(m_3,m_5,m_1,m_2,m_4).
\end{equation} Under the action of $\sigma$,fluxes within the same cycle are identified, i.e. $m_{i_1}=\dots=m_{i_{\sigma_1}}\,, \dots\,, m_{i_{n-\sigma_k+1}}=\dots=m_{i_n}$. The same rule applies for the topological fugacities $z_i$.

The next step is to determine the dressing factors under $\sigma$. The Casimir invariants form the polynomial ring of $\mathbb{C}^n$. The degree of each coordinate is scaled by a factor of $2$ and the Hilbert series $\frac{1}{(1-t^2)^n}$ is exactly the dressing factor of $\urm(1)^n$. The Hilbert series of the invariant ring under $\Sfrak_n$ follows from the Molien formula:
\begin{align}
\begin{aligned}
   \HS\left((\mathbb{C}^n)^{\Sfrak_n}\right)
   &= \frac{1}{n!}\sum_{\sigma\in \Sfrak_n} \frac{1}{\text{det}(1-t^2\cdot \sigma)}
   \\&= \frac{1}{n!}\sum_{\sigma\in \Sfrak_n}\frac{1}{(1-t^{2\sigma_1})\cdots(1-t^{2\sigma_k})} 
   \\&= \frac{1}{(1-t^2)\cdots(1-t^{2n})} \;. 
\end{aligned}
\label{dressingweyl}
\end{align}
which is the dressing factor for the unbroken $\urm(n)$. Hence, on the fixed loci of $\sigma$, the dressing factors turns into:
\begin{equation}
    P_{\urm(1)^n}=\frac{1}{(1-t^2)^n} \quad \overset{\sigma}{\longrightarrow} \quad \frac{1}{\text{det}(1-t^2\cdot\sigma)}=\frac{1}{(1-t^{2\sigma_1})\cdots(1-t^{2\sigma_k})}
\end{equation}

Now the monopole formula under the action of $\sigma$ is expressed. The fugacities $z_i$ and $Z(\vec B)$ are set to $1$ for brevity, however recall that the $z_i$ have the same behaviour as $t^{\Delta(m_1,\dots, m_n;\vec{B})}$ under $\Sfrak_n$ and $Z(\vec B)$ is invariant. The monopole formula under the element $\sigma$ becomes:
\begin{equation}
\label{fixedloci}
\HS\left(\Ccal(\text{\Quiver{fig:q4}
})\right)^{\sigma}
=\sum_{\vec{B}} \sum_{m_{i_1}=\ldots=m_{i_{\sigma_1}}}\dots\sum_{m_{i_{n-\sigma_k+1}}=\ldots=m_{i_n}} 
\left(\prod_{l=1}^{k} \frac{1}{(1-t^{2\sigma_l})}\right)
P_B(\vec{B})\cdot t^{\Delta(m_1,\dots,m_n;\vec{B})} \,.
\end{equation}

According to Molien formula, the monopole formula applied to $\mathcal C\left(\text{\Quiver{fig:q4}}\right)/\Sfrak_n$ is the average over the $\Sfrak_n$ action:
\begin{equation}
\label{molien}
\HS\left(\Ccal\left(\text{\Quiver{fig:q4}}/\Sfrak_n\right)\right)=\frac{1}{n!}\sum_{\sigma\in \Sfrak_n}\HS\left(\Ccal\left(\text{\Quiver{fig:q4}}\right)\right)^{\sigma}.
\end{equation}
This generalised Molien sum is realised through cycle index for special cases in \cite{Hanany:2018cgo} and also agrees with the Hilbert series of the wreathed quiver \cite{Bourget:2020bxh}.

\paragraph{Proof of the $\Sfrak_n$ relation.}

In this paragraph, the claim \eqref{eq2} is proved at the level of the Hilbert series. For claim \eqref{eq2} to be true, the following must hold:
\begin{equation}
\HS\left(\Ccal\left(\text{\Quiver{fig:q4}}/\Sfrak_n\right)\right)=\HS\left(\Ccal\left(\text{\Quiver{fig:q3}}\right)\right).
\label{hs43}
\end{equation}

To begin with, the monopole formula of \Quiver{fig:q3} can be written as:
\begin{equation}
\label{monopoleQ1QB}
   \HS\left(\Ccal(\text{\Quiver{fig:q3}})\right)=\sum_{\vec{B}} \sum_{m_1\geq\dots\geq m_n} P_{\urm(n)}(m_1,\dots,m_n) P_B (\vec{B})\cdot t^{\Delta(m_1,\dots,m_n;\vec{B})}.
\end{equation}
where the magnetic fluxes are restricted to the principle Weyl chamber, and $P_{\urm(n)}(m_1,\dots,m_n)$ are the dressing factors of $\urm(n)$. The value of $P_{\urm(n)}$ is determined by the number of identical magnetic fluxes \cite{Cremonesi:2013lqa}. If the values of the fluxes $(m_1,\dots,m_n)$ satisfy $m_{i_1}=\dots=m_{i_{\sigma_1}}>\dots> m_{i_{n-\sigma_k+1}}=\dots=m_{i_n}$, then $P_{\urm(n)}(m_1,\dots,m_n)=\frac{1}{(1-t^2)\dots(1-t^{2\sigma_1})}\dots \frac{1}{(1-t^2)\dots(1-t^{2\sigma_k})}$.
A short-hand notation for the fixed loci of vector of magnetic fluxes under $\sigma$ is introduced as:
\begin{subequations}
\begin{alignat}{2}
&(m,\sigma): &\qquad
m_{i_1}&=\ldots=m_{i_{\sigma_1}}\,, \ldots\,, m_{i_{n-\sigma_k+1}}=\ldots=m_{i_n} \,,\\
&(m;\sigma): &\qquad 
m_{i_1}&=\ldots=m_{i_{\sigma_1}}> \ldots> m_{i_{n-\sigma_k+1}}=\ldots=m_{i_n} 
\;,  \; \substack{ \text{\small{with all possible orderings}} \\ 
\text{\small{of the cycles in $\sigma$ \,.}}
}
\end{alignat}
\end{subequations}
For example, $\left(m,(1\ 3)(2\ 5\ 4)\right)$ means $m_1=m_3,\ m_2=m_5=m_4$, and $\left(m;(1\ 3)(2\ 5\ 4)\right)$ means $m_1=m_3>m_2=m_5=m_4\ \bigcup\  m_2=m_5=m_4>m_1=m_3$. In addition $\sum_{(m,\sigma)}$ and $\sum_{(m;\sigma)}$ denotes the sum of the magnetic flux with respect to these conditions.

Without restricting the monopole formula \eqref{monopoleQ1QB} to the principal Weyl chamber, it can be expressed as:
\begin{equation}
\label{monopoleQ1QBweyl}
    \HS\left(\Ccal(\text{\Quiver{fig:q3}})\right)=\sum_{\vec{B}} \sum_{\sigma\in \Sfrak_n} \frac{\sigma_1!\cdots\sigma_k!}{n!} \sum_{(m;\sigma)} \frac{1}{(1-t^2)\cdots(1-t^{2\sigma_1})}\cdots \frac{1}{(1-t^2)\cdots(1-t^{2\sigma_k})} P_B (\vec{B})\cdot t^{\Delta(m_1,\dots,m_n;\vec{B})}.
\end{equation}
With \eqref{dressingweyl}, the monopole formula \eqref{monopoleQ1QBweyl} is rewritten as:
\begin{align}
\label{lastQ1QB}
    \HS\left(\Ccal(\text{\Quiver{fig:q3}})\right)&=\sum_{\vec{B}} \sum_{\sigma\in \Sfrak_n} \frac{\sigma_1!\cdots\sigma_k!}{n!} \sum_{(m;\sigma)} \left(\prod_{\sigma_i\in\sigma} 
    \left(\frac{1}{\sigma_i!}
    \sum_{\rho'\in \Sfrak_{\sigma_i}}
    \frac{1}{(1-t^{2\rho'_1})\cdots(1-t^{2\rho'_{l'}})}
    \right)
    \right) P_B (\vec{B})\cdot t^{\Delta(m_1,\dots,m_n;\vec{B})}
    \notag \\
    &=\frac{1}{n!} \sum_{\vec{B}} \sum_{\sigma\in \Sfrak_n} \sum_{\rho\in \Sfrak_{\sigma_1}\times\ldots\times \Sfrak_{\sigma_k}}\sum_{(m;\sigma)}\frac{1}{(1-t^{2\rho_1})\cdots(1-t^{2\rho_{l}})} P_B (\vec{B})\cdot t^{\Delta(m_1,\dots,m_n;\vec{B})}
    \notag \\
    &=\frac{1}{n!} \sum_{\vec{B}} \sum_{\rho\in \Sfrak_n} \sum_{\sigma\ni\rho}\sum_{(m;\sigma)} \frac{1}{(1-t^{2\rho_1})\cdots(1-t^{2\rho_{l}})} P_B (\vec{B})\cdot t^{\Delta(m_1,\dots,m_n;\vec{B})}.
\end{align}
Here $(\rho'_1,\dots,\rho'_{l'})$ are $\sigma_i$-partition corresponding to the sizes of cycles of $\rho'\in \Sfrak_{\sigma_i}$, and $(\rho_1,\dots,\rho_{l})$ are $n$-partition corresponding to the sizes of cycles of $\rho\in\Sfrak_{\sigma_1}\times\ldots\times \Sfrak_{\sigma_k}$, the minimal permutation subgroup containing $\sigma$. A further bit of notation is that the sum over all $\sigma\in\Sfrak_n$ whose minimal permutation subgroup contains $\Sfrak_{\rho_1}\times\ldots\times \Sfrak_{\rho_l}$ is denoted $\sum_{\sigma\ni\rho}$. Moreover, if follows that $\sum_{\sigma\in \Sfrak_n} \sum_{\rho\in \Sfrak_{\sigma_1}\times\ldots\times \Sfrak_{\sigma_k}}$ is equivalent to $\sum_{\rho\in \Sfrak_n} \sum_{\sigma\ni\rho}$.
Also note that:
\begin{equation}
\label{subpartition}
    \sum_{(m,\rho)} t^{\Delta(m_1,\dots,m_n;\vec{B})}=\sum_{\sigma\ni\rho}\sum_{(m;\sigma)} t^{\Delta(m_1,\dots,m_n;\vec{B})},
\end{equation}
for example, $\sum_{\left(m,(1\ 3)(2\ 5\ 4)\right)}=\sum_{m_1=m_3,m_2=m_5=m_4}=\sum_{m_1=m_2=m_3=m_4=m_5}+\sum_{m_1=m_3>m_2=m_5=m_4}$\\$+\sum_{m_1=m_3<m_2=m_5=m_4}$.
\\
Combining \eqref{fixedloci}, \eqref{molien}, \eqref{lastQ1QB} and \eqref{subpartition}, the following HS is obtained:
\begin{align}
    \HS\left(\Ccal(\text{\Quiver{fig:q3}})\right)&=\frac{1}{n!} \sum_{\vec{B}} \sum_{\rho\in \Sfrak_n} \sum_{(m,\rho)} \frac{1}{(1-t^{2\rho_1})\cdots(1-t^{2\rho_{l}})} P_B (\vec{B})\cdot t^{\Delta(m_1,\dots,m_n;\vec{B})}
    \notag \\
    &=\frac{1}{n!}\sum_{\vec{B}} \sum_{\sigma\in \Sfrak_n} \sum_{(m,\sigma)} \frac{1}{(1-t^{2\sigma_1})}\cdots\frac{1}{(1-t^{2\sigma_k})} P_B(\vec{B})\cdot t^{\Delta(m_1,\dots,m_n;\vec{B})} \notag \\
    &= \HS\left(\Ccal(\text{\Quiver{fig:q4}}/\Sfrak_n)\right),
\end{align}
here the first line to the second line is just a relabelling from $\rho$ to $\sigma$. Hence, Claim \eqref{eq2} is proved at the level of the Hilbert series.

\paragraph{Ungauging.}
One can always ungauge on $\mathcal Q_B$ if it is non-empty. However, if $\mathcal{Q}_B$ is empty, one can set an arbitrary $m_i=0$ and take a $\frac{1}{1-t^2}$ factor away from the whole dressing factor. The proof proceeds analogous to the above. Hence, the Claim \eqref{eq1} is also verified on the level of the Hilbert series.

\subsection{\texorpdfstring{Generalised Molien sum of $\Z_q^n$ on the monopole formula}{Zq action on the monopole formula}}
\label{sec:monopolezqn}

In this section, the assumption that the edges between the complete graph and $\mathcal{Q}_B$ and $\mathcal{Q}_B$ itself are simply laced is made. Apart from the outer-automorphism symmetry on the complete graph quiver, there is another discrete action $\Z_{q}^{n}$ acting on the Cartan algebra $\urm(1)^n$ of the gauge algebra. After the $\Z_{q}^{n}$ quotient, the edge multiplicity of edges in the complete graph increase $q$-fold, and the edge connecting complete graph and $\mathcal{Q}_B$ pick up a lace $q$ pointing to $\mathcal{Q}_B$. The edges in the complete graph can also be thought of as becoming double $q$-laced. However, for Abelian nodes, one cannot distinguish between the edge multiplicity and charge multiplicity at the level of the Coulomb branch. If $\mathcal{Q}_B$ is empty or unframed, a $\urm(1)\subset\urm(1)^n$ is decoupled from complete graph, then the action reduced to $\Z_{q}^{n-1}$.

At the level of Hilbert series:
\begin{equation}
\HS\left(\Ccal(\text{\Quiver{fig:zqafter}})\right) 
    =\begin{cases}
\HS\left(\Ccal(\text{\Quiver{fig:zqbefore}}/\Z_{q}^{n-1})\right) &\text{when $\mathcal{Q}_B$ is empty or unframed}\\
\HS\left(\Ccal(\text{\Quiver{fig:zqbefore}}/\Z_{q}^{n})\right) &\text{when $\mathcal{Q}_B$ is framed}
\end{cases} \,.
\end{equation}
To show this, the refinement fugacities are turned back on. The $\Z_{q}^{n}$ acts on these fugacities in the following way:
\begin{align}
\Z_{q}^{n}: \qquad (z_1,\dots,z_n)\to (\omega_{q}^{k_1} z_1,\dots,\omega_{q}^{k_n} z_n) \,,
\label{eq:Z_q^n_on_fug}
\end{align}
where $\omega_{q}$ is the primitive $q$-th root of unity and $(k_1,\dots,k_n)\in\Z_q^n$. Similar to \cite{Mekareeya:2022spm,Nawata:2023rdx}, the Molien formula yields:
\begin{align}
    \HS\left(\Ccal(\text{\Quiver{fig:zqbefore}}/\Z_{q}^{n})\right)&=\frac{1}{q^n}\sum_{k_1,\dots,k_n}\sum_{m_1,\dots,m_n} P(m_1,\dots,m_n) \cdot \omega_{q}^{m_1\cdot k_1} z_1^{m_1} \cdots \omega_{q}^{m_n\cdot k_n} z_n^{m_n} \cdot t^{\Delta(m_1,\dots, m_n)}
    \notag \\
    &=\sum_{m_1,\dots,m_n} P(m_1,\dots,m_n) \cdot z_1^{q\cdot m_1} \cdots z_n^{q\cdot m_n} \cdot t^{\Delta\left(q\cdot m_1,\dots, q\cdot m_n\right)}
    \notag \\
    &=\HS\left(\Ccal(\text{\Quiver{fig:zqafter}})\right).
    \label{moliensumzq}
\end{align}
Considering the ungauging process: if $\mathcal{Q}_B$ is framed, there is no need to ungauge; if $\mathcal{Q}_B$ is empty or unframed, one can set one $m_i=0$ (or other choices so long as it cancels the same shift invariance), as a result, the $\Z_{q}$ action on $z_i$ is trivial and the overall action becomes $\Z_{q}^{n-1}$:
\begin{equation}
    \HS\left(\Ccal(\text{\Quiver{fig:zqbefore}}/\Z_{q}^{n-1})\right)\vert_{m_i=0}=\HS\left(\Ccal(\text{\Quiver{fig:zqafter}})\right)\vert_{m_i=0}.
\end{equation}
Hence, Claims \eqref{eqzq} and \eqref{eqzqb} are verified on the level of the Hilbert series.

As we can see from the proof, this $\Z_q$ quotient has two origin: one is to increase the edge multiplicity $q$-times, the other is to increase the charge of hypermultiplets $q$-times. On the Coulomb branch geometry these two constructions have no difference, but can be distinguished by deformation and the corresponding Higgs branch.

\section{Abelianisation}
\label{sec:abel}

In this section the discrete action on the Coulomb branch of a 3d $\Ncal=4$ theory with unitary gauge groups via the Abelianisation method introduced in \cite{Bullimore:2015lsa} is discussed. From the monopole formula it is easy to read the representation of the monopole operators under continuous global symmetry, but it is hard to find the exact representation under the discrete action. Abelianisation provides a systemic way to find out the discrete action induced by quiver outer-automorphism.

Start with an Abelian theory of rank $n$, i.e, the gauge group is $\urm(1)^n$, with hypermultiplets with charge $\rho\oplus\bar{\rho}$.

The monopole operators transform under the topological $\urm(1)_{top}^n$. For each topological charge $A\in \Z^n$, there exists a (bare) monopole operator $v_{A}$. The Coulomb branch chiral ring is generated by the set of bare monopole operators $v_{A}$ and the set of Casimir invariants $\phi_i$ of gauge group, which are subject to relations determined by the charge matrix $\rho$.

The simplest example is the theory of a single $\urm(1)$ with no hypermultiplet. The (bare) monopole operator can be written as a holomorphic function of $v_1$ and $v_{-1}$: for a positive integer $k$, $v_k=v_1^k$ and $v_{-k}=v_{-1}^k$. Hence, the chiral ring generators are $v_1,v_{-1},\phi$. The relation between generators is $v_1v_{-1}=1$. The ring $\C[v_1,v_{-1},\phi]/\langle v_1v_{-1}=1\rangle$ is exactly $T^*\C^\times$.

For a non-Abelian theory with unitary gauge group $G$ of rank $n$, if the representation of hypermultiplets contains at least all the roots of $G$, one can always find a corresponding Abelian theory whose gauge group is the Cartan subgroup of the non-Abelian gauge group \cite{Teleman:2018wac}. The charges of the hypermultiplets in this Abelian theory should add up to the charges of the representations of hypermultiplets under the Cartan subgroup, i.e.\ the bare monopole operators in both theories have the same conformal dimension contribution. The difference between the two Coulomb branch chiral rings is a quotient of the Weyl group $\mathcal{W}_G$.

For the most general case, the Weyl group action on the Coulomb branch is:
\begin{equation}
    \phi_i\longleftrightarrow\phi_{\sigma(i)}\; , \qquad  v_{(n_1,\dots,n_k)}\longleftrightarrow v_{(n_{\sigma(1)},\dots,n_{\sigma(k)})} \; .
\end{equation}

\section{\texorpdfstring{Examples of $\Sfrak_n$ quotients}{Examples of Sn quotients}}
\label{sec:SnExamples}
The $\Sfrak_n$ gauging on quivers which are or contain certain complete graphs are computed and the corresponding $\Sfrak_n$ on their Coulomb branches are studied. This action is studied through the Hilbert series computed with the monopole formula and through Abelianisation for each example.

\subsection{\texorpdfstring{From Kleinian $A$ to Kleinian $D$ -- An $\Sfrak_2$ gauging}{From Kleinian A to Kleinian D -- An S2 gauging}}
\label{sec:AtoD}

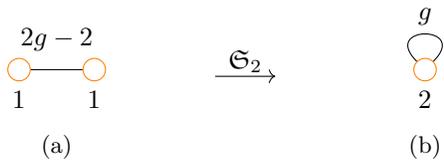
\begin{figure}[H]
    \centering
    
   \begin{subfigure}[t]{0.2\textwidth}
      \centering
    \begin{tikzpicture}
\node[gauge,label=below:{$1$},color=orange!100] (0) at (0,0) {};
            \node[gauge,label=below:{$1$},color=orange!100] (1) at (1,0) {};
            \node[label=above:{$2g-2$}] (3) at (0.5,0) {};
            \draw (0)--(1);
    \end{tikzpicture}
    \caption{}
    \label{fig:quiverA}
   \end{subfigure}
   \begin{subfigure}[t]{0.1\textwidth}
 \centering
 \raisebox{0.8\height}{
    \begin{tikzpicture}
        \draw[->]        (0,0)   -- (0.8,0);
        \node[] at (0.4,0.2) {$\Sfrak_2$};
    \end{tikzpicture}}
    \end{subfigure}
    \begin{subfigure}[t]{0.2\textwidth}
 \centering
    \begin{tikzpicture}
        \node[gauge, label=below:{$2$},color=orange!100] (2) []{};
        \draw (2) to [out=135, in=45,looseness=8] node[pos=0.5,above]{$g$} (2);
    \end{tikzpicture}
    \caption{}
    \label{fig:quiverD}
   \end{subfigure}
   
    \caption{\subref{fig:quiverA}: The Abelian quiver $\CG_{2,2g-2}$, whose Coulomb branch is the $A_{2g-3}$ surface singularity. \subref{fig:quiverD}: The quiver $\ML_{2,g}$, whose Coulomb branch is the $D_{g+1}$ Kleinian singularity.}
    \label{fig:quiverAD}
\end{figure}

Consider the Abelian quiver $\text{\Quiver{fig:quiverA}}=\CG_{2,2g-2}$ of Figure \ref{fig:quiverA}, whose Coulomb branch is the $A$-type singularity $A_{2g-3}\cong\C^2/\Z_{2g-2}$. The Coulomb branch of the quiver \Quiver{fig:quiverD} in Figure \ref{fig:quiverD} is the $D$-type singularity $D_{g+1}\cong\C^2/\mathrm{Dic}_{g-1}$. Embracing \eqref{eq2}, one is led to an interesting relation between the two quivers of \Figref{fig:quiverAD}: as a complete graph, \Quiver{fig:quiverA} has an $\Sfrak_2$ outer-automorphism. By gauging this $\Sfrak_2$, in view of \eqref{eq2}, \Quiver{fig:quiverD} is obtained, a quiver with a single $\urm(2)$ gauge node and with $g$ hypermultiplets in the adjoint representation. Hence, the following relationship between the Coulomb branches of these two quivers is found:
\begin{equation}
\Ccal(\text{\Quiver{fig:quiverD}}) = \Ccal(\text{\Quiver{fig:quiverA}})/\Sfrak_2.
\label{ADs2}
\end{equation}
This, in fact, reproduces the well-known relation between the two Kleinian singularities:

\begin{equation}
D_{g+1}\cong A_{2g-3}/\Sfrak_2 \, ,
\end{equation}
but now it is phrased as an operation on quivers.

\subsubsection*{Monopole formula}
Let us examine the relationship \eqref{ADs2} using the Hilbert series, which for these singularities are:
\begin{subequations}
\begin{align}
\HS\left(A_{2g-3}\right)&=\PE\left[t^2+\left(q+\frac{1}{q}\right)t^{2g-2}-t^{4g-4}\right] \label{Atype}\\
\HS\left(D_{g+1}\right)&=\PE\left[t^4+t^{2g-2}+t^{2g}-t^{4g}\right], \label{Dtype}
\end{align}
\end{subequations}
where $q$ is the fugacity of the topological $\urm(1)_J$. The global symmetry on $A_{2g-3}$ is $\urm(1)_J$, except for the special case of $g=2$, where the global symmetry is enhanced to $\surm(2)$. There is no continuous global symmetry on $D_{g+1}$, since the $\Sfrak_2$ action breaks the $\urm(1)_J$ symmetry. 

Now the result is verified further from the monopole formula and generalised Molien sum.  The magnetic fluxes of \Quiver{fig:quiverA} are denoted as $m_1$ and $m_2$, and the overall $\urm(1)$ is ungauged by taking $m_2=0$. The monopole formula is:
\begin{equation}
    \HS\left(\Ccal(\text{\Quiver{fig:quiverA}})\right)=\frac{1}{(1-t^2)^2}\sum_{m_1=-\infty}^\infty q^{m_1+m_2}\cdot t^{(2g-2)\vert m_1-m_2 \vert}\vert_{m_2=0}=\PE\left[t^2+\left(q+\frac{1}{q}\right)t^{2g-2}-t^{4g-4}\right],
\end{equation}
reproducing the result in \eqref{Atype}. Using the same notation, the monopole formula of \Quiver{fig:quiverD} is:
\begin{equation}
    \HS\left(\Ccal(\text{\Quiver{fig:quiverD}})\right)=\frac{1}{1-t^4}+\frac{1}{(1-t^2)^2}\sum_{m_1>m_2}^\infty t^{(2g-2)\vert m_1-m_2 \vert}\vert_{m_2=0}=\PE\left[t^4+t^{2g-2}+t^{2g}-t^{4g}\right],
\end{equation}
reproducing the result in \eqref{Dtype}.
Let the non-trivial action of $\Sfrak_2$ be denoted as $(1\ 2)$. Using \eqref{fixedloci}, the monopole formula on the fixed loci $m_1=m_2=0$ of $(1\ 2)$ is:
\begin{align}
    \HS\left(\Ccal(\text{\Quiver{fig:quiverA}})\right)^{(1\ 2)}
    &=\frac{1-t^2}{1-t^4}\left(t^{(2g-2)\vert m_1-m_2\vert}\right)\vert_{m_1=m_2=0}
    \notag\\
    &=\frac{1}{1+t^2}
\end{align}
By applying the generalised Molien sum \eqref{molien}, there is agreement that:
\begin{align}
    \frac{1}{2}\left(\HS\left(\Ccal(\text{\Quiver{fig:quiverA}})\right)\vert_{q=1}+\HS\left(\Ccal(\text{\Quiver{fig:quiverA}})\right)^{(1\ 2)}\right)&=\frac{1}{2}\left(\frac{1-t^{4g-4}}{(1-t^2)(1-t^{2g-2})^2}+\frac{1}{1+t^2}\right)
    \notag \\
    &=\PE\left[t^4+t^{2g-2}+t^{2g}-t^{4g}\right]
    \notag \\
    &=\HS\left(\Ccal(\text{\Quiver{fig:quiverD}})\right).
\end{align}
From the PE, the numbers and representations/charges of the generators and relations at each degree are read off. From \eqref{Atype}, $A_{2g-3}$ has one generator at degree $2$ with charge $0$, two generators at degree $2g-2$ with charge $+1$ and $-1$, and one relation between them at degree $4g-4$ with charge $0$. From \eqref{Dtype}, $D_{g+1}$ has three generators at degree $4$, $2g-2$, and $2g$ respectively; and one relation between them at degree $4g$. The explicit generators and relations for the $A_{2g-3}$ and $D_{g+1}$ singularities are given in Table \ref{tab:A-type} and Table \ref{tab:D-type}, respectively.

\begin{table}[H]
    \centering

\begin{subtable}{0.485\textwidth}
\centering
\scalebox{1}{
\begin{tabular}{ccc}
\toprule 
Generators& $\urm(1)_J$ Charge &Degree  \\ \midrule

$\phi$ & $0$ & $2$  \\  
 $u$ & $1$ &$2g-2$  \\  
  $v$ & $-1$ &$2g-2$ \\ \addlinespace  \midrule
Relation & & \\ \midrule
 $\phi^{2g-2}+uv=0$ & $0$ &$4g-4$  \\  \bottomrule
\end{tabular}}
\caption{$A_{2g-3}$ singularity}
\label{tab:A-type}
\end{subtable}
\begin{subtable}{0.485\textwidth}
\centering
\scalebox{1}{
\begin{tabular}{ccc}
\toprule
 Generators & Degree  \\ \midrule

 $x$&$4$  \\  
 $w$ &$2g-2$ \\ 
 $z$&$2g$  \\ \midrule 
 Relation & \\ \midrule
 $z^2-xw^2+x^g=0$&$4g$  \\  \bottomrule
\end{tabular}}
\caption{$D_{g+1}$ singularity}
\label{tab:D-type}
\end{subtable}
   \caption{Generators and relations: \subref{tab:A-type} for Kleinian $A_{2g-3}$; and \subref{tab:D-type} for Kleinian $D_{g+1}$.}
   \label{tab:A_and_D}
\end{table}

Now let us construct the $D_{g+1}$ surface singularity starting from the Kleinian $A_{2g-3}$ singularity. The action of $\Sfrak_2$ on the generators of $A_{2g-3}$ is given by
\begin{equation}
    \phi\to-\phi\, ,\quad u\to v \,,\quad v\to u \,.
\label{actionAD}
\end{equation}
There are four fundamental invariants under this $\Sfrak_2$ action: $\phi^2$, $u+v$, $\phi(u-v)$, and $uv$. Together with the relations there are three independent generators: $\phi^2=x$, $u+v=2w$, and $\phi(u-v)=2z$. These have the same degrees as the generators of $D_{g+1}$. Moreover, the relation between $x$, $w$, and $z$ is indeed $z^2-xw^2+x^g=0$, which coincides with the relation of the $D_{g+1}$ singularity, cf.\ Table \ref{tab:D-type}.

\subsubsection*{Abelianisation}
Now let us verify the generators and relations of the monopole operators via Abelianisation, as described in Section~\ref{sec:abel}.
For the Abelian theory $\text{\Quiver{fig:quiverA}}=\CG_{2,2g-2}$, the Casimir operators of the gauge group $\urm(1)^2$ is labelled as $\phi_1$ and $\phi_2$, and the general monopole operators as $v_{(n_1,n_2)}$, where $(n_1,n_2)\in \Z^2$.  The chiral ring relations between these operators are:
\begin{subequations}
\begin{align}
v_{(n_1,n_2)}v_{(m_1,m_2)}&=v_{(n_1+m_1,n_2+m_2)}(\phi_1-\phi_2)^{(2g-2)\text{ABS}_+(n_1-n_2,m_1-m_2)}  \,,\\
    \text{with} \quad 
    \text{ABS}_+(n,m)&\coloneqq \max(n,0) + \max(m,0) -\max (n+m,0) \,.
\end{align}    
\end{subequations}
The defining Coulomb branch relations are:
\begin{subequations}
    \begin{align}
        v_{(1,1)}v_{(-1,-1)} &=1 \label{free1} \;,\\
        v_{(1,0)}v_{(-1,0)}&=(\phi_1-\phi_2)^{2g-2} \label{1stA} \;,\\
        v_{(0,1)}v_{(0,-1)}&=(\phi_1-\phi_2)^{2g-2} \label{2ndA} \;,\\
        v_{(1,0)}v_{(0,1)}&=v_{(1,1)} (\phi_1-\phi_2)^{2g-2}\label{1stdegree} \;,\\
        v_{(-1,0)}v_{(0,-1)}&=v_{(-1,-1)} (\phi_1-\phi_2)^{2g-2}\label{2nddegree} \;,\\
        v_{(kn_1,kn_2)}&=v_{(n_1,n_2)}^k \label{holo} \;.
    \end{align}
\end{subequations}
Here, \eqref{holo} reflects the holomorphicity of the chiral ring.
From \eqref{free1} it is clear the freely acting diagonal $\urm(1)$ contributes to the relation of $T^*\C^\times$: $\C/[v_{(1,1)},v_{(-1,-1)},\phi_1+\phi_2]/\langle v_{(1,1)}v_{(-1,-1)}=1 \rangle$. Next, \eqref{1stA} is the relation of the $A_{2g-3}$ singularity: $\C/[v_{(1,0)},v_{(-1,0)},\phi_1-\phi_2]/\langle v_{(1,0)}v_{(-1,0)}=(\phi_1-\phi_2)^{2g-2} \rangle$. The generators $v_{(1,0)},\ v_{(-1,0)},\ \phi_1-\phi_2$ can be identified with $u,\ v,\ \phi$, respectively, as in Table \ref{tab:A-type}. The relation \eqref{2ndA} is then derived from \eqref{free1} and \eqref{1stA}
\begin{equation}
v_{(0,1)}v_{(0,-1)}=v_{(1,1)}v_{(-1,0)}v_{(-1,-1)}v_{(1,0)}=(\phi_1-\phi_2)^{2g-2}\,.
\end{equation}
Hence, the overall Coulomb branch of \Quiver{fig:quiverA} is $T^*\C^\times \times A_{2g-3}$, in which $T^*\C^\times$ can be decoupled via the ungauging process. After the ungauging there is an identification $v_{n_1+k,n_2+k}\sim v_{n_1,n_2}$.

According to the statement above, the Coulomb branch of non-Abelian theory \Quiver{fig:quiverD} is a quotient of the Coulomb branch of \Quiver{fig:quiverA} by the Weyl group $\Sfrak_2$. The nontrivial action of $\Sfrak_2$ is simply permuting the two $\urm(1)$s in the Cartan subgroup:
\begin{equation}
    v_{(n_1,n_2)}\longleftrightarrow v_{(n_2,n_1)}\;,\quad \phi_1\longleftrightarrow \phi_2\;,
\end{equation}
which agrees with \eqref{actionAD}.
The free part $T^*\C^\times$ is invariant under this action, and the singular part $A_{2g-3}$ becomes $D_{g+1}$. Therefore, the overall Coulomb branch of \Quiver{fig:quiverD} is $T^*\C^\times \times D_{g+1}$.

\subsection{\texorpdfstring{$A_3$ affine quiver with $\Sfrak_2$ gauging}{a3 nilpotent}}
\label{sec:a3s2}

A similar analysis is repeated for the affine $A_3$ quiver. The Coulomb branch is the closure of the minimal nilpotent orbit of $\mathfrak{sl}_4$ which is denoted $a_3$.
\begin{figure}[H]
    \centering
    
   \begin{subfigure}[t]{0.2\textwidth}
      \centering
    \begin{tikzpicture}
\node[gauge,label=below:{$1$},color=orange!100] (0) at (0,0) {};
            \node[gauge,label=below:{$1$},color=cyan!100] (1) at (1,0) {};
            \node[gauge,label=above:{$1$},color=orange!100] (2) at (1,1) {};
            \node[gauge,label=above:{$1$},color=cyan!100] (3) at (0,1) {};
            \draw (0)--(1)--(2)--(3)--(0);
    \end{tikzpicture}
    \caption{}
    \label{fig:quiverA3}
   \end{subfigure}
   \begin{subfigure}[t]{0.1\textwidth}
 \centering
 \raisebox{0.8\height}{
    \begin{tikzpicture}
        \draw[->]        (0,0)   -- (0.8,0);
        \node[] at (0.4,0.2) {$\Sfrak_2$};
    \end{tikzpicture}}
    \end{subfigure}
    \begin{subfigure}[t]{0.2\textwidth}
 \centering
    \begin{tikzpicture}
    \node[gauge,label=below:{$1$},color=cyan!100] (0) at (0,0) {};
        \node[gauge,label=below:{$2$},color=orange!100] (1) at (1,0) {};
        \node[gauge,label=below:{$1$},color=cyan!100] (2) at (2,0) {};
        \draw (1) to [out=135, in=45,looseness=8] node[pos=0.5,above]{} (1);
        \draw[] (0)--(1) (1)--(2);
    \end{tikzpicture}
    \caption{}
    \label{fig:quiverA3s2}
   \end{subfigure}
      \begin{subfigure}[t]{0.1\textwidth}
 \centering
 \raisebox{0.8\height}{
    \begin{tikzpicture}
        \draw[->]        (0,0)   -- (0.8,0);
        \node[] at (0.4,0.2) {$\Sfrak_2$};
    \end{tikzpicture}}
    \end{subfigure}
    \begin{subfigure}[t]{0.2\textwidth}
 \centering
    \begin{tikzpicture}
    \node[gauge,label=below:{$2$},color=cyan!100] (0) at (0,0) {};
        \node[gauge,label=below:{$2$},color=orange!100] (1) at (1,0) {};
        \draw (0) to [out=135, in=45,looseness=8] node[pos=0.5,above]{} (0);
        \draw (1) to [out=135, in=45,looseness=8] node[pos=0.5,above]{} (1);
        \draw[] (0)--(1);
    \end{tikzpicture}
    \caption{}
    \label{fig:quiverA3s22}
   \end{subfigure}
   
    \caption{\subref{fig:quiverA3}: Quiver \Quiver{fig:quiverA3} is the affine $A_3$ quiver. The Coulomb branch is the closure of minimal nilpotent orbit $a_3$. The \textcolor{orange}{orange} and \textcolor{cyan}{cyan} parts are two $\CG_{2,0}$ sub-graphs. \subref{fig:quiverA3s2}: Non-Abelian quiver \Quiver{fig:quiverA3s2}, whose Coulomb branch is the closure of next-to-minimal nilpotent orbit $\overline{n.min. B_2}$. \subref{fig:quiverA3s22}: Non-Abelian quiver \Quiver{fig:quiverA3s22}.}
    \label{fig:quiverA3s2s}
\end{figure}

An $\Sfrak_2$ quotient may be taken on this quiver if a pair of $\urm(1)$s diagonally opposite are treated as a sub-complete graph of $\CG_{2,0}$ on which the $\Sfrak_2$ gauging occurs. Note that this is the only possibility, as the alternative, which takes a pair of $\urm(1)$ on the same side as $\CG_{2,1}$ for the gauging, has odd edge multiplicity. Importantly, the magnetic quiver for $\overline{n.min. B_2}$ is found, which is consistent with the result of Kostant and Brylinski \cite{1992math......4227B} and also consistent with previous studies on discrete actions on Coulomb branches of $3d$ $\mathcal N=4$ quiver gauge theories \cite{Bourget:2020bxh}.
Furthermore, since there are two pairs of diagonally opposite $\urm(1)$s as two sub-complete graphs of $\CG_{2,0}$, an $\Sfrak_2\times\Sfrak_2$ quotient may also be taken, the quivers are shown in \Figref{fig:quiverA3s2s}.

\subsubsection*{Monopole formula}
Let us derive the chiral ring relations of these three quivers via Hilbert series and show how the $\Sfrak_2$ and $\Sfrak_2\times\Sfrak_2$ quotients act on the generators. The unrefined Hilbert series

read:
\begin{subequations}
\begin{align}
\HS\left(\mathcal C\left(\text{\Quiver{fig:quiverA3}}\right)\right)&=\frac{(1 + t^2) (1 + 8 t^2 + t^4)}{(1-t^2)^6} \label{hsa3}\\
\HS\left(\mathcal C\left(\text{\Quiver{fig:quiverA3s2}}\right)\right)&=\frac{(1 + t^2) (1 + 3 t^2 + t^4)}{(1-t^2)^6} \label{hsa3s2}\\
\HS\left(\mathcal C\left(\text{\Quiver{fig:quiverA3s22}}\right)\right)&=\frac{1 + 3 t^2 + 11 t^4 + 10 t^6 + 11 t^8 + 3 t^{10} + t^{12}}{(1-t^2)^3(1-t^4)^3}. \label{}
\end{align}
\end{subequations}
The coefficient of $t^2$ term indicates global symmetry on the Coulomb branches are $\surm(4)$, $\sorm(5)\cong\usprm(4)$ and $\surm(2)\times\surm(2)$ respectively.

It is straightforward to show that these Hilbert series follow from the Molien sum relation \eqref{molien}:
\begin{subequations}
\begin{align}
\HS\left(\mathcal C\left(\text{\Quiver{fig:quiverA3}}\right)\right)^{(1\ 3)}=\HS\left(\mathcal C\left(\text{\Quiver{fig:quiverA3}}\right)\right)^{(2\ 4)}&=\frac{1 + t^2}{(1-t^2)^4}\label{}\\
\HS\left(\mathcal C\left(\text{\Quiver{fig:quiverA3}}\right)\right)^{(1\ 3)(2\ 4)}&=\frac{1+t^4}{(1+t^2)(1-t^4)^2}\label{}\\
\frac{1}{2}\left(\HS\left(\mathcal C\left(\text{\Quiver{fig:quiverA3}}\right)\right)+\HS\left(\mathcal C\left(\text{\Quiver{fig:quiverA3}}\right)\right)^{(1\ 3)}\right)&=\frac{1}{2}\left(\HS\left(\mathcal C\left(\text{\Quiver{fig:quiverA3}}\right)\right)+\HS\left(\mathcal C\left(\text{\Quiver{fig:quiverA3}}\right)\right)^{(2\ 4)}\right) \notag\\
&=\frac{1}{2}\left(\frac{(1 + t^2) (1 + 8 t^2 + t^4)}{(1-t^2)^6}+\frac{1 + t^2}{(1-t^2)^4}\right) \notag\\
&=\frac{(1 + t^2) (1 + 3 t^2 + t^4)}{(1-t^2)^6} \notag\\
&=\HS\left(\mathcal C\left(\text{\Quiver{fig:quiverA3s2}}\right)\right)\label{}\\
\frac{1}{4}\left(\HS\left(\mathcal C\left(\text{\Quiver{fig:quiverA3}}\right)\right)+\HS\left(\mathcal C\left(\text{\Quiver{fig:quiverA3}}\right)\right)^{(1\ 3)}\right. &\left.+\HS\left(\mathcal C\left(\text{\Quiver{fig:quiverA3}}\right)\right)^{(2\ 4)}+\HS\left(\mathcal C\left(\text{\Quiver{fig:quiverA3}}\right)\right)^{(1\ 3)(2\ 4)}\right) \notag\\
&=\frac{1}{4}\left(\frac{(1 + t^2) (1 + 8 t^2 + t^4)}{(1-t^2)^6}+2\frac{1 + t^2}{(1-t^2)^4}+\frac{1+t^4}{(1+t^2)(1-t^4)^2}\right) \notag\\
&=\frac{1 + 3 t^2 + 11 t^4 + 10 t^6 + 11 t^8 + 3 t^{10} + t^{12}}{(1-t^2)^3(1-t^4)^3} \notag\\
&=\HS\left(\mathcal C\left(\text{\Quiver{fig:quiverA3s22}}\right)\right).
\end{align}
\end{subequations}
where $(1\ 3)$ represents the permutation action between the \textcolor{orange}{orange} $\urm(1)^2$ and $(2\ 4)$ represents the permutation action between the \textcolor{cyan}{cyan} $\urm(1)^2$.

Now let us extract the information of the independent generators and relations of the first two quivers via the refined PL:
\begin{subequations}
\begin{align}
PL(\eqref{hsa3})&=[1,0,1]t^2-[0,1,0]t^4+O(t^5) \,,\\
PL(\eqref{hsa3s2})&=[2,0]t^2-([0,0]+[0,1])t^4+O(t^5) \,.
\end{align}
\end{subequations}
As a result, the explicit generators and relations are derived as shown in Tables \ref{tab:a3relations} and \ref{tab:a3s2relations}.

\begin{table}[H]
\centering
\scalebox{1}{
\begin{tabular}{ccc}
\toprule
Generators & Reps of $\surm(4)$ & Degree\\ \midrule 
$M_i^j$ & $[1,0,1]+[0,0,0]$ & $2$ \\ \addlinespace \midrule
Relations & & \\ \midrule
$\text{Tr}(M)=0$ & $[0,0,0]+[0,2,0]+[1,0,1]$ & $2$ \\  
$M_i^jM_k^l-M_i^lM_k^j=0\ \left(\text{rk}(M)\leq1\right)$ & $[0,1,0]$ & $4$ \\ \bottomrule
\end{tabular}}
\caption{Generators and relations of Coulomb branch of \Quiver{fig:quiverA3}, which is $a_3\cong\C[M_i^j]/\langle \text{tr}(M)=0,\ \text{rk}(M)\leq1 \rangle$, where $i,j=1,2,3,4$.}
\label{tab:a3relations}
\end{table}

\begin{table}[H]
\centering
\scalebox{1}{
\begin{tabular}{ccc}
\toprule
Generators & Reps of $\usprm(4)$ & Degree\\ \midrule 
$S_{ij}$ & $[2,0]$ & $2$ \\ \addlinespace \midrule
Relations & & \\ \midrule
$\Omega^{jk} S_{ij} S_{kl}=0$ & $[0,0]+[0,1]$ & $4$ \\ \bottomrule
\end{tabular}}
\caption{Generators and relations of Coulomb branch of \Quiver{fig:quiverA3s2}, which is $a_3/\Sfrak_2$. The global symmetry is $\sorm(5)\cong\usprm(4)$. $\Omega$ is the standard skew-symmetric matrix.}
\label{tab:a3s2relations}
\end{table}
In this case, it is hard to guess the orbifold action on the generators from the Hilbert series alone.

\subsubsection*{Abelianisation}
Alternatively, Abelianisation can be used to inspect the quotients in Figure \ref{fig:quiverA3s2s}.
For the Abelian theory $\text{\Quiver{fig:quiverA3}}$, the Casimir operators of the \textcolor{orange}{orange} $\urm(1)^2$ are labelled as $\phi_1$, $\phi_3$, and for the \textcolor{cyan}{cyan} $\urm(1)^2$ as $\phi_2,\phi_4$. A general monopole operator is denoted by $v_{(n_1,n_2,n_3,n_4)}$, where $(n_1,n_3)\in \Z^2$ are the magnetic fluxes for the \textcolor{orange}{orange} $\urm(1)^2$ gauge nodes and $(n_2,n_4)\in \Z^2$ are the magnetic fluxes for the \textcolor{cyan}{cyan} $\urm(1)^2$ gauge nodes. The chiral ring relations are:
\begin{subequations}
    \begin{align}
v_{(1,1,1,1)}v_{(-1,-1,-1,-1)}&=1 \label{free1A3} \;,\\
        v_{(1,0,0,0)}v_{(-1,0,0,0)}&=(\phi_4-\phi_1)(\phi_1-\phi_2)\label{1stA3}\;,\\
        v_{(0,1,0,0)}v_{(0,-1,0,0)}&=(\phi_1-\phi_2)(\phi_2-\phi_3) \label{2ndA3} \;,\\
        v_{(0,0,1,0)}v_{(0,0,-1,0)}&=(\phi_2-\phi_3)(\phi_3-\phi_4) \label{3rdA3} \;,\\
        v_{(0,0,0,1)}v_{(0,0,0,-1)}&=(\phi_3-\phi_4)(\phi_4-\phi_1) \label{4thA3} \;,\\
        v_{(1,0,0,0)}v_{(0,1,0,0)}&=v_{(1,1,0,0)}(\phi_1-\phi_2) \label{1stdegreeA3} \;,\\
        v_{(0,1,0,0)}v_{(0,0,1,0)}&=v_{(0,1,1,0)}(\phi_2-\phi_3) \label{2nddegreeA3} \;,\\
        v_{(0,0,1,0)}v_{(0,0,0,1)}&=v_{(0,0,1,1)}(\phi_3-\phi_4) \label{3rddegreeA3}\;, \\
        v_{(0,0,0,1)}v_{(1,0,0,0)}&=v_{(1,0,0,1)}(\phi_4-\phi_1) \label{4thdegreeA3} \;,\\
        v_{(1,0,0,0)}v_{(0,1,0,0)}v_{(0,0,1,0)}&=v_{(1,1,1,0)}(\phi_1-\phi_2)(\phi_2-\phi_3) \label{5thdegreeA3} \;,\\
        v_{(0,1,0,0)}v_{(0,0,1,0)}v_{(0,0,0,1)}&=v_{(0,1,1,1)}(\phi_2-\phi_3)(\phi_3-\phi_4) \label{6thdegreeA3} \;,\\
        v_{(0,0,1,0)}v_{(0,0,0,1)}v_{(1,0,0,0)}&=v_{(1,0,1,1)}(\phi_3-\phi_4)(\phi_4-\phi_1) \label{7thdegreeA3} \;,\\
v_{(0,0,0,1)}v_{(1,0,0,0)}v_{(0,1,0,0)}&=v_{(1,1,0,1)}(\phi_4-\phi_1)(\phi_1-\phi_2) \label{8thdegreeA3} \;,\\
v_{(1,0,0,0)}v_{(0,1,0,0)}v_{(0,0,1,0)}v_{(0,0,0,1)}&=v_{(1,1,1,1)}(\phi_1-\phi_2)(\phi_2-\phi_3)(\phi_3-\phi_4)(\phi_4-\phi_1) \label{9thdegreeA3} \;,\\
    v_{(kn_1,kn_2)}&=v_{(n_1,n_2)}^k  \;.
    \end{align}
\end{subequations}
From \eqref{free1A3} it is clear that the freely acting diagonal $\urm(1)$ contributes to the relation of $T^*\C^\times$: 
\newline$\C[v_{(1,1,1,1)},v_{(-1,-1,-1,-1)},\phi_1+\phi_2+\phi_3+\phi_4]/\langle v_{(1,1,1,1)}v_{(-1,-1,-1,-1)}=1 \rangle$. After decoupling this free $\urm(1)$, there is the identification $v_{(n_1+k,n_2+k,n_3+k,n_4+k)}\sim v_{(n_1,n_2,n_3,n_4)}$. By rearranging the remaining generators into matrix form:
\begin{equation}
\begin{pmatrix}
    \phi_1-\phi_2&v_{(0,1,0,0)}&v_{(0,1,1,0)}&v_{(1,0,0,0)}\\
    v_{(0,-1,0,0)}&\phi_2-\phi_3&v_{(0,0,1,0)}&v_{(0,0,1,1)}\\
    v_{(0,-1,-1,0)}&v_{(0,0,-1,0)}&\phi_3-\phi_4&v_{(0,0,0,1)}\\
    v_{(-1,0,0,0)}&v_{(0,0,-1,-1)}&v_{(0,0,0,-1)}&\phi_4-\phi_1
\end{pmatrix},
\end{equation}
The chiral ring relations (\ref{1stA3}--\ref{9thdegreeA3}) hold by requiring the rank of the matrix is no greater than $1$. Hence the generators and relations can be identified with $M_i^j$ in Table \ref{tab:a3relations}.

The non-trivial $\Sfrak_2$ action which  permutes the \textcolor{orange}{orange} $\urm(1)^2$ is given by
\begin{equation}
v_{(n_1,n_2,n_3,n_4)}\longleftrightarrow v_{(n_3,n_2,n_1,n_4)}\, ,\qquad \phi_1\longleftrightarrow \phi_3 \;,
\end{equation}
and the non-trivial $\Sfrak_2$ action permuting the \textcolor{cyan}{cyan} $\urm(1)^2$ reads
\begin{equation}
v_{(n_1,n_2,n_3,n_4)}\longleftrightarrow v_{(n_1,n_4,n_3,n_2)}\, ,\qquad  \phi_2\longleftrightarrow \phi_4 \;.
\end{equation}

Having studied these simple examples, let us turn to another nice consequence of the statements in Section \ref{sec:loops_complete_graphs}, where in the following example a known hypersurface is studied, but in a different light.

\subsection{\texorpdfstring{Symplectic hypersurfaces in 2 quaternionic dimensions and $\Sfrak_2$ gauging}{Symplectic hypersurfaces in 2 quaternionic dimensions and S2 gauging}}
Symplectic singularities which are hypersurfaces are very rare.
According to a conjecture by \cite{Lehn_2011,Yamagishi_2020} these are
\begin{enumerate}
\item Klein singularities (quaternionic dimension 1),
\item Intersections of Slodowy slices in the $\sprm(g)$ nilpotent cone (dimension 2),
\item The transverse slice between the regular and minimal orbits in $G_2$ (dimension 3).
\end{enumerate}
It is easy to construct quivers such that their moduli spaces are symplectic hypersurfaces \cite{ipmuhanany}.

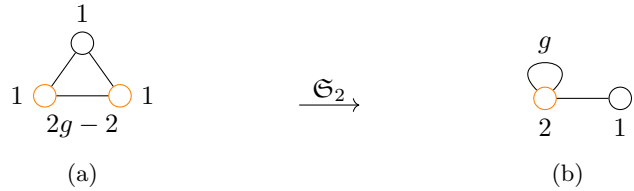
\begin{figure}[H]
    \centering
    
   \begin{subfigure}[t]{0.3\textwidth}
      \centering
    \begin{tikzpicture}
\node[gauge,label=left:{$1$},color=orange!100] (0) at (0,0) {};
            \node[gauge,label=right:{$1$},color=orange!100] (1) at (1,0) {};
            \node[gauge,label=above:{$1$}] (2) at (0.5,0.7) {};
            \node[label=below:{$2g-2$}] at (0.5,0) {};
            \draw (0)--(1);
            \draw (0)--(2);
            \draw (1)--(2);
    \end{tikzpicture}
    \caption{}
    \label{fig:quiver211tri}
   \end{subfigure}
   \begin{subfigure}[t]{0.1\textwidth}
 \centering
 \raisebox{0.8\height}{
    \begin{tikzpicture}
        \draw[->]        (0,0)   -- (0.8,0);
        \node[] at (0.4,0.2) {$\Sfrak_2$};
    \end{tikzpicture}}
    \end{subfigure}
    \begin{subfigure}[t]{0.3\textwidth}
 \centering
    \begin{tikzpicture}
\node[gauge, label=below:{$2$},color=orange!100] (2) []{};
        \draw (2) to [out=135, in=45,looseness=8] node[pos=0.5,above]{$g$} (2);
        \node[gauge,label=below:{$1$}] (1) at (1,0) {};
        \draw (2)--(1);
    \end{tikzpicture}
    \caption{}
    \label{fig:quiver21loop}
   \end{subfigure}
   
    \caption{\subref{fig:quiver211tri}: Quiver \Quiver{fig:quiver211tri} is composed of two parts: the complete graph $\CG_{2,2g-2}$ in \textcolor{orange}{orange}, and the background $\mathcal{Q}_B$, given by the uncoloured $\urm(1)$ node. They are connected by edges of multiplicity $1$. \subref{fig:quiver21loop}: Quiver \Quiver{fig:quiver21loop} is composed of \Quiver{fig:quiverD}, shown in \textcolor{orange}{orange}, connecting to the same background quiver $Q_B$.}
    \label{fig:quivertriloop}
\end{figure}

For case 2, the family of these symplectic hypersurfaces are the Coulomb branches of the 
multiloop quiver \Quiver{fig:quiver21loop} on the RHS of \Figref{fig:quivertriloop}.
This family of hypersurfaces was studied in \cite{Hanany:2010qu} as the Higgs branch of a certain $4d$ $\mathcal N=2$ $A_1$ Class $\mathcal{S}$ theory on a Riemann surface of genus $g$ and 1 puncture.

The Coulomb branch of this quiver was identified in \cite{Finkelberg:2020cb} as $\mathcal{S}^{[2g]}_{[2g-2,1,1]}$, the intersection of the nilpotent cone of $\sprm(g)$ with the slodowy slice to the orbit $[2g-2,1,1]$. Incidentally, this moduli space has a Higgs branch realisation by the $\Sfrak_2$ quotient of the affine $D_{g+1}$ quiver:
\begin{equation}
        \mathcal{S}^{[2g]}_{[2g-2,1,1]}=\mathcal{H}\left(
        \vcenter{\hbox{
        \begin{tikzpicture}
        \node[gauge, label=below:{$2$},color=orange!100] (0) []{};
        \draw (0) to [out=135, in=45,looseness=8] node[pos=0.5,above]{} (0);
        \node[gauge,label=below:{$2$}] (1) at (0.6,0) {};
        \node[gauge,label=below:{$2$}] (2) at (1.6,0) {};
        \node[gauge,label=right:{$1$}] (4) at (2.1,.4) {};
        \node[gauge,label=right:{$1$}] (5) at (2.1,-0.4) {};
        \node[] (3) at (1.1,0) {$\cdots$};
        \draw (0)--(1) (2)--(4) (2)--(5);
        \draw [decorate,decoration={brace,amplitude=5pt}] (0.5,0.3)--(1.7,0.3);
        \node at (1.1,0.7) {$g-2$};
        \end{tikzpicture}}}\right)=        \Ccal\left(
        \vcenter{\hbox{
        \begin{tikzpicture}
        \node[gauge, label=below:{$2$},color=orange!100] (2) []{};
        \draw (2) to [out=135, in=45,looseness=8] node[pos=0.5,above]{$g$} (2);
        \node[gauge,label=below:{$1$}] (1) at (1,0) {};
        \draw (2)--(1);
        \end{tikzpicture}}}\right).
\label{eq:dgs2}
\end{equation}
Notice that the \textcolor{orange}{orange} $\urm(2)$ node of \Quiver{fig:quiver21loop} has loops. According to the claim \eqref{eq2}, the Coulomb branch of \Quiver{fig:quiver21loop} admits a double-cover, whose quiver \Quiver{fig:quiver211tri} is shown on the LHS of \Figref{fig:quivertriloop}:
\begin{equation}
\Ccal\left(\text{\Quiver{fig:quiver211tri}}\right)\cong\Ccal\left(\text{\Quiver{fig:quiver211tri}}\right)/\Sfrak_2.
\end{equation}
In \Quiver{fig:quiver211tri}, the \textcolor{orange}{orange} $\urm(1)$ nodes on the bottom together with the $2g\!-\!2$ edges can be treated as a complete graph. There is a $\Sfrak_2$ symmetry permuting these two nodes. By gauging this $\Sfrak_2$ symmetry, it becomes the \textcolor{orange}{orange} $\urm(2)$ node with $g$ adjoints in quiver \Quiver{fig:quiver21loop}.

\subsubsection*{Monopole formula}

Let us derive the chiral ring relations of these two quivers from Hilbert series and show how this $\Sfrak_2$ quotient acts on the generators. Firstly, let us look at the refined Hilbert series of these two quivers:
\begin{subequations}
\begin{align}
\HS\left(\mathcal C\left(\text{\Quiver{fig:quiver211tri}}\right)\right)&=\left(1+t^2-\left(q+\frac{1}{q}\right)[1]t^{2g+1}-([2]+1)t^{4g-2}+([2]+1)t^{4g}\right.
\notag \\
&\quad \left. +\left(q+\frac{1}{q}\right)[1]t^{6g-3}-t^{8g-4}-t^{8g-2}\right)\cdot \PE\left[([2]+1)t^2+\left(q+\frac{1}{q}\right)[1]t^{2g-1}\right] \label{tri11g}\\
\HS\left(\mathcal C\left(\text{\Quiver{fig:quiver21loop}}\right)\right)&=\PE\left[[2]t^2+[1]t^{2g-1}-t^{4g}\right], \label{loop11g}
\end{align}
\end{subequations}
where $q$ is the fugacity for $\urm(1)$, and $[n]$ refers to the character for the $n+1$ dimensional representation\footnote{Here, for a rank $k$ Lie group, the Dynkin label $[n_1,\dots,n_k]$ is used to represent both the irrep of corresponding highest weight and its character.} of $\surm(2)$. 
The Coulomb branch global symmetry of \Quiver{fig:quiver211tri} is $\urm(2)\cong \surm(2)\times\urm(1)$; while the Coulomb branch global symmetry of \Quiver{fig:quiver21loop} is $\surm(2)$. The $\Sfrak_2$ quotient breaks the $\urm(1)$ global symmetry and leaves the $\surm(2)$ intact.

It is straightforward to verify that the monopole formula satisfies the generalised Molien sum \eqref{molien}:
\begin{equation}
    \HS\left(\mathcal C\left(\text{\Quiver{fig:quiver21loop}}\right)\right)=\frac{1}{2}\left( \HS\left(\mathcal C\left(\text{\Quiver{fig:quiver211tri}}\right)\right)+ \HS\left(\mathcal C\left(\text{\Quiver{fig:quiver211tri}}\right)\right)^{(1\ 2)}\right)\vert_{q=1} \; .
\end{equation}
To extract the information about the independent generators and relations for each variety, the PL of the HS is taken:
\begin{subequations}
\begin{align}
\PL\left(\eqref{tri11g}\right)&=([2]+1)t^2-t^4+\left(q+\frac{1}{q}\right)[1]\ t^{2g-1}-\left(q+\frac{1}{q}\right)[1]\ t^{2g+1} \\
&\qquad \qquad \qquad -([2]+1)t^{4g-2}+\text{O}(t^{4g-1})  \;, \notag\\
\PL\left(\eqref{loop11g}\right)&=[2]\ t^2+[1]\ t^{2g-1}-t^{4g} \;.
\end{align}
\end{subequations}
Building on this, the explicit generators and relations are derived as shown in Tables \ref{tab:trirelation} and \ref{tab:hyprelation}.
\begin{table}[H]
\centering
\scalebox{1}{
\begin{tabular}{ccc}
\toprule
Generators & Reps of $\left(\urm(1),\surm(2)\right)$ & Degree\\ \midrule 
$\Phi$ & $\left(0,[0]\right)$ & $2$ \\  
$M_{ij}$ & $\left(0,[2]\right)$ & $2$ \\  
$X_{i}$ & $\left(1,[1]\right)$ & $2g-1$ \\  
$Y_{i}$ & $\left(-1,[1]\right)$ & $2g-1$ \\ \addlinespace \midrule
Relations & & \\ \midrule
$\epsilon^{ik}\epsilon^{jl} M_{ij} M_{kl}-f_1\Phi^2=0$ & $\left(0,[0]\right)$ & $4$ \\  
$\epsilon^{ik} M_{ij} X_k-f_2\Phi X_j=0$ & $\left(1,[1]\right)$ & $2g+1$ \\  
$\epsilon^{ik} M_{ij} Y_k-f_2\Phi Y_k=0$ & $\left(-1,[1]\right)$ &  $2g+1$ \\  
$X_i Y_j-f_3\Phi^{2g-2} M_{ij}=0$ & $\left(0,[2]\right)$ & $4g-2$ \\  
$\epsilon^{ij} X_iY_j-f_4\Phi^{2g-1}=0$ & $\left(0,[0]\right)$ & $4g-2$ \\  \bottomrule
\end{tabular}}
\caption{Generators and relations of Coulomb branch of \Quiver{fig:quiver211tri}, with $f_{1,2,3,4}$ constants that  depend on $g$.}
\label{tab:trirelation}
\end{table}

\begin{table}[H]
\centering
\scalebox{1}{
\begin{tabular}{ccc}
\toprule
Generators & Reps of $\surm(2)$ & Degree\\ \midrule

$M_{ij}$ & $[2]$ & $2$ \\  
$B_{i}$ & $[1]$ & $2g-1$ \\ \addlinespace \midrule
Relation & & \\ \midrule
$\left(\epsilon^{ik}\epsilon^{jl} M_{ij} M_{kl}\right)^g-f \epsilon^{ik}\epsilon^{jl} M_{ij} B_k B_l=0$ & $[0]$ & $4g$ \\  \bottomrule
\end{tabular}}
\caption{Generators and relations of Coulomb branch of \Quiver{fig:quiver21loop}, where $f$ is a constant depending on $g$.}
\label{tab:hyprelation}
\end{table}

Next, let us construct $\Ccal\left(\text{\Quiver{fig:quiver21loop}}\right)$ from $\Ccal\left(\text{\Quiver{fig:quiver211tri}}\right)$. The $\Sfrak_2$ action on the generators of $\Ccal\left(\text{\Quiver{fig:quiver211tri}}\right)$ is given by:
\begin{equation}
    \Phi\to-\Phi\, ,\quad X_i\to Y_i \, ,\quad Y_i\to X_i \,.
    \label{hyperaction}
\end{equation}
There are five fundamental invariants under the $\Sfrak_2$ action: $M_{ij}$, $\Phi^2$, $X_i+Y_i$, $\Phi(X_i-Y_i)$, and $X_iY_j$. Due to the relations, there are only two independent generators: $M_{ij}$ and $X_i+Y_i=B_i$. Moreover, the relation between $M_{ij}$ and $B_i$ is indeed given by $\left(\epsilon^{ik}\epsilon^{jl} M_{ij} M_{kl}\right)^g-f \epsilon^{ik}\epsilon^{jl} M_{ij} B_k B_l=0$.

\subsubsection*{Abelianisation}

For the Abelian theory $\text{\Quiver{fig:quiver211tri}}$, the Casimir operators of the bottom $\urm(1)^2$ are denoted as $\phi_1$, $\phi_2$, and for the top $\urm(1)$ as $\phi_3$. Likewise, a generic monopole operator is denoted by $v_{(n_1,n_2,n_3)}$, with $(n_1,n_2,n_3)\in \Z^3$. The chiral ring relations for these operators are:
\begin{align} v_{(n_1,n_2,n_3)}v_{(m_1,m_2,m_3)}=&v_{(n_1+m_1,n_2+m_2,n_3+m_3)}(\phi_1-\phi_2)^{(2g-2)\text{ABS}_+(n_1-n_2,m_1-m_2)}\notag \\
    &\cdot(\phi_2-\phi_3)^{\text{ABS}_+(n_2-n_3,m_2-m_3)}(\phi_3-\phi_1)^{\text{ABS}_+(n_3-n_1,m_3-m_1)} \;.
\end{align}
The defining Coulomb branch relations are thus given by:
\begin{subequations}
    \begin{align}
        v_{(1,1,1)}v_{(-1,-1,-1)}&=1 \label{freehp} \;,\\
        v_{(1,0,0)}v_{(-1,0,0)}&=(\phi_1-\phi_2)^{2g-2}(\phi_3-\phi_1) \label{1sthp}\;,\\
        v_{(0,1,0)}v_{(0,-1,0)}&=(\phi_1-\phi_2)^{2g-2}(\phi_2-\phi_3) \label{2ndhp}\;,\\
        v_{(0,0,1)}v_{(0,0,-1)}&=(\phi_2-\phi_3)(\phi_3-\phi_1) \label{3rdhp}\;,\\
        v_{(1,0,0)}v_{(0,1,0)}&=v_{(1,1,0)}(\phi_1-\phi_2)^{2g-2} \label{4thhp}\;,\\
        v_{(-1,0,0)}v_{(0,-1,0)}&=v_{(-1,-1,0)}(\phi_1-\phi_2)^{2g-2} \label{5thhp}\;,\\
        v_{(1,0,0)}v_{(0,0,1)}&=v_{(1,0,1)}(\phi_3-\phi_1) \label{6thhp}\;,\\
        v_{(-1,0,0)}v_{(0,0,-1)}&=v_{(-1,0,-1)}(\phi_3-\phi_1) \label{7thhp}\;,\\
        v_{(0,1,0)}v_{(0,0,1)}&=v_{(0,1,1)}(\phi_2-\phi_3) \label{8thhp}\;,\\
        v_{(0,-1,0)}v_{(0,0,-1)}&=v_{(0,-1,-1)}(\phi_2-\phi_3) \label{9thhp}\;,\\
        v_{(1,0,0)}v_{(0,1,0)}v_{(0,0,1)}&=v_{(1,1,1)}(\phi_1-\phi_2)^{2g-2}(\phi_2-\phi_3)(\phi_3-\phi_1) \label{1stdegreehp} \;,\\
        v_{(-1,0,0)}v_{(0,-1,0)}v_{(0,0,-1)}&=v_{(-1,-1,-1)}(\phi_1-\phi_2)^{2g-2}(\phi_2-\phi_3)(\phi_3-\phi_1)\;,\label{2nddegreehp}\\
        v_{(kn_1,kn_2,kn_3)}&=v_{(n_1,n_2,n_3)}^k \label{holohp} \;.
    \end{align}
\end{subequations}
Here \eqref{holohp} reflects the holomorphicity of the chiral ring.
The freely acting diagonal $\urm(1)$ is seen from \eqref{freehp}, which contributes to the relation of $T^*\C^\times$: $\C/[v_{(1,1,1)},v_{(-1,-1,-1)},\phi_1+\phi_2+\phi_3]/\langle v_{(1,1,1)}v_{(-1,-1,-1)}=1 \rangle$. After decoupling this $T^*\C^\times$, there is the identification $v_{(n_1+k,n_2+k,n_3+k)}\sim v_{(n_1,n_2,n_3)}$. Here $\phi_1-\phi_2,\ v_{(0,0,1)},\ v_{(0 ,0,-1)},\ \phi_1+\phi_2-2\phi_3$ can be rearranged and identified with $\Phi,\ M_{ij}$ in Table \ref{tab:trirelation}. Also there is an arrangement of $v_{(0,1,0)},\ v_{(0,-1,0)},\ v_{(1,0,0)},\ v_{(-1,0,0)}$ into $X_i,\ Y_i$. Under this identification, the relations in Table \ref{tab:trirelation} from (\ref{1sthp}--\ref{9thhp}) are recovered.

According to the statement above, the Coulomb branch of the non-Abelian theory \Quiver{fig:quiver21loop} is a quotient of the Coulomb branch of \Quiver{fig:quiver211tri} by the Weyl group $\Sfrak_2$. The non-trivial $\Sfrak_2$ action is simply permuting the two bottom $\urm(1)$s:
\begin{equation}
    v_{(n_1,n_2,n_3)}\longleftrightarrow v_{(n_2,n_1,n_3)},\ \phi_1\longleftrightarrow \phi_2,
\end{equation}
which agrees with \eqref{hyperaction}. This action leads to the Coulomb branch of \Quiver{fig:quiver21loop}.

\subsection{\texorpdfstring{Complete graph of three nodes -- $\CG_{3,2g-2}$ quiver with $\Sfrak_3$ gauging}{Complete graph of three nodes CG(3,2g-2) with S3 gauging}}
\label{sec:a2s3}

The quiver $\text{\Quiver{fig:quivercg3}}=\CG_{3,2g-2}$ is the magnetic quiver of the $AD$ theory $(A_2,A_{6g-7})$. 

There is a clear $\Sfrak_3$ automorphism acting on \Quiver{fig:quivercg3}. From \eqref{eq2}, by gauging this $\Sfrak_3$, the quiver \Quiver{fig:quiver3g} is found which is a single $\urm(3)$ node with $g$ adjoints, shown in \Figref{fig:quiver3cg3}.
\begin{figure}[H]
    \centering
    
   \begin{subfigure}[t]{0.3\textwidth}
      \centering
    \begin{tikzpicture}
\node[gauge,label=left:{$1$},color=orange!100] (0) at (0,0) {};
            \node[gauge,label=right:{$1$},color=orange!100] (1) at (1.5,0) {};
            \node[gauge,label=above:{$1$},color=orange!100] (2) at (0.75,1.05) {};
            \node[label=below:{$2g-2$},color=orange!100] (3) at (0.75,0) {};
            \draw (0)--(1);
            \draw (0)--(2)node[pos=0.5,above,sloped]{$2g\!-\!2$};
            \draw (1)--(2)node[pos=0.5,above,sloped]{$2g\!-\!2$};
    \end{tikzpicture}
    \caption{}
    \label{fig:quivercg3}
   \end{subfigure}
   \begin{subfigure}[t]{0.1\textwidth}
 \centering
 \raisebox{0.8\height}{
    \begin{tikzpicture}
        \draw[->]        (0,0)   -- (0.8,0);
        \node[] at (0.4,0.2) {$\Sfrak_3$};
    \end{tikzpicture}}
    \end{subfigure}
    \begin{subfigure}[t]{0.3\textwidth}
 \centering
    \begin{tikzpicture}
        \node[gauge, label=below:{$3$},color=orange!100] (2) []{};
        \draw (2) to [out=135, in=45,looseness=8] node[pos=0.5,above]{$g$} (2);
    \end{tikzpicture}
    \caption{}
    \label{fig:quiver3g}
   \end{subfigure}
   
    \caption{\subref{fig:quivercg3}: Abelian quiver $\CG_{3,2g-2}$. \subref{fig:quiver3g}: Quiver $\ML_{3,g}$.}
    \label{fig:quiver3cg3}
\end{figure}

Consequently, there exists the following relation between the Coulomb branches of the two quivers:
\begin{equation}
    \Ccal(\text{\Quiver{fig:quiver3g}})=\Ccal(\text{\Quiver{fig:quivercg3}})/\Sfrak_3.
\end{equation}
Again, Hilbert series and Abelianisation techniques are use in turn to shed light on the fine details.

\subsubsection*{Monopole formula}
Let us examine the generators and relations via their Coulomb branch Hilbert series:
\begin{subequations}
\begin{align}
\HS\left(\mathcal C
\left(\text{\Quiver{fig:quivercg3}}\right)\right)&=\left(1+2t^{4g-4}-\left(z_1+z_2+z_1z_2+\frac{1}{z_1}+\frac{1}{z_2}+\frac{1}{z_1z_2}\right)t^{8g-8}+2t^{12g-12}+t^{16g-16}\right)
\notag \\
&\quad \cdot\PE\left[2t^2+\left(z_1+z_2+z_1z_2+\frac{1}{z_1}+\frac{1}{z_2}+\frac{1}{z_1z_2}\right)t^{4g-4}-2t^{4g-4}\right] \;, \label{cg3}\\
\HS\left(\mathcal C\left(\text{\Quiver{fig:quiver3g}}\right)\right)&=\frac{1+2t^{4g-2}+2t^{4g}+t^{8g-2}}{(1-t^4)(1-t^6)(1-t^{4g-4})^2} \;, \label{3g}
\end{align}
\end{subequations}
here $z_1$ and $z_2$ are the fugacities for the topological $\urm(1)_1\times\urm(1)_2$. The Coulomb branch global symmetry of \Quiver{fig:quivercg3} is $\urm(1)_1\times\urm(1)_2$. The $\Sfrak_3$ quotient breaks the $\urm(1)_1\times\urm(1)_2$ global symmetry, leaving behind no continuous Coulomb branch global symmetry for \Quiver{fig:quiver3g}.

Again, it is straightforward to verify that the generalised Molien sum \eqref{molien} is satisfied:
\begin{align}
    \HS\left(\mathcal C\left(\text{\Quiver{fig:quiver3g}}\right)\right)&=\frac{1}{2}\left( \HS\left(\mathcal C\left(\text{\Quiver{fig:quivercg3}}\right)\right)+ \HS\left(\mathcal C\left(\text{\Quiver{fig:quivercg3}}\right)\right)^{(1\ 2)}+ \HS\left(\mathcal C\left(\text{\Quiver{fig:quivercg3}}\right)\right)^{(2\ 3)}+ \HS\left(\mathcal C\left(\text{\Quiver{fig:quivercg3}}\right)\right)^{(3\ 1)} \right. \notag \\
    &\left.\quad+ \HS\left(\mathcal C\left(\text{\Quiver{fig:quivercg3}}\right)\right)^{(1\ 2\ 3)}+ \HS\left(\mathcal C\left(\text{\Quiver{fig:quivercg3}}\right)\right)^{(1\ 3\ 2)}\right)\vert_{z_1=z_2=1}.
\end{align}
As before, the PL enables us to exact the information on the independent generators and relations:
\begin{subequations}
\begin{align}
\PL\left(\eqref{cg3}\right)&=2t^2+\left(z_1+z_2+z_1z_2+\frac{1}{z_1}+\frac{1}{z_2}+\frac{1}{z_1z_2}\right)t^{4g-4}
\notag \\
&\quad -\left(3+z_1+z_2+z_1z_2+\frac{1}{z_1}+\frac{1}{z_2}+\frac{1}{z_1z_2}\right)t^{8g-8}+\text{O}(t^{8g-7})\\
\PL\left(\eqref{3g}\right)&=t^4+t^6+2t^{4g-4}+2t^{4g-2}+2t^{4g}-3t^{8g-4}-3t^{8g-2}-3t^{8g}+\text{O}(t^{8g+1})\;.
\end{align}
\end{subequations}
The explicit generators and relations are summarised in Tables \ref{tab:cg3relation} and \ref{tab:3grelation}.

\begin{table}[H]
\centering
\scalebox{1}{
\begin{tabular}{ccc}
\toprule 
Generators & Charge of $\urm(1)_1\times\urm(1)_2$ & Degree \\ \midrule 
$\Phi_1$, $\Phi_2$, $\Phi_3$ & $(0,0)$ & $2$ \\ 
$X_1$ & $(1,0)$ & $4g-4$\\
$X_2$ & $(0,1)$ & $4g-4$\\
$X_3$ & $(-1,-1)$ & $4g-4$\\ 
$Y_1$ & $(-1,0)$ & $4g-4$\\ 
$Y_2$ & $(0,-1)$ & $4g-4$\\ 
$Y_3$ & $(1,1)$ & $4g-4$\\ \addlinespace \midrule
Relations & & \\ \midrule
$\Phi_1+\Phi_2+\Phi_3=0$ & $(0,0)$ & $2$ \\
$X_1Y_1-(\Phi_2\Phi_3)^{2g-2}=0$ & $(0,0)$ & $8g-8$ \\
$X_2Y_2-(\Phi_3\Phi_1)^{2g-2}=0$ & $(0,0)$ & $8g-8$ \\
$X_3Y_3-(\Phi_1\Phi_2)^{2g-2}=0$ & $(0,0)$ & $8g-8$ \\
$Y_2Y_3-X_1\Phi_1^{2g-2}=0$ & $(1,0)$ & $8g-8$ \\
$X_2X_3-Y_1\Phi_1^{2g-2}=0$ & $(-1,0)$ & $8g-8$ \\
$Y_1Y_3-X_2\Phi_2^{2g-2}=0$ & $(0,1)$ & $8g-8$ \\
$X_3X_1-Y_2\Phi_2^{2g-2}=0$ & $(0,-1)$ & $8g-8$ \\
$Y_2Y_1-X_3\Phi_3^{2g-2}=0$ & $(-1,-1)$ & $8g-8$ \\
$X_1X_2-Y_3\Phi_3^{2g-2}=0$ & $(1,1)$ & $8g-8$ \\ \bottomrule
\end{tabular}}
\caption{Generators and relations of Coulomb branch of \Quiver{fig:quivercg3}. At degree $2$, instead of two independent generators, three generators are used together with $1$ relation to demonstrate the $\Sfrak_3$ symmetry. When $g=2$, it reduces to the generators and relations of $a_2\cong\C[M_i^j]/\langle \text{tr}(M)=0,\ \text{rk}(M)\leq1 \rangle$, where $i,j=1,2,3$.}
\label{tab:cg3relation}
\end{table}

\begin{table}[H]
\centering
\scalebox{1}{
\begin{tabular}{ccc}
\toprule 
Generators & Degree \\ \midrule 

$p_2$ & $4$ \\
$p_3$ & $6$ \\
$u_1$, $u_2$ & $4g-4$ \\
$v_1$, $v_2$ & $4g-2$ \\
$w_1$, $w_2$ & $4g$ \\ \addlinespace \midrule
Relations & \\ \midrule
$r_1,r_2,r_3$ & $8g-4$ \\
$r_1',r_2',r_3'$ & $8g-2$ \\
$r_1'',r_2'',r_3''$ & $8g$ \\ \bottomrule
\end{tabular}}
\caption{Generators and relations of Coulomb branch of \Quiver{fig:quiver3g}.} 
\label{tab:3grelation}
\end{table}

Next, one can construct $\Ccal\left(\text{\Quiver{fig:quiver3g}}\right)$ from $\Ccal\left(\text{\Quiver{fig:quivercg3}}\right)$ explicitly. The $\Sfrak_3$ action on the generators of $\Ccal\left(\text{\Quiver{fig:quivercg3}}\right)$ is given by:
\begin{equation}
    \Phi_i\to-\Phi_{\sigma(i)} \;,\quad  X_i\to X_{\sigma(i)}\;,\quad Y_i\to Y_{\sigma(i)} \;.
\label{actions3}
\end{equation}
This action is induced from the $\Sfrak_3$ subgroup of outer-automorphism of $\mathrm{SL}(3,\C)$, for more details see Appendix \ref{app:s4}. The generators are given by: $p_2=\Phi_1^2+\Phi_2^2+\Phi_3^2$, $p_3=\Phi_1\Phi_2(\Phi_1-\Phi_2)+\Phi_2\Phi_3(\Phi_2-\Phi_3)+\Phi_3\Phi_1(\Phi_3-\Phi_1)$, $u_1=X_1+X_2+X_3$, $u_2=Y_1+Y_2+Y_3$, $v_1=X_1\Phi_1+X_2\Phi_2+X_3\Phi_3$, $v_2=Y_1\Phi_1+Y_2\Phi_2+Y_3\Phi_3$, $w_1=X_1(\Phi_2-\Phi_3)+X_2(\Phi_3-\Phi_1)+X_3(\Phi_1-\Phi_2)$ and $w_2=Y_1(\Phi_2-\Phi_3)+Y_2(\Phi_3-\Phi_1)+Y_3(\Phi_1-\Phi_2)$.
The $9$ relations $r_1$, $r_2$, $r_3$, $r_1'$, $r_2'$, $r_3'$, $r_1''$, $r_2''$ and $r_3''$ after the quotient can be worked out explicitly for small $g$.

\subsubsection*{Abelianisation}
For the Coulomb branch of $\text{\Quiver{fig:quivercg3}}\cong\CG_{3,2g-2}$, the Casimir operators of gauge group $\urm(1)^3$ are labelled as $\phi_1$, $\phi_2$ and $\phi_3$, and the generic monopole operator as $v_{(n_1,n_2,n_3)}$, with $(n_1,n_2,n_3)\in \Z^3$. The chiral ring relations between these operators are:
\begin{align}
v_{(n_1,n_2,n_3)}v_{(m_1,m_2,m_3)}=&v_{(n_1+m_1,n_2+m_2,n_3+m_3)}(\phi_1-\phi_2)^{(2g-2)\text{ABS}_+(n_1-n_2,m_1-m_2)} \\
    &\cdot(\phi_2-\phi_3)^{(2g-2)\text{ABS}_+(n_2-n_3,m_2-m_3)}(\phi_3-\phi_1)^{(2g-2)\text{ABS}_+(n_3-n_1,m_3-m_1)} \notag
\end{align}
which allows for the extraction of the determining Coulomb branch relations:
\begin{subequations}
    \begin{align}
        v_{(1,1,1)}v_{(-1,-1,-1)}&=1 \label{freea2} \;,\\
        v_{(1,0,0)}v_{(-1,0,0)}&=(\phi_1-\phi_2)^{2g-2}(\phi_3-\phi_1)^{2g-2} \label{1sta2}\;,\\
        v_{(0,1,0)}v_{(0,-1,0)}&=(\phi_1-\phi_2)^{2g-2}(\phi_2-\phi_3)^{2g-2} \label{2nda2}\;,\\
        v_{(0,0,1)}v_{(0,0,-1)}&=(\phi_2-\phi_3)^{2g-2}(\phi_3-\phi_1)^{2g-2}\label{3rda2}\;,\\
        v_{(1,0,0)}v_{(0,1,0)}&=v_{(1,1,0)}(\phi_1-\phi_2)^{2g-2}\label{4tha2}\;,\\
        v_{(0,1,0)}v_{(0,0,1)}&=v_{(0,1,1)}(\phi_2-\phi_3)^{2g-2}\label{5tha2}\;,\\
        v_{(0,0,1)}v_{(1,0,0)}&=v_{(1,0,1)}(\phi_3-\phi_1)^{2g-2}\label{6tha2}\;,\\
        v_{(-1,0,0)}v_{(0,-1,0)}&=v_{(-1,-1,0)}(\phi_1-\phi_2)^{2g-2}\label{7tha2}\;,\\
        v_{(0,-1,0)}v_{(0,0,-1)}&=v_{(0,-1,-1)}(\phi_2-\phi_3)^{2g-2}\label{8tha2}\;,\\
        v_{(0,0,-1)}v_{(-1,0,0)}&=v_{(-1,0,-1)}(\phi_3-\phi_1)^{2g-2}\label{9tha2}\;,\\
        v_{(1,0,0)}v_{(0,1,0)}v_{(0,0,1)}&=v_{(1,1,1)}(\phi_1-\phi_2)^{2g-2}(\phi_2-\phi_3)^{2g-2}(\phi_3-\phi_1)^{2g-2}\;,\\
        v_{(-1,0,0)}v_{(0,-1,0)}v_{(0,0,-1)}&=v_{(-1,-1,-1)}(\phi_1-\phi_2)^{2g-2}(\phi_2-\phi_3)^{2g-2}(\phi_3-\phi_1)^{2g-2} \;,\\
        v_{(kn_1,kn_2,kn_3)}&=v_{(n_1,n_2,n_3)}^k\;.
    \end{align}
\end{subequations}
The freely acting $\urm(1)$ is the diagonal $\urm(1)$ as seen from \eqref{freea2}, which contributes to $T^*\C^{\times}=\C[v_{(1,1,1)},v_{(-1,-1,-1)},\phi_1+\phi_2+\phi_3]/\langle v_{(1,1,1)}v_{(-1,-1,-1)}=1 \rangle$. After decoupling this $T^*\C^\times$ there is the identification $v_{(n_1+k,n_2+k,n_3+k)}\sim v_{(n_1,n_2,n_3)}$. The generators $v_{(1,0,0)}$, $v_{(0,1,0)}$, $v_{(-1,-1,0)}$, $v_{(-1,0,0)}$, $v_{(0,-1,0)}$, $v_{(1,1,0)}$, $\phi_1-\phi_2$, $\phi_2-\phi_3$, $\phi_3-\phi_1$ can be identified with $X_1$, $X_2$, $X_3$, $Y_1$, $Y_2$, $Y_3$, $\Phi_3$, $\Phi_1$, $\Phi_2$ respectively in Table \ref{tab:cg3relation}. The relations (\ref{1sta2} -- \ref{9tha2}) can be repackaged into matrix form with constraint of rank no larger than $1$:
\begin{equation}
\begin{pmatrix}
(\phi_1-\phi_2)^2 & v_{(0,1,0)} & v_{(-1,0,0)}\\
v_{(0,-1,0)} & (\phi_2-\phi_3)^2 & v_{(0,0,1)}\\
v_{(1,0,0)} & v_{(0,0,-1)} & (\phi_3-\phi_1)^2
\end{pmatrix},
\end{equation}
they are exactly the generators and relations of $a_2/\Z_{2g-2}^2$, as in \eqref{Ma2k4}.

The $\Sfrak_3$ Weyl group action on the Abelian Coulomb branch is:
\begin{equation}
\phi_i\longleftrightarrow\phi_{\sigma(i)} \; ,\qquad v_{(n_1,n_2,n_3)}\longleftrightarrow v_{(n_{\sigma(1)},n_{\sigma(2)},n_{\sigma(3)})},
    \label{abels3}
\end{equation}
which agrees with \eqref{actions3}. This action leads to the Coulomb branch of \Quiver{fig:quiver3g}.

\subsubsection*{\texorpdfstring{Quotient by an $\Sfrak_2$ subgroup}{Quotient by an S2 subgroup}}

\begin{figure}[H]
    \centering
    
   \begin{subfigure}[t]{0.3\textwidth}
      \centering
    \begin{tikzpicture}
\node[gauge,label=left:{$1$},color=orange!100] (0) at (0,0) {};
            \node[gauge,label=right:{$1$},color=orange!100] (1) at (1.5,0) {};
            \node[gauge,label=above:{$1$},color=orange!100] (2) at (0.75,1.05) {};
            \node[label=below:{$2g-2$},color=orange!100] (3) at (0.75,0) {};
            \draw (0)--(1);
            \draw (0)--(2)node[pos=0.5,above,sloped]{$2g\!-\!2$};
            \draw (1)--(2)node[pos=0.5,above,sloped]{$2g\!-\!2$};
    \end{tikzpicture}
   \end{subfigure}
   \begin{subfigure}[t]{0.1\textwidth}
 \centering
 \raisebox{0.8\height}{
    \begin{tikzpicture}
        \draw[->]        (0,0)   -- (0.8,0);
        \node[] at (0.4,0.2) {$\Sfrak_2$};
    \end{tikzpicture}}
    \end{subfigure}
    \begin{subfigure}[t]{0.3\textwidth}
 \centering
    \begin{tikzpicture}
        \node[gauge, label=below:{$2$},color=orange!100] (1) at (0,0) {};
        \node[gauge, label=below:{$1$},color=orange!100] (2) at (1.5,0) {};
        \draw (1)--(2);
        \node[label=above:{$2g\!-\!2$},color=orange!100] (3) at (0.8,-0.15) {};
        \draw (1) to [out=135, in=45,looseness=8] node[pos=0.5,above]{$g$} (1);
    \end{tikzpicture}
   \end{subfigure}
   
    \caption{Magnetic quivers of the $\Sfrak_2\in\Sfrak_3$ quotient on the Coulomb branch of $\CG_{3,2g-2}$.}
    \label{fig:quivercg3s2}
\end{figure}
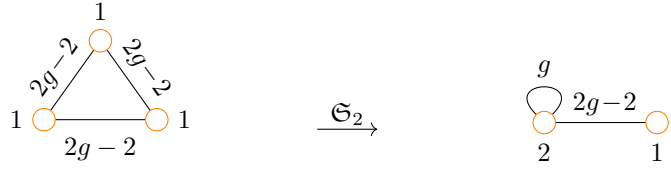

A quotient by an $\Sfrak_2\subset\Sfrak_3$ symmetry can also be applied on the Coulomb branch of $\text{\Quiver{fig:quivercg3}}\cong\CG_{3,2g-2}$. The quivers are shown in \Figref{fig:quivercg3s2}.

The $\Sfrak_2$ Weyl group action on the Abelian Coulomb branch is a subgroup action of $\Sfrak_3$ in \eqref{abels3}:
\begin{equation}
\phi_i\longleftrightarrow\phi_{\sigma(i)} \;,\quad v_{(n_1,n_2,n_3)}\longleftrightarrow v_{(n_{\sigma(1)},n_{\sigma(2)},n_3)},
\end{equation}
where $\sigma\in\Sfrak_2\cong\{\text{Id},\ (1\ 2)\}$. The Hilbert series are detailed in Section \ref{sec:a2s4}.

\subsection{\texorpdfstring{3d mirror of $D_p(\surm(n))$ theories with $\Sfrak_m$ gauging}{Dp(SU(n)) theories with Sm gauging}}
\label{sec:dpsun}

The complete graph can be found as a subquiver in the $3d$ mirror of $D_p(\surm(n))$. The $D_p(\surm(n))$ theories can be realised as $6d\ \mathcal{N}=(2,0)$ theories compactified on a sphere with one Type I irregular puncture and one regular $(1^n)$ maximal puncture. The procedure to derive magnetic quivers for these theories are given in \cite{giacomelli:2021new} and the resulting magnetic quiver takes the form of \Quiver{fig:q4}. In particular, the irregular puncture gives the complete graph part, and the full puncture gives the form of $Q_B$ which is a linear quiver of $(1)-\cdots-(n-1)$ in this case.

Let us restrict to the case where $p\geq n$, here in the complete graph part of the magnetic quivers of $D_p(\surm(n))$ theories there are $m=\text{GCD}(p,n)$ nodes with edge multiplicity $\frac{n(p-n)}{m^2}$. The $\mathcal{Q}_B$ is given, as before, by a linear quiver $(1)-\cdots-(n-1)$. Each $\urm(1)$ node in the complete graph is connected to the $\urm(n-1)$ in $\mathcal{Q}_B$ by $\frac{n}{m}$ bi-fundamentals. There are also $\frac{(n-m)(p-n-m)}{2m}$ free hypermultiplets together.

Then equation \eqref{eq2} implies one can gauge any $\Sfrak_{\sigma_1}\times\dots\times \Sfrak_{\sigma_k}$ permutation subgroup of $\Sfrak_n$ to the 3d mirror of $D_p(\surm(n))$ theories.  For $D_{n(2g-1)}(\surm(n))$ theories, after gauging $\Sfrak_n$, a magnetic quiver for Class $\mathcal{S}$ theories with genus $g$ and one maximal regular $(1^n)$ puncture is found. 

Some examples are listed in the following.

\begin{figure}[h]
    \centering
   \begin{subfigure}[t]{0.45\textwidth}
    \centering
    \begin{tikzpicture}
        \node[gauge,label=above:{$1$}] (0) at (0,0) {};
        \node[gauge,label=below:{$1$}] (1) at (0.86,-0.5) {};
        \node[gauge,label=above:{$1$}] (2) at (0.86,0.5) {};
        \node[gauge,label=below:{$5$}] (4) at (1.86,0) {};
        \node[gauge,label=below:{$4$}] (5) at (2.86,0) {};
        \node[gauge,label=below:{$3$}] (3) at (3.86,0) {};
        \node[gauge,label=below:{$2$}] (6) at (4.86,0) {};
        \node[gauge,label=below:{$1$}] (7) at (5.86,0) {};
            %\draw (0) to [out=135, in=225,looseness=8] node[pos=0.5,left]{$2$} (0);
        \draw[transform canvas={yshift=-1pt}] (0)--(1);
        \draw[transform canvas={yshift=1pt}] (0)--(1);
        \draw[transform canvas={yshift=-0.8pt,
             xshift=0.6pt}] (0)--(2);
        \draw[transform canvas={yshift=0.8pt,
             xshift=-0.6pt}] (0)--(2);
        \draw[transform canvas={
             xshift=1pt}] (1)--(2);
        \draw[transform canvas={
             xshift=-1pt}] (1)--(2);
        \draw[transform canvas={yshift=-1pt}] (0)--(4);
        \draw[transform canvas={yshift=1pt}] (0)--(4);
        \draw[transform canvas={yshift=-0.8pt,
             xshift=0.6pt}] (1)--(4);
        \draw[transform canvas={yshift=0.8pt,
             xshift=-0.6pt}] (1)--(4);
        \draw[transform canvas={yshift=-1pt}] (2)--(4);
        \draw[transform canvas={yshift=1pt}] (2)--(4);
        \draw[-] (4)--(5)--(3)--(6)--(7);
        \end{tikzpicture}
    \caption{$D_9(\surm(6))$ magnetic quiver}
    \label{D9SU6}
   \end{subfigure}
    \begin{subfigure}[t]{0.45\textwidth}
   \centering
   \begin{tikzpicture}
        \node[gauge,label=above:{$1$}] (0) at (0,0) {};
        \node[gauge,label=below:{$1$}] (1) at (0.86,-0.5) {};
        \node[gauge,label=above:{$1$}] (2) at (0.86,0.5) {};
       
        \node[gauge,label=below:{$2$}] (6) at (1.86,0) {};
        \node[gauge,label=below:{$1$}] (7) at (2.86,0) {};
            
        \draw[transform canvas={yshift=-1pt}] (0)--(1);
        \draw[transform canvas={yshift=1pt}] (0)--(1);
        \draw[transform canvas={yshift=-0.8pt,
             xshift=0.6pt}] (0)--(2);
        \draw[transform canvas={yshift=0.8pt,
             xshift=-0.6pt}] (0)--(2);
        \draw[transform canvas={
             xshift=1pt}] (1)--(2);
        \draw[transform canvas={
             xshift=-1pt}] (1)--(2);
        \draw[-] (0)--(6) (1)--(6) (2)--(6);
        \draw[-] (6)--(7);
        \end{tikzpicture}
    \caption{$D_9(\surm(3))$ magnetic quiver}
    \label{D9SU3}
   \end{subfigure}

      \begin{subfigure}[t]{0.45\textwidth}
    \centering
    \begin{tikzpicture}
            \node[gauge,label=below:{$1$}] (0) at (0,0) {};
            \node[gauge,label=below:{$1$}] (1) at (1,-1) {};
            \node[gauge,label=above:{$1$}] (2) at (0,1) {};
            \node[gauge,label=above:{$1$}] (3) at (1,2) {};
            \node[gauge,label=below:{$7$}] (4) at (2,0.5) {};
            \node[gauge,label=below:{$6$}] (5) at (3,0.5) {};
            \node (dots) at (4,0.5) {$\cdots$};
            \node[gauge,label=below:{$2$}] (6) at (5,0.5) {};
            \node[gauge,label=below:{$1$}] (7) at (6,0.5) {};
            \draw[transform canvas={yshift=-1.4pt}] (0)--(1);
            \draw[transform canvas={yshift=1.4pt}] (0)--(1);
            \draw[transform canvas={xshift=-1pt}] (0)--(2);
            \draw[transform canvas={xshift=1pt}] (0)--(2);
            \draw[transform canvas={xshift=-1pt}] (1)--(3);
            \draw[transform canvas={xshift=1pt}] (1)--(3);
            \draw[transform canvas={yshift=1.4pt,xshift=-0.1pt}] (2)--(3);
            \draw[transform canvas={yshift=-1.4pt,xshift=0.1pt}] (2)--(3);
            \draw[transform canvas={yshift=0.7pt,
             xshift=-0.7pt}] (0)--(3);
            \draw[transform canvas={yshift=-0.7pt,
             xshift=0.7pt}] (0)--(3);
             \draw[transform canvas={yshift=-0.7pt,
             xshift=-0.7pt}] (1)--(2);
            \draw[transform canvas={yshift=0.7pt,
             xshift=0.7pt}] (1)--(2);
             \draw[transform canvas={yshift=-1.4pt}] (3)--(4);
            \draw[transform canvas={yshift=1.4pt}] (3)--(4);
            \draw[transform canvas={yshift=-1pt}] (2)--(4);
            \draw[transform canvas={yshift=1pt}] (2)--(4);
            \draw[transform canvas={yshift=-1pt}] (0)--(4);
            \draw[transform canvas={yshift=1pt}] (0)--(4);
            \draw[transform canvas={yshift=-1.4pt}] (1)--(4);
            \draw[transform canvas={yshift=1.4pt}] (1)--(4);
             
             \draw[-] (4)--(5)--(dots)--(6)--(7);
        \end{tikzpicture}
    \caption{$D_{12}(\surm(8))$ magnetic quiver}
    \label{D12SU8}
   \end{subfigure}
    \begin{subfigure}[t]{0.45\textwidth}
   \centering
   \begin{tikzpicture}
            \node[gauge,label=below:{$1$}] (0) at (0,0) {};
            \node[gauge,label=below:{$1$}] (1) at (1,-1) {};
            \node[gauge,label=above:{$1$}] (2) at (0,1) {};
            \node[gauge,label=above:{$1$}] (3) at (1,2) {};
            \node[gauge,label=below:{$3$}] (4) at (2,0.5) {};
            \node[gauge,label=below:{$2$}] (5) at (3,0.5) {};
            \node[gauge,label=below:{$1$}] (6) at (4,0.5) {};
            
            \draw[transform canvas={yshift=-1.4pt}] (0)--(1);
            \draw[transform canvas={yshift=1.4pt}] (0)--(1);
            \draw[transform canvas={xshift=-1pt}] (0)--(2);
            \draw[transform canvas={xshift=1pt}] (0)--(2);
            \draw[transform canvas={xshift=-1pt}] (1)--(3);
            \draw[transform canvas={xshift=1pt}] (1)--(3);
            \draw[transform canvas={yshift=1.4pt,xshift=-0.1pt}] (2)--(3);
            \draw[transform canvas={yshift=-1.4pt,xshift=0.1pt}] (2)--(3);
            \draw[transform canvas={yshift=0.7pt,
             xshift=-0.7pt}] (0)--(3);
            \draw[transform canvas={yshift=-0.7pt,
             xshift=0.7pt}] (0)--(3);
             \draw[transform canvas={yshift=-0.7pt,
             xshift=-0.7pt}] (1)--(2);
            \draw[transform canvas={yshift=0.7pt,
             xshift=0.7pt}] (1)--(2);
           
             \draw[-] (0)--(4) (1)--(4) (2)--(4) (3)--(4);
             \draw[-] (4)--(5)--(6);
        \end{tikzpicture}
    \caption{$D_{12}(\surm(4))$ magnetic quiver}
    \label{D12SU4}
   \end{subfigure}
    \caption{\subref{D9SU6}: Magnetic Quiver for $D_9(\surm(6))$. \subref{D9SU3}: Magnetic Quiver for $D_9(\surm(3))$. \subref{D12SU8}: Magnetic Quiver for $D_{12}(\surm(8))$. \subref{D12SU4}: Magnetic Quiver for $D_{12}(\surm(4))$.}
    \label{D9}
\end{figure}

\Quiver{D9SU6} and \Quiver{D9SU3} give the magnetic quivers for $D_9(\surm(6))$ and $D_9(\surm(3))$ theories. Both of them can be quotiented by $\Sfrak_2$ or $\Sfrak_3$. For $D_9(\surm(3))$ theory, the results after the quotient are shown in \Figref{D93}. The same operation can be done for $D_9(\surm(6))$.

\Quiver{D12SU8} and \Quiver{D12SU4} gives the magnetic quivers for $D_{12}(\surm(8))$ and $D_{12}(\surm(4))$ theories. Both of them can be quotiented by $\Sfrak_2$, $\Sfrak_2\times\Sfrak_2$, $\Sfrak_3$, or $\Sfrak_4$. For $D_{12}(\surm(4))$ theory, the results after the quotient are shown in \Figref{D124}. The same operation can be done for $D_{12}(\surm(8))$.

\begin{figure}[H]
    \centering
    \begin{subfigure}[t]{0.4\textwidth}
   \centering
   \begin{tikzpicture}
        \node[gauge,label=left:{$1$},color=orange!100] (0) at (0,0) {};
        \node[gauge,label=below:{$1$},color=orange!100] (1) at (0.86,-0.5) {};
        \node[gauge,label=above:{$1$},color=orange!100] (2) at (0.86,0.5) {};
       
        \node[gauge,label=below:{$2$}] (6) at (1.86,0) {};
        \node[gauge,label=below:{$1$}] (7) at (2.86,0) {};
            
        \draw[transform canvas={yshift=-1pt}] (0)--(1);
        \draw[transform canvas={yshift=1pt}] (0)--(1);
        \draw[transform canvas={yshift=-0.8pt,
             xshift=0.6pt}] (0)--(2);
        \draw[transform canvas={yshift=0.8pt,
             xshift=-0.6pt}] (0)--(2);
        \draw[transform canvas={
             xshift=1pt}] (1)--(2);
        \draw[transform canvas={
             xshift=-1pt}] (1)--(2);
        \draw[-] (0)--(6) (1)--(6) (2)--(6);
        \draw[-] (6)--(7);
        \end{tikzpicture}
    \caption{}
    \label{D9SU3S1}
   \end{subfigure}
       \begin{subfigure}[t]{0.3\textwidth}
    \centering
    \begin{tikzpicture}
        \node[gauge,label=below:{$1$},color=orange!100] (0) at (0,0.5) {};
        
        \node[gauge,label=left:{$2$},color=orange!100] (1) at (0,1.5) {};
        \draw (1) to [out=135, in=45,looseness=8] node[pos=0.5,above]{$2$} (1);
        \draw[transform canvas={xshift=-1pt}] (0)--(1);
        \draw[transform canvas={xshift=1pt}] (0)--(1);
        \node[gauge,label=below:{$2$}] (4) at (1,1) {};
        \node[gauge,label=below:{$1$}] (5) at (2,1) {};
        \draw (4)--(5) (0)--(4) (1)--(4);
    \end{tikzpicture}
    \caption{}
    \label{D9SU3S2}
    \end{subfigure}
        \begin{subfigure}[t]{0.4\textwidth}
    \centering
    \begin{tikzpicture}
        \node[gauge,label=below:{$3$},color=orange!100] (1) at (0,0) {};
        \draw (1) to [out=225, in=135,looseness=8] node[pos=0.5,left]{$2$} (1);
        \node[gauge,label=below:{$2$}] (4) at (1,0) {};
        \node[gauge,label=below:{$1$}] (5) at (2,0) {};
        \draw (1)--(4)--(5);
    \end{tikzpicture}
    \caption{}
    \label{D9SU3S3}
    \end{subfigure}
    \caption{\subref{D9SU3S1}: Magnetic Quiver for $D_9(\surm(3))$. \subref{D9SU3S2}: $\Sfrak_2$ quotient on Coulomb branch of \Quiver{D9SU3S1}. \subref{D9SU3S3}: $\Sfrak_3$ quotient on Coulomb branch of \Quiver{D9SU3S1}.}
    \label{D93}
\end{figure}

\begin{figure}[H]
    \centering
    \begin{subfigure}[t]{0.4\textwidth}
   \centering
   \begin{tikzpicture}
            \node[gauge,label=below:{$1$},color=orange!100] (0) at (0,0) {};
            \node[gauge,label=below:{$1$},color=orange!100] (1) at (1,-1) {};
            \node[gauge,label=above:{$1$},color=orange!100] (2) at (0,1) {};
            \node[gauge,label=above:{$1$},color=orange!100] (3) at (1,2) {};
            \node[gauge,label=below:{$3$}] (4) at (2,0.5) {};
            \node[gauge,label=below:{$2$}] (5) at (3,0.5) {};
            \node[gauge,label=below:{$1$}] (6) at (4,0.5) {};
            \draw[transform canvas={yshift=-1.4pt}] (0)--(1);
            \draw[transform canvas={yshift=1.4pt}] (0)--(1);
            \draw[transform canvas={xshift=-1pt}] (0)--(2);
            \draw[transform canvas={xshift=1pt}] (0)--(2);
            \draw[transform canvas={xshift=-1pt}] (1)--(3);
            \draw[transform canvas={xshift=1pt}] (1)--(3);
            \draw[transform canvas={yshift=1.4pt,xshift=-0.1pt}] (2)--(3);
            \draw[transform canvas={yshift=-1.4pt,xshift=0.1pt}] (2)--(3);
            \draw[transform canvas={yshift=0.7pt,
             xshift=-0.7pt}] (0)--(3);
            \draw[transform canvas={yshift=-0.7pt,
             xshift=0.7pt}] (0)--(3);
             \draw[transform canvas={yshift=-0.7pt,
             xshift=-0.7pt}] (1)--(2);
            \draw[transform canvas={yshift=0.7pt,
             xshift=0.7pt}] (1)--(2);
             \draw (0)--(4) (1)--(4) (2)--(4) (3)--(4);
             \draw[-] (4)--(5)--(6);
        \end{tikzpicture}
    \caption{}
    \label{D12SU4S1}
   \end{subfigure}
       \begin{subfigure}[t]{0.3\textwidth}
    \centering
    \begin{tikzpicture}
            \node[gauge,label=above:{$2$},color=orange!100] (0) at (0,0) {};
            \node[gauge,label=below:{$1$},color=orange!100] (1) at (0.86,-0.5) {};
            \node[gauge,label=above:{$1$},color=orange!100] (2) at (0.86,0.5) {};
            \node[gauge,label=below:{$3$}] (4) at (2,0) {};
            \node[gauge,label=below:{$2$}] (5) at (3,0) {};
            \node[gauge,label=below:{$1$}] (6) at (4,0) {};
            \draw (0) to [out=135, in=225,looseness=8] node[pos=0.5,left]{$2$} (0);
            \draw[transform canvas={yshift=-1pt}] (0)--(1);
            \draw[transform canvas={yshift=1pt}] (0)--(1);
            \draw[transform canvas={yshift=-0.8pt,
             xshift=0.6pt}] (0)--(2);
            \draw[transform canvas={yshift=0.8pt,
             xshift=-0.6pt}] (0)--(2);
            \draw[transform canvas={
             xshift=1pt}] (1)--(2);
            \draw[transform canvas={
             xshift=-1pt}] (1)--(2);
             \draw (0)--(4) (1)--(4) (2)--(4);
             \draw[-] (4)--(5)--(6);
        \end{tikzpicture}
    \caption{}
    \label{D12SU4S2}
    \end{subfigure}
        \begin{subfigure}[t]{0.4\textwidth}
    \centering
    \begin{tikzpicture}
           
        \node[gauge,label=left:{$2$},color=orange!100] (0) at (0,0) {};
        \draw (0) to [out=225, in=315,looseness=8] node[pos=0.5,below]{$2$} (0);
        \node[gauge,label=left:{$2$},color=orange!100] (1) at (0,1) {};
        \draw (1) to [out=135, in=45,looseness=8] node[pos=0.5,above]{$2$} (1);
        \draw[transform canvas={xshift=-1pt}] (0)--(1);
        \draw[transform canvas={xshift=1pt}] (0)--(1);
        \node[gauge,label=below:{$3$}] (4) at (1.14,0.5) {};
            \node[gauge,label=below:{$2$}] (5) at (2.04,0.5) {};
            \node[gauge,label=below:{$1$}] (6) at (3.14,0.5) {};
        \draw (0)--(4) (1)--(4);
        \draw[-] (4)--(5)--(6);
        \end{tikzpicture}
    \caption{}
    \label{D12SU4S2S2}
    \end{subfigure}
     \begin{subfigure}[t]{0.4\textwidth}
    \centering
    \begin{tikzpicture}
           
        \node[gauge,label=below:{$1$},color=orange!100] (0) at (0,0) {};
       
        \node[gauge,label=left:{$3$},color=orange!100] (1) at (0,1) {};
        \draw (1) to [out=135, in=45,looseness=8] node[pos=0.5,above]{$2$} (1);
        \draw[transform canvas={xshift=-1pt}] (0)--(1);
        \draw[transform canvas={xshift=1pt}] (0)--(1);
        \node[gauge,label=below:{$3$}] (4) at (1.14,0.5) {};
            \node[gauge,label=below:{$2$}] (5) at (2.14,0.5) {};
            \node[gauge,label=below:{$1$}] (6) at (3.14,0.5) {};
        \draw (0)--(4) (1)--(4);
        \draw[-] (4)--(5)--(6);
        \end{tikzpicture}
    \caption{}
    \label{D12SU4S3}
    \end{subfigure}
     \begin{subfigure}[t]{0.4\textwidth}
    \centering
    \begin{tikzpicture}
          
        \node[gauge,label=below:{$4$},color=orange!100] (0) at (0,0) {};
        \draw (0) to [out=225, in=135,looseness=8] node[pos=0.5,left]{$2$} (0);
        \node[gauge,label=below:{$3$}] (4) at (1,0) {};
            \node[gauge,label=below:{$2$}] (5) at (2,0) {};
            \node[gauge,label=below:{$1$}] (6) at (3,0) {};
        \draw (0)--(4);
        \draw[-] (4)--(5)--(6);
        \end{tikzpicture}
    \caption{}
    \label{D12SU4S4}
    \end{subfigure}

    \caption{\subref{D12SU4S1}: Magnetic Quiver for $D_{12}(\surm(4))$. \subref{D12SU4S2}: $\Sfrak_2$ quotient on Coulomb branch of \Quiver{D12SU4S1}. \subref{D12SU4S2S2}: $\Sfrak_2\times\Sfrak_2$ quotient on Coulomb branch of \Quiver{D12SU4S1}.  \subref{D12SU4S3}: $\Sfrak_3$ quotient on Coulomb branch of \Quiver{D12SU4S1}.  \subref{D12SU4S4}: $\Sfrak_4$ quotient on Coulomb branch of \Quiver{D12SU4S1}.}
    \label{D124}
\end{figure}

The unrefined Coulomb branch Hilbert series for \Quiver{D9SU3S1} and the $\mathfrak{S}_2$ and $\mathfrak{S}_3$ quotients \Quiver{D9SU3S2} and \Quiver{D9SU3S3} are computed using monopole formula and Molien sums. Note that in the $\mathfrak{S}_3$ quotient, the element $(1\ 2)$, $(1\ 3)$, and $(2\ 3)$ give the same contribution, as do the pair $(1\ 2\ 3)$ and $(1\ 3\ 2)$. The Hilbert Series after the action of an element in $\mathfrak{S}_n$ only depends on the conjugacy class of the element.

\begin{subequations}
\begin{align}
    \HS\left(\Ccal(\text{\Quiver{D9SU3S1}} )\right)&=\frac{\left(\begin{aligned}1&+4t^2+10t^4+32t^6+65t^8+104t^{10}+153t^{12}+150t^{14}\\&+153t^{16}+104t^{18}+65t^{20}+32t^{22}+10t^{24}+4t^{26}+t^{28}\end{aligned}\right)}{(1-t^4)^{-1}(1-t^2)^6(1-t^6)^5},
        \\
    \HS\left(\Ccal(\text{\Quiver{D9SU3S2}} /\Sfrak_2)\right)&=\frac{1}{2}\left( \HS\left(\Ccal(\text{\Quiver{D9SU3S1}})\right) +\HS\left(\Ccal(\text{\Quiver{D9SU3S1}})\right)^{(1\ 2)} \right)
    \notag \\
    &=\frac{1}{2} 
        \left( \HS\left(\Ccal(\text{\Quiver{D9SU3S1}})\right)+\frac{(1-t^4)(1+2t^2+4t^4+8t^6+7t^8+8t^{10}+4t^{12}+2t^{14}+t^{16})}{(1-t^2)^6(1-t^6)^3} \right)
    \notag \\
        &=\frac{\left(\begin{aligned}1&+3t^2+7t^4+19t^6+34t^8+52t^{10}+71t^{12}+70t^{14}\\&+71t^{16}+52t^{18}+34t^{20}+19t^{22}+7t^{24}+3t^{26}+t^{28}\end{aligned}\right)}{(1-t^4)^{-1}(1-t^2)^6(1-t^6)^5},
    \\
   \HS\left(\Ccal(\text{\Quiver{D9SU3S3}} /\Sfrak_3)\right)
   &=\frac{1}{6}\left( \HS\left(\Ccal(\text{\Quiver{D9SU3S1}})\right) +\HS\left(\Ccal(\text{\Quiver{D9SU3S1}})\right)^{(1\ 2)} +\HS\left(\Ccal(\text{\Quiver{D9SU3S1}})\right)^{(2\ 3)}\right.
   \notag \\
   &\quad +\left.\HS\left(\Ccal(\text{\Quiver{D9SU3S1}})\right)^{(1\ 3)} +\HS\left(\Ccal(\text{\Quiver{D9SU3S1}})\right)^{(1\ 2\ 3)} +\HS\left(\Ccal(\text{\Quiver{D9SU3S1}})\right)^{(1\ 3\ 2)}  \right)
   \notag \\
   &=\frac{1}{6}\left(\HS\left(\Ccal(\text{\Quiver{D9SU3S1}})\right)+3\frac{(1-t^4)(1+2t^2+4t^4+8t^6+7t^8+8t^{10}+4t^{12}+2t^{14}+t^{16})}{(1-t^2)^6(1-t^6)^3}+2\frac{1-t^4}{(1-t^2)^7}
   \right)
   \notag \\
   &=\frac{(1-t^8)(1+2t^2+3t^4+5t^6+8t^8+11t^{10}+14t^{12}+11t^{14}+8t^{16}+5t^{18}+3t^{20}+2t^{22}+t^{24})}{(1-t^2)^6(1-t^6)^5}.
\end{align}
\end{subequations}
Both \Quiver{D9SU3S1} and \Quiver{D9SU3S3} admit a well-defined 4d construction, in the sense that they are magnetic quivers for such 4d $\mathcal{N}=2$ SCFTs. In contrast, it remains to be explored if \Quiver{D9SU3S2} also is a magnetic quiver of some 4d theory (compactified from 6d). An intuitive guess is that all these theories can be compactified on a genus $g$ Riemann surface with certain number of punctures, the genus $g$ contributes to the number of loops, and the punctures should encode the partition data of the subgroups of $\Sfrak_3$. The investigation of the discrete quotient in the 4d point of view is left to future work.

\subsection{\texorpdfstring{Pure $\urm(1)^3$ with $\Sfrak_3$ gauging -- D-brane physics}{pure gauge theory}}

In this section a well-studied case is used to show the claim and Molien sum technique also work for bad quivers. The quiver before discrete gauging consists of three disconnected $\urm(1)$ nodes. After $\Sfrak_2$ gauging is one $\urm(2)$ node with one adjoint and one $\urm(1)$ node. After $\Sfrak_3$ gauging is $\urm(3)$ node with one adjoint.
\begin{figure}[H]
    \centering
    \begin{subfigure}[t]{0.25\textwidth}
   \centering
   \begin{tikzpicture}
        \node[gauge,label=below:{$1$},color=orange!100] (0) at (0,0) {};
            \node[gauge,label=below:{$1$},color=orange!100] (1) at (1,0) {};
            \node[gauge,label=above:{$1$},color=orange!100] (2) at (0.5,0.7) {};
        \end{tikzpicture}
    \caption{}
    \label{u1u1u1}
   \end{subfigure}
        \begin{subfigure}[t]{0.25\textwidth}
    \centering
    \begin{tikzpicture}
        \node[gauge, label=below:$2$,color=orange!100] (2) []{};
        \draw (2) to [out=135, in=45,looseness=8] node[pos=0.5,above]{$1$} (2);
        \node[gauge,label=below:{$1$},color=orange!100] (1) at (1,0) {};
    \end{tikzpicture}
    \caption{}
    \label{u1u2}
    \end{subfigure}
           \begin{subfigure}[t]{0.25\textwidth}
    \centering
    \begin{tikzpicture}
        \node[gauge, label=below:$3$,color=orange!100] (3) []{};
        \draw (3) to [out=135, in=45,looseness=8] node[pos=0.5,above]{$1$} (3);
    \end{tikzpicture}
    \caption{}
    \label{u3}
    \end{subfigure}
    \caption{\subref{u1u1u1}: $\CG_{3,0}$. \subref{u1u2}: $\ML_{2,1}$ with $\CG_{1,0}$. \subref{u3}: $\ML_{3,1}$.}
    \label{pureu1}
\end{figure}

In this case, the supersymmetry is enhanced to $\mathcal{N}=8$, the Coulomb branch receives no quantum correction. The discrete gauging between each quiver simply reflects the classical description: $(\mathbb{R}^3\times S^1)^r/\mathcal{W}_G$, where $\mathcal{W}_G$ is the Weyl group of the gauge group. In \cite{Braverman:2016pwk}, it is expressed as the cotangent bundle of the maximal torus of the Langlands (GNO) dual group of the gauge group: $T^*T^\vee/\mathcal{W}$. Applying this description to the quivers above, the following are found:
\begin{subequations}
\begin{align}
    \Ccal\left(\CG_{3,0}\right)&=T^*(\mathbb{C}^\times)^3\\
    \Ccal\left(\CG_{3,0}/\Sfrak_2\right)=\Ccal\left(\ML_{2,1}\times\CG_{1,0}\right)&=T^*(\mathbb{C}^\times)^2/\Sfrak_2\times T^*\mathbb{C}^\times\\
    \Ccal\left(\CG_{3,0}/\Sfrak_3\right)=\Ccal\left(\ML_{3,1}\right)&=T^*(\mathbb{C}^\times)^3/\Sfrak_3
\end{align}
\end{subequations}

The monopole formula gives a divergence when applied to these quivers:
\begin{subequations}
\begin{align}
    \HS\left(\Ccal\left(\CG_{3,0}\right)\right)&=\frac{1}{(1-t^2)^3}\sum_{m_1,m_2,m_3}z_1^{m_1}z_2^{m_2}z_3^{m_3}\\
    \HS\left(\Ccal\left(\CG_{3,0}/\Sfrak_2\right)\right)&=\frac{1}{1-t^2}\sum_{m_3}z_3^{m_3}\left(\frac{1}{(1-t^2)(1-t^4)}\sum_{m_1=m_2}z^{m_1+m_2}+\frac{1}{(1-t^2)^2}\sum_{m_1>m_2}z^{m_1+m_2}\right)\\
    \HS\left(\Ccal\left(\CG_{3,0}/\Sfrak_3\right)\right)&=\frac{1}{(1-t^2)(1-t^4)(1-t^6)}\sum_{m_1=m_2=m_3}z^{m_1+m_2+m_3}
    \notag \\
    &\quad +\frac{1}{(1-t^2)^2(1-t^4)}\left(\sum_{m_1>m_2=m_3}+\sum_{m_1=m_2>m_3}\right)z^{m_1+m_2+m_3}
    \notag \\
    &\quad +\frac{1}{(1-t^2)^3}\sum_{m_1>m_2>m_3}z^{m_1+m_2+m_3}
\end{align}
\end{subequations}
The reason is that there are no less than one freely acting $\urm(1)$s in each quiver. However, the divergent monopole formula is still a valid tool. For example, the Coulomb branch of pure $\urm(1)$ theory is $\mathbb{C}[\phi,u_{-},u_{+}]/\langle u_{-}u_{+}-1\rangle$. The monopole formula for this theory is $\frac{1}{1-t^2}\sum_{m_1}z^{m_1}$, This can be interpreted as: for any fixed monopole charge $m_1$, there is one monopole operator $u_+^{m_1}\phi^n=u_-^{-m_1}\phi^n$ at degree $2n$. From the geometry point of view, there is a $\mathbb{C}$ fibring over $\mathbb{C}^\times$.

After demonstrating the $\Sfrak_n$ quotient, now let us demonstrate the $\Z_q$ quotient.

\section{\texorpdfstring{Examples of $\Z_q$ quotient}{Examples of Zq quotient}}
\label{sec:ZqQuot}

The $\Z_q$ gauging on quivers which are or contain certain complete graphs are computed and the corresponding $\Z_q$ on their Coulomb branches are studied. This action is studied through the Hilbert series computed with the monopole formula for each example.

\subsection{\texorpdfstring{Kleinian $A$ with $\Z_q$ quotient}{Kleinian A with Zq quotient}}
Now let us look at the simplest example of Claim \eqref{eqzq}. The $A_{k-1}\cong \C^2/\Z_k$ singularity may be realised through the Coulomb branch of an Abelian quiver $\CG_{2,k}$, as \Quiver{fig:quiverAk} in \Figref{fig:quiverAkAkq}.
\begin{figure}[H]
    \centering
   \begin{subfigure}[t]{0.2\textwidth}
      \centering
    \begin{tikzpicture}
\node[gauge,color=orange!100,label=below:{$1$}] (0) at (0,0) {};
            \node[gauge,color=orange!100,label=below:{$1$}] (1) at (1,0) {};
            \node[label=above:{$k$}] (3) at (0.5,0) {};
            \draw (0)--(1);
    \end{tikzpicture}
    \caption{}
    \label{fig:quiverAk}
   \end{subfigure}
   \begin{subfigure}[t]{0.1\textwidth}
 \centering
 \raisebox{0.8\height}{
    \begin{tikzpicture}
        \draw[->]        (0,0)   -- (0.8,0);
        \node[] at (0.4,0.25) {$\Z_q$};
    \end{tikzpicture}}
    \end{subfigure}
    \begin{subfigure}[t]{0.2\textwidth}
 \centering
    \begin{tikzpicture}
\node[gauge,color=orange!100,label=below:{$1$}] (0) at (0,0) {};
            \node[gauge,color=orange!100,label=below:{$1$}] (1) at (1,0) {};
            \node[label=above:{$qk$}] (3) at (0.5,0) {};
            \draw (0)--(1);
    \end{tikzpicture}
    \caption{}
    \label{fig:quiverAkq}
   \end{subfigure}
    \caption{\subref{fig:quiverAk}: Quiver $\CG_{2,k}$, whose Coulomb branch is the $A_{k-1}$ singularity. \subref{fig:quiverAkq}: Quiver $\CG_{2,kq}$ of \eqref{eq1}, whose Coulomb branch is the $A_{kq-1}$ singularity.}
    \label{fig:quiverAkAkq}
\end{figure}
Since $A_{kq-1}$ is a $\Z_q$ quotient of $A_{k-1}$, the Coulomb branch of $\CG_{2,kq}$ is a $\Z_q$ quotient of the Coulomb branch of $\CG_{2,k}$:
\begin{equation}
\Ccal\left(\CG_{2,kq}\right)=\Ccal\left(\CG_{2,k}\right)/\Z_q.
\end{equation}
The $\Z_q$ action on $\Ccal\left(\CG_{2,k}\right) = A_{k-1}\cong\C[\phi,u,v]/\langle \phi^k-uv \rangle$ is given by:
\begin{equation}
    u\to \omega_q \cdot u,\ v\to \omega^{-1}_q \cdot v \;,
\end{equation}
and the fundamental invariants are $u^k=u'$, $v^k=v'$, subject to $uv=\phi^k$. This agrees with $\Ccal\left(\CG_{2,kq}\right) = A_{kq-1}\cong\C[\phi,u',v']/\langle \phi^{kq}-u'v' \rangle$. The $\Z_q$ action can also be implemented in the Hilbert series: it acts on the $\urm(1)_J$ fugacity as in \eqref{eq:Z_q^n_on_fug}:
\begin{equation}
    z\to \omega_q \cdot z \;.
\label{eq:zq_map_fug}
\end{equation}
Combined with the Molien sum \eqref{moliensumzq}, the Hilbert Series of $\Ccal\left(\CG_{2,kq}\right)$ is obtained from the Hilbert Series of $\Ccal\left(\CG_{2,k}\right)$:
\begin{align}
    \HS(A_{kq-1})&=\frac{1}{q}\sum_{i=1}^{q}\HS(A_{k-1})\vert_{z\to \omega_q^i \cdot z}
    \notag \\
    &=\frac{1}{q}\sum_{i=1}^{q}\PE\left[t^2+(\omega_q^i \cdot z+\frac{1}{\omega_q^i \cdot z})t^k-t^{2k}\right]
    \notag \\
    &=\PE\left[t^2-(z^q+\frac{1}{z^q})t^{kq}-t^{2kq}\right].
\end{align}

From the quivers in \Figref{fig:quiverAkAkq}, the $\Z_q$ quotient can be implemented through the edge multiplicity $k$. On the Coulomb branch level, instead of $q$-times the edge multiplicity, the action \eqref{eq:zq_map_fug} can also be interpreted as $q$-times the charge (double $q$ lace) of hypermultiplets under the gauge $\urm(1)^2$. The edge multiplicity and double lace on Coulomb branch cannot be distinguished; the difference is manifest on the Higgs branch, which is left to future work. Moreover, the two potential theory also differ in their higher symmetries, as recently analysed in \cite{Mekareeya:2022spm,Nawata:2023rdx,Bhardwaj:2023zix}.

\subsection{\texorpdfstring{$\mathbb{H}^2$ with $\Z_q$ quotient}{H2 with Zq quotient}}
As an extension of the $A$-type singularity to $4$ complex dimensions, there is $\C^4/\Z_q$ with the action of $\Z_q$: $\text{diag}(\omega_q,\omega_q,\omega_q^{-1},\omega_q^{-1})$. The magnetic quiver of this singularity is \Quiver{fig:quiverC4zq}. The Coulomb branch of \Quiver{fig:quiverC4zq} can be seen as a $\Z_q$ quotient on the Coulomb branch of \Quiver{fig:quiverH2} by the Claim \eqref{eqzqb}:
\begin{equation}
    \Ccal\left(\text{\Quiver{fig:quiverC4zq}}\right)=\Ccal\left(\text{\Quiver{fig:quiverH2}}\right)/\Z_q
\end{equation}
here the middle node is treated as the background quiver.
\begin{figure}[H]
    \centering
   \begin{subfigure}[t]{0.2\textwidth}
      \centering
    \begin{tikzpicture}
\node[gauge,color=orange!100,label=below:{$1$}] (0) at (0,0) {};
            \node[gauge,label=below:{$1$}] (1) at (1,0) {};
            \node[gauge,color=orange!100,label=below:{$1$}] (2) at (2,0) {};
            \draw (0)--(1)--(2);
    \end{tikzpicture}
    \caption{}
    \label{fig:quiverH2}
   \end{subfigure}
   \begin{subfigure}[t]{0.1\textwidth}
 \centering
 \raisebox{0.8\height}{
    \begin{tikzpicture}
        \draw[->]        (0,0)   -- (0.8,0);
        \node[] at (0.4,0.25) {$\Z_q$};
    \end{tikzpicture}}
    \end{subfigure}
    \begin{subfigure}[t]{0.2\textwidth}
 \centering
    \begin{tikzpicture}
\node[gauge,color=orange!100,label=below:{$1$}] (0) at (0,0) {};
            \node[gauge,label=below:{$1$}] (1) at (1,0) {};
            \node[gauge,color=orange!100,label=below:{$1$}] (2) at (2,0) {};
            \node[label=above:{$q$}] (3) at (0.5,0) {};
            \node[label=above:{$q$}] (3) at (1.5,0) {};
            \draw (0)--(1)--(2);
            \draw (0.45,0.1)--(0.6,0)--(0.45,-0.1);
            \draw (1.55,0.1)--(1.4,0)--(1.55,-0.1);
    \end{tikzpicture}
    \caption{}
    \label{fig:quiverC4zq}
   \end{subfigure}
    \caption{\subref{fig:quiverH2}: Quiver \Quiver{fig:quiverH2} is an Abelian linear quiver of three nodes. The Coulomb branch is $\mathbb{H}^2\cong\C^4$. \subref{fig:quiverC4zq}: Quiver \Quiver{fig:quiverC4zq} has two $q$ laced edges pointing to the middle $\urm(1)$. The Coulomb branch is $\C^4/\Z_q$.}
    \label{fig:quiverH2C4zq}
\end{figure}

\begin{figure}[H]
    \centering
   \begin{subfigure}[t]{0.25\textwidth}
      \centering
    \begin{tikzpicture}
\node[gauge,label=below:{$1$}] (0) at (0,0) {};
            \node[gauge,label=below:{$1$}] (1) at (1,0) {};

            \node[] (2) at (1.55,0) {$\cdots$};
            \node[gauge,label=below:{$1$}] (3) at (2,0) {};
            \node[gauge,label=below:{$1$}] (4) at (3,0) {};
            \draw (0)--(1) (3)--(4);
            \draw[-] (1)--(1.25,0) (1.75,0)--(3);
    \end{tikzpicture}
    \caption{}
    \label{fig:quiverHn}
   \end{subfigure}
   \begin{subfigure}[t]{0.1\textwidth}
 \centering
 \raisebox{0.8\height}{
    \begin{tikzpicture}
        \draw[->]        (0,0)   -- (0.8,0);
        \node[] at (0.4,0.25) {$\Z_q$};
    \end{tikzpicture}}
    \end{subfigure}
    \begin{subfigure}[t]{0.25\textwidth}
 \centering
    \begin{tikzpicture}
    \node[gauge,label=below:{$1$}] (0) at (0,0) {};
            \node[gauge,label=below:{$1$}] (1) at (1,0) {};

            \node[] (2) at (1.55,0) {$\cdots$};
            \node[gauge,label=below:{$1$}] (3) at (2,0) {};
            \node[gauge,label=below:{$1$}] (4) at (3,0) {};
            \node[label=above:{$q$}] (5) at (0.5,0) {};
            \node[label=above:{$q$}] (6) at (2.5,0) {};
            \draw (0)--(1) (3)--(4);
            \draw (0.45,0.1)--(0.6,0)--(0.45,-0.1);
            \draw (1.55+1,0.1)--(1.4+1,0)--(1.55+1,-0.1);
            \draw[-] (1)--(1.25,0) (1.75,0)--(3);
    \end{tikzpicture}
    \caption{}
    \label{fig:quiverhnq}
   \end{subfigure}
    \caption{\subref{fig:quiverHn}: Quiver \Quiver{fig:quiverHn} is an linear Abelian quiver composed of $n+1$ nodes. The Coulomb branch is $h_{n,1}=\mathbb{H}^n\cong\C^{2n}$. \subref{fig:quiverhnq}: Quiver \Quiver{fig:quiverhnq} has two $q$ laced edges pointing to the middle, as leftmost and rightmost. The Coulomb branch is $h_{n,q}=\C^{2n}/\Z_q$. The $h_{n,q}$ appears an elementary slice in the affine Grassmannian, and in more general symplectic singularities.}
    \label{fig:quiverHnhnq}
\end{figure}
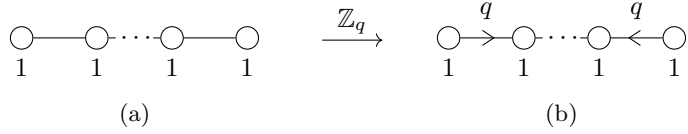

This construction can be further extended to $2n$ complex dimensions: i.e.\ $\C^{2n}/\Z_q$ with the action $\text{diag}(\omega_q,\cdots,\omega_q,\omega_q^{-1},\cdots,\omega_q^{-1})$. The Coulomb branch construction is shown in \Figref{fig:quiverHnhnq} and the Coulomb branches obey the relation:
\begin{equation}
    \Ccal\left(\text{\Quiver{fig:quiverhnq}}\right)=\Ccal\left(\text{\Quiver{fig:quiverHn}}\right)/\Z_q.
\end{equation}

In this case, however, the leftmost and rightmost nodes cannot be seen as identical, so the quiver does not fit the Claim \eqref{eqzqb}. But, since the proof of Claim \eqref{eqzqb} only assumes the subquiver to be an Abelian instead of a complete graph, it can be applied to any simply-laced Abelian subquiver.

\subsection{\texorpdfstring{$a_2$ with $\Z_{q}^2$ gauging}{a2 with Zq2 gauging}}
\label{sec:a2zq}
Another interesting example of \eqref{eqzq} is the quotient on $\CG_{3,1}$. The Coulomb branch of $\CG_{3,1}$ is $a_2$, the closure of minimal nilpotent orbit of $\mathfrak{sl}_3$. The coordinate ring of this Coulomb branch is written as $\C[M^i_j]/\langle \Tr(M)=0,\ 
\rk(M)\leq 1 \rangle$, where $i,j=1,2,3$. According to Claim \eqref{eqzq}, there is the following relationship:
\begin{equation}
\Ccal\left(
        \vcenter{\hbox{
        \begin{tikzpicture}
            \node[gauge,color=orange!100,label=below:{$1$}] (0) at (0,0) {};
            \node[gauge,color=orange!100,label=below:{$1$}] (1) at (1,0) {};
            \node[gauge,color=orange!100,label=above:{$1$}] (2) at (0.5,0.7) {};
            \draw (0)--(1) (1)--(2) (0)--(2);
            \node[label=above:{$q$}] (3) at (0.1,0.12) {};
            \node[label=above:{$q$}] (4) at (0.9,0.12) {};
            \node[label=above:{$q$}] (5) at (0.5,-0.6) {};
        \end{tikzpicture}
        }}\right)=
\Ccal\left(
    \vcenter{\hbox{
        \begin{tikzpicture}
            \node[gauge,color=orange!100,label=below:{$1$}] (0) at (0,0) {};
            \node[gauge,color=orange!100,label=below:{$1$}] (1) at (1,0) {};
            \node[gauge,color=orange!100,label=above:{$1$}] (2) at (0.5,0.7) {};
            \draw (0)--(1) (1)--(2) (0)--(2);
        \end{tikzpicture}
    }}\right)/\Z_q^2.
\end{equation}
The Coulomb branch ring of $\CG_{3,q}$ is the ring of invariants $\C[M_i^j]^{\Z_{q}\times\Z_{q}}/\langle \Tr(M)=0,\ \rk(M)\leq 1 \rangle$. The $\Z_{q}\times\Z_{q}$ action is given by: $M_i^j\to \omega^{k_i-k_j}_q M_i^j$, with $(k_i,k_j)\in\Z_q^2$.

\section{\texorpdfstring{Examples of combined $\Sfrak_n$ and $\Z_q$ quotients}{Examples of combined Sn and Zq quotients}}
\label{sec:CombinedQuot}
After studying the $\Sfrak_n$ and $\Z_q$ quotients individually, it is time to explore the combined action.

\subsection{Kleinian review: $Dih$ or $Dic$}
\label{sec:Kleinian AD}

In Section \ref{sec:AtoD}, we showed the $\Sfrak_2$ quotient relation between $A_{2g-3}$ and $D_{g+1}$ through quivers. The Claim \eqref{zqsnrelation} indicates that $\Ccal({\CG_{2,k}})\cong A_{k-1}$ and $\Ccal({\ML_{2,qk+1}})\cong D_{qk+2}$ are related by a $\Z_{2q}\!\rtimes\!\Sfrak_2\cong Dih_{2q}$ quotient. However, this claim might be confusing without specifying the module of the group action. For example, when $k=1$, the two Coulomb branches $A_0\cong\C^2$ and $D_{q+2} \cong \C^2/Dic_q$ are related by the non-split extension $\Z_{2q}.\Sfrak_2=Dic_{q}$ instead of the split extension $\Z_{2q}\!\rtimes\!\Sfrak_2=Dih_{2q}$. To fit the claim, we should take the module of the generators $\text{span}[u,v,\phi]$ instead of $\C^2$, then the actions are:
\begin{equation}
    \sigma=\begin{pmatrix}
        \omega_{2q}&0&0\\
        0&\omega^{-1}_{2q}&0\\
        0&0&1
    \end{pmatrix},\quad \tau=\begin{pmatrix}
        0&1&0\\
        1&0&0\\
        0&0&-1
    \end{pmatrix},
\end{equation}
which exactly generate the group $Dih_{2q}=\langle \sigma,\tau \vert \sigma^{2q}=\tau^2=\sigma\tau\sigma\tau=\text{Id} \rangle$.

\subsection{\texorpdfstring{$A_2$ affine quiver with $\Sfrak_4$ gauging}{A2 quiver}}
\label{sec:a2s4}

In this section the combined discrete quotient on the Coulomb branch of the affine $A_2$ quiver $\CG_{3,1}$ is studied. As in Section \ref{sec:a2zq}, a $\Z_q^2$ quotient on $a_2$ is performed, which takes the $\CG_{3,1}$ quiver to the $\CG_{3,q}$ quiver. Recalling Section \ref{sec:a2s3}, a further $\Sfrak_3$ quotient on the Coulomb branch of $\CG_{3,2g-2}$ is performed, which results in the $\ML_{3,2}$ quiver -- a single $\urm(3)$ gauge node with $2$ adjoint hypermultiplets.

Let us consider the simplest example of the combined action. A $\Sfrak_2$ or $\Sfrak_3$ symmetry can be quotiented from $\CG_{3,2}$, the corresponding quivers are shown in Table \ref{tab:a_2}. According to Claim \eqref{eqzqsn}, the Coulomb branches are $a_2/\Z_2^2\!\rtimes\!\Sfrak_2$ and $a_2/\Z_2^2\!\rtimes\!\Sfrak_3$ respectively. There are surprising examples due to the coincidences of the group action: $\Z_2^2\!\rtimes\!\Sfrak_2\cong Dih_4$ and $\Z_2^2\!\rtimes\!\Sfrak_3\cong\Sfrak_4$. The latter leads to a recently studied symplectic singularity \cite{2023arXiv230807398F}. The details are discussed below.

In terms of gauging outer automorphism symmetries both $\mathfrak S_2$ and $\mathfrak S_3$ could be gauged on the Coulomb branch of $\CG_{3,2}$, but not on $\CG_{3,1}$ due to its odd edge multiplicity.

The Hilbert series of all possible combinations of $\mathbb Z_2^2$ and $\mathfrak S_2$ and $\mathfrak S_3$ gaugings on $a_2$ are computed. The quivers, unrefined HS, and the volume of the Coulomb branches are provided in Table \ref{tab:a_2}.

\input{Tablea2}

Note that the ratios of the volumes\footnote{The (inverse) volume of the Coulomb branches can be calculated from Hilbert series as $\mathrm{vol}^{-1}=\lim_{t\to 1} (1-t)^d \HS$, where $d$ is the complex dimension of the Coulomb branch. If two space are related by a $G$ quotient, then the ratio between the volumes should be the order $\vert G \vert$.} of the moduli spaces is also consistent with the gaugings performed. Indeed, $\frac{\mathrm{vol}^{-1}\left(a_2/\Z_2^2\right)}{\mathrm{vol}^{-1}\left(a_2\right)}=4$, $\frac{\mathrm{vol}^{-1}\left(a_2/\Z_2^2\rtimes\Sfrak_2\right)}{\mathrm{vol}^{-1}\left(a_2\right)}=8$, and $\frac{\mathrm{vol}^{-1}\left(a_2/\Z_2^2\rtimes\Sfrak_3\right)}{\mathrm{vol}^{-1}\left(a_2\right)}=24$.

In addition, full details of the computation of the Hilbert Series of these quotients using the Molien sum \eqref{molien} and \eqref{moliensumzq} are provided. Here, the fugacities $(z^\prime_1, z^\prime_2, z^\prime_3)$ for $\CG_{3,1}$, and rescaled fugacities $(z_1,z_2,z_3)=({z^\prime_1}^2,{z^\prime_2}^2,{z^\prime_3}^2)$ for $\CG_{3,2}$ are turned on. The quiver is ungauged by setting $m_3=0$ and $z^\prime_3=z_3=1$:
\begin{subequations}
\begin{align}
    \HS(a_2)&=\frac{1}{(1-t^2)^2}\sum_{m_1,m_2}{z^\prime_1}^{m_1}{z^\prime_2}^{m_2}t^{\vert m_1 \vert+\vert m_2 \vert+\vert m_1-m_2 \vert}
    \notag \\
    &=\frac{1 + 2 t^2 -(z^\prime_1+z^\prime_2+z^\prime_1 z^\prime_2+\frac{1}{z^\prime_1}+\frac{1}{z^\prime_2}+\frac{1}{z^\prime_1z^\prime_2})t^4 + 2 t^{6} + t^{8}}{(1-z^\prime_1 t^2)(1-z^\prime_2 t^2)(1-z^\prime_1 z^\prime_2 t^2)(1-\frac{t^2}{z^\prime_1})(1-\frac{t^2}{z^\prime_2})(1-\frac{t^2}{z^\prime_1z^\prime_2})}
    \\
    \HS(a_2/\mathbb Z_2^2)&=\frac{1}{4}\sum_{i=1}^2\sum_{j=1}^2\HS(a_2)\vert_{z_1'\to\omega_2^i\cdot z_1',z_2'\to\omega_2^j\cdot z_2'}\notag \\
    &=\frac{1}{(1-t^2)^2}\sum_{m_1,m_2}{z^\prime_1}^{2m_1}{z^\prime_2}^{2m_2}t^{2\vert m_1 \vert+2\vert m_2 \vert+2\vert m_1-m_2 \vert}\vert_{{z^\prime_1}^2=z_1,{z^\prime_2}^2=z_2}
    \notag \\
    &=\frac{1 + 2 t^4 -(z_1+z_2+z_1z_2+\frac{1}{z_1}+\frac{1}{z_2}+\frac{1}{z_1z_2})t^8 + 2 t^{12} + t^{16}}{(1-z_1 t^4)(1-z_2 t^4)(1-z_1 z_2 t^4)(1-\frac{t^4}{z_1})(1-\frac{t^4}{z_2})(1-\frac{t^4}{z_1z_2})(1+t^2)^{-2}}
    \\
    \HS(a_2/\mathbb Z_2^2\rtimes\Sfrak_2)
    &=\frac{1}{2}\left(\HS(a_2/\mathbb Z_2^2)+\HS(a_2/\mathbb Z_2^2)^{(1\ 2)}\right)
    \notag \\
    &=\frac{1}{2}\left(\HS(a_2/\mathbb Z_2^2)+\frac{1}{1-t^4}\sum_{m_1=m_2}z_1^{m_1}z_2^{m_2}t^{2\vert m_1 \vert+2\vert m_2 \vert+2\vert m_1-m2 \vert} \right)\vert_{z_1=z_2=z}
    \notag \\
    &=\frac{1 + t^2 + 2 t^4 + (2+z+\frac{1}{z}) t^6 + 2 t^8 + t^{10} + t^{12}}{(1-zt^4)(1-z^2t^4)(1-\frac{1}{z}t^4)(1-\frac{1}{z^2}t^4)}\\
    \label{eqs3}\HS(a_2/\mathbb Z_2^2\rtimes\Sfrak_3)&=\frac{1}{6}\left( \HS(a_2/\mathbb Z_2^2) + \HS(a_2/\mathbb Z_2^2)^{(1\ 2)} + \HS(a_2/\mathbb Z_2^2)^{(1\ 3)} + \HS(a_2/\mathbb Z_2^2)^{(2\ 3)} \right.
    \notag \\
    &\qquad \left. + \HS(a_2/\mathbb Z_2^2)^{(1\ 2\ 3)} + \HS(a_2/\mathbb Z_2^2)^{(1\ 3\ 2)} \right)
    \notag \\
    &=\frac{1}{6}\left( \HS(a_2/\mathbb Z_2^2)+\frac{1}{1-t^4}(\sum_{m_1=m_2}+\sum_{m_1=0,m_2}+\sum_{m_1,m_2=0})z_1^{m_1}z_2^{m_2}t^{2\vert m_1 \vert +2\vert m_2 \vert+2\vert m_1-m2 \vert}\right.
    \notag \\
    &\qquad \left. + \frac{1-t^2}{1-t^6} + \frac{1-t^2}{1-t^6}\right)\vert_{z_1=z_2=1}
    \notag \\
    &=\frac{1+2t^6+2t^8+t^{14}}{(1-t^4)^3(1-t^6)}\,.
\end{align}
\end{subequations}

Note that in \eqref{eqs3}, the elements $(1\ 2)$, $(1\ 3)$ and $(2\ 3)$ give the same contribution, so do the elements $(1\ 2\ 3)$ and $(1\ 3\ 2)$. In general, the elements in the same conjugacy of $\Sfrak_n$ class give the same contribution.

Finally, a comment on the $a_2/\mathbb Z_2^2\rtimes \Sfrak_3$. This case raises a special interest due to the accidental identification of the discrete symmetry $\Z_2^2\rtimes\Sfrak_3$ with $\Sfrak_4$.
This symplectic singularity is also found and constructed explicitly as part of the nilpotent cone of $\mathfrak{e}_8$ in \cite{2023arXiv230807398F}, as the slice between $E_8(a_6)$ and $E_8(b_6)$. The original construction is given in Appendix \ref{app:s4}. By comparing the Hilbert series with the generators and relations, the claim that the quiver and the Hilbert Series presented in the last row of Table \ref{tab:a_2} are indeed the magnetic quiver and Hilbert Series for $a_2/\Sfrak_4$, respectively is made.

In addition, the Hasse diagram and decorated quiver description of $a_2/\Sfrak_4$ is given in \Figref{hassea2s4}.
\begin{figure}[H]
    \centering
    \begin{tikzpicture}
                \node at (-0.5,-0.75) {$C_2$};
                \node at (-0.5,-2.25) {$\mu_2$};
                \node[hasse] (t) at (0,0) {};
                \node[hasse] (m) at (0,-1.5) {};
                \node[hasse] (b) at (0,-3) {};
                \draw (t)--(m)--(b);
                \node[gauge,label=below:{$1$}] (0) at (1,-0.2) {};
                
            \node[gauge,label=below:{$1$}] (1) at (2,-0.2) {};
            \node[gauge,label=above:{$1$}] (2) at (1.5,0.5) {};
            \draw[transform canvas={yshift=-1pt}] (0)--(1);
            \draw[transform canvas={yshift=1pt}] (0)--(1);
            \draw[transform canvas={yshift=-0.8pt,
             xshift=0.6pt}] (0)--(2);
            \draw[transform canvas={yshift=0.8pt,
             xshift=-0.6pt}] (0)--(2);
            \draw[transform canvas={yshift=0.8pt,
             xshift=0.6pt}] (1)--(2);
            \draw[transform canvas={yshift=-0.8pt,
             xshift=-0.6pt}] (1)--(2);
                \draw[purple] (0) circle (0.25cm);
                \draw[purple] (1) circle (0.25cm);
                \draw[purple] (2) circle (0.25cm);
          
            \node[gauge,label=below:{$1$}] (0) at (1,-1.5) {};
		\node[gauge,label=below:{$1$}] (1) at (2,-1.5) {};
		\draw[transform canvas={yshift=1pt}] (0)--(1);
        \draw[transform canvas={yshift=-1pt}] (0)--(1);
		\node at (1.5,-1.5+0.3) {\scriptsize $2,2$};
        \draw[purple] (0) circle (0.25cm);
        \draw[purple] (1) circle (0.25cm);
        \draw (1.4,-1.5-.2)--(1.6,-1.5)--(1.4,-1.5+.2);
        
        \node[gauge,label=below:{$1$}] (1) at (1.5,-3) {};
        \draw[purple] (1) circle (0.25cm);
            \end{tikzpicture}
    \caption{Hasse diagram of $a_2/\Sfrak_4$.}
    \label{hassea2s4}
\end{figure}
In the diagram, $C_2$ is the Kleinian singularity $A_3$ with an intrinsic symmetry $\Sfrak_2$, and $\mu_2$ is a non-normal slice with normalisation $A_3$\footnote{The slice $C_2$ can be generalised to $C_p$. The ring of $C_p$ is locally the same as $A_{2p-1}$, $\mathbb{C}[st,s^{2p},t^{2p}]$. Geometrically there is a monodromy map send $st\to-st$. It appears as the top slice of the $a_2/\Z_{p}^2\!\rtimes\!\Sfrak_3$ Hasse diagram. The non-normal slice $\mu_2$ can be generalised to $\mu_p$, whose normalisation is $A_{2p-1}$. The ring of $\mu_p$ is $\mathbb{C}[s^2t^2,s^3t^3,s^{2p},t^{2p},s^{2p+1}t,st^{2p+1}]$. It appears as the bottom slice of the $a_2/\Z_{p}^2\!\rtimes\!\Sfrak_3$ Hasse diagram.}. The decorated quiver with a simply-laced bouquet is introduced and discussed in \cite{Bourget_2022instanton,Bourget_2022dim6}. The investigation of the decorated quiver with a non-simply-laced bouquet or complete graphs is left for future work.

\subsection{\texorpdfstring{$A_3$ affine quiver with $\Sfrak_2$ and $\Z_2$ gauging}{a3 nilpotent}}

A similar analysis for the affine $A_3$ quiver: $\CG_{4,1}$ is performed. The Coulomb branch is the closure of the minimal nilpotent orbit of $A_3$, which is denoted $a_3$.
Hence, for $a_3$, quotients by $\Sfrak_2$, $\Sfrak_2^2$, $\mathbb Z_2^3$, and products thereof may be taken. The quotient of $\Sfrak_2$ and $\Sfrak_2^2$ are discussed above in Section \ref{sec:a3s2}. The resulting quivers, unrefined HS, and volume of the resulting moduli space are summarised in Table \ref{tab:a_3}.

\input{Tablea3}

Again, one verifies that the (inverse) volume ratios agree with the order of the quotient group: $\frac{\mathrm{vol}^{-1}\left(a_3/\Sfrak_2\right)}{\mathrm{vol}^{-1}\left(a_3\right)}=2$, $\frac{\mathrm{vol}^{-1}\left(a_3/\Sfrak_2\times\Sfrak_2\right)}{\mathrm{vol}^{-1}\left(a_3\right)}=4$, $\frac{\mathrm{vol}^{-1}\left(a_3/\Z_2^3\right)}{\mathrm{vol}^{-1}\left(a_3\right)}=8$, $\frac{\mathrm{vol}^{-1}\left(a_3/\Z_2^3\rtimes\Sfrak_2\right)}{\mathrm{vol}^{-1}\left(a_3\right)}=16$, and $\frac{\mathrm{vol}^{-1}\left(a_3/\Z_2^3\rtimes\Sfrak_2\!\times\!\Sfrak_2\right)}{\mathrm{vol}^{-1}\left(a_3\right)}=32$.

\subsection{\texorpdfstring{$D_4$ affine quiver with $\Sfrak_4$ gauging}{d4 nilpotent}}
The Coulomb branch of the $D_4$ affine quiver is $d_4$, the closure of the minimal nilpotent orbit of $\mathfrak{so}_8$. The quiver has a natural $\Sfrak_4$ action that permutes the $4$ $\urm(1)$ nodes. By gauging this $\Sfrak_4$ and its subgroups, the orbifolds of $d_4$ by the subgroup that is gauged can be obtained, the quivers are labelled by $(1)$ in Table \ref{tab:d_4}. This construction is well-studied in \cite{Hanany:2018vph,Hanany:2018dvd,Bourget:2020bxh,Gledhill:2021cbe}. Here a different realisation of the orbifold of $d_4$ by $\Sfrak_4$ and its subgroups is provided by using both the $\Z_q$ and the $S_n$ quotients, the quivers are labelled by $(2)$ in the Table \ref{tab:d_4}. The symplectic singularity in the last row of this table is also found and constructed explicitly as part of the nilpotent cone of $\mathfrak{f}_4$ in \cite{2023arXiv230807398F}, as the slice between $E_8(a_6)$ and $E_8(b_6)$.

\input{Tabled4}

A particularly interesting case is the symplectic singularity $d_4/\Sfrak_4$, where the action matches for both cases $(1)$ and $(2)$, shown in the last two rows of Table \ref{tab:d_4}. Therefore there are two different constructions of this singularity. One is a unitary quiver with an adjoint loop, and the other is a non-simply laced unitary quiver with adjoint. The Hasse diagrams are shown in \eqref{hassed4s4type1} and \eqref{hassed4s4type2}, where one can observe that it is computed in two different ways, starting from each corresponding quiver.
The brane system for the quiver in the last row will be discussed in an upcoming paper.

The Hasse diagram can be derived from the decorated quiver technique:

\begin{figure}[H]
    \centering
    \raisebox{-.5\height}{
    \begin{tikzpicture}
                \node at (-0.5,-0.75) {$A_1$};
                \node at (-0.9,-2.25) {$m$};
                \node at (1,-2.25) {$2A_1$};
                \node at (-0.9,-3.75) {$m$};
                \node at (0.9,-3.75) {$A_1$};
                \node at (-0.5,-5.25) {$a_2$};
                \node[hasse] (t) at (0,0) {};
                \node[hasse] (m1) at (0,-1.5) {};
                \node[hasse] (m2l) at (-1,-3) {};
                \node[hasse] (m2r) at (1,-3) {};
                \node[hasse] (m3) at (0,-4.5) {};
                \node[hasse] (b) at (0,-6) {};
                \draw (t)--(m1)--(m2l)--(m3)--(b) (m2r)--(m3);
                \draw[transform canvas={xshift=-2pt}](m1)--(m2r);
                \draw[transform canvas={xshift=2pt}](m1)--(m2r);
            %quiver1
            \node[gauge,label=below:{$2$}] (0) at (1-0.3,0.5) {};
            \node[gauge,label=above:{$1$}] (1) at (0-0.3,1.5) {};
            \node[gauge,label=above:{$1$}] (2) at (0.35,1.5) {};
            \node[gauge,label=above:{$1$}] (3) at (1.05,1.5) {};
            \node[gauge,label=above:{$1$}] (4) at (2-0.3,1.5) {};
		\draw (0)--(1) (0)--(2) (0)--(3) (0)--(4);
        \draw[purple] (1) circle (0.25cm);
        \draw[purple] (2) circle (0.25cm);
        \draw[purple] (3) circle (0.25cm);
        \draw[purple] (4) circle (0.25cm);
            %quiver2
            \node[gauge,label=below:{$2$}] (0) at (1.5,-1.5) {};
            \node[gauge,label=above:{$1$}] (2) at (1,1-1.5) {};
            \node[gauge,label=above:{$1$}] (3) at (1.85,1-1.5) {};
            \node[gauge,label=above:{$1$}] (4) at (2.5,1-1.5) {};
        \draw (1.2,-1-0.25)--(1.2,-0.9)--(1.45,-1.1);
        \draw[transform canvas={xshift=1pt}] (0)--(2);
        \draw[transform canvas={xshift=-1pt}] (0)--(2);
		\draw (0)--(3) (0)--(4);
        \draw[purple] (2) circle (0.25cm);
        \draw[purple] (3) circle (0.25cm);
        \draw[purple] (4) circle (0.25cm);
        %quiver3
        \node[gauge,label=below:{$2$}] (0) at (1.5+1,-3.2) {};
            \node[gauge,label=above:{$1$}] (2) at (1+1,1-3.2) {};
            \node[gauge,label=above:{$1$}] (4) at (2+1,1-3.2) {};
        \draw[transform canvas={xshift=1pt}] (0)--(2);
        \draw[transform canvas={xshift=-1pt}] (0)--(2);
        \draw (2.2,-2.7-0.25)--(2.2,-2.6)--(2.45,-2.8);
		\draw[transform canvas={xshift=1pt}] (0)--(4);
        \draw[transform canvas={xshift=-1pt}] (0)--(4);
        \draw (2.8,-2.7-0.25)--(2.8,-2.6)--(2.55,-2.8);
        \draw[purple] (2) circle (0.25cm);
        \draw[purple] (4) circle (0.25cm);
        %quiver4
                \node[gauge,label=below:{$2$}] (0) at (-1.5-1,-3.2) {};
            \node[gauge,label=above:{$1$}] (2) at (-1-1,1-3.2) {};
            \node[gauge,label=above:{$1$}] (4) at (-2-1,1-3.2) {};
        \draw[transform canvas={xshift=0pt}] (0)--(2);
        \draw[transform canvas={xshift=0pt}] (0)--(4);
		\draw[transform canvas={xshift=2pt}] (0)--(4);
        \draw[transform canvas={xshift=-2pt}] (0)--(4);
        \draw (-2.8,-2.7-0.25)--(-2.8,-2.6)--(-2.55,-2.8);
        \draw[purple] (2) circle (0.25cm);
        \draw[purple] (4) circle (0.25cm);
        %quiver5
                \node[gauge,label=below:{$2$}] (0) at (1.7,-5.2) {};
            \node[gauge,label=above:{$1$}] (4) at (1.7,1-5.2) {};
        \draw[transform canvas={xshift=-3pt}] (0)--(4);
        \draw[transform canvas={xshift=-1pt}] (0)--(4);
		\draw[transform canvas={xshift=1pt}] (0)--(4);
        \draw[transform canvas={xshift=3pt}] (0)--(4);
        \draw (1.5,-4.8)--(1.7,-4.6)--(1.9,-4.8);
        \draw[purple] (4) circle (0.25cm);
            \end{tikzpicture}
            }
    \caption{Hasse diagram and decorated quiver of $d_4/\Sfrak_4(1)$.}
    \label{hassed4s4type1}
\end{figure}
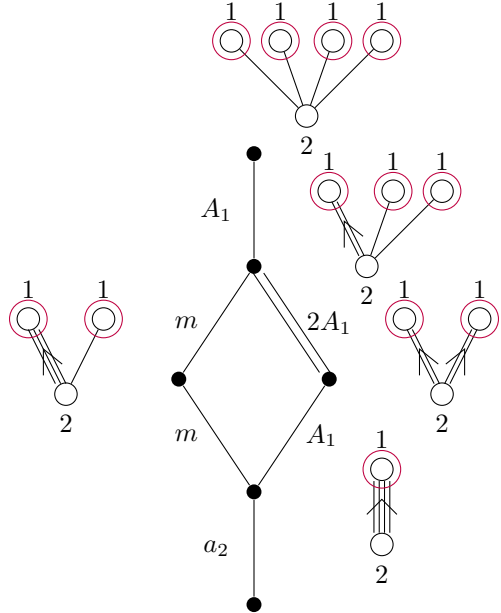

\begin{figure}[H]
    \centering
    \raisebox{-.5\height}{
    \begin{tikzpicture}
                \node at (-0.5,-0.75) {$A_1$};
                \node at (-0.9,-2.25) {$m$};
                \node at (1,-2.25) {$2A_1$};
                \node at (-0.9,-3.75) {$m$};
                \node at (0.9,-3.75) {$A_1$};
                \node at (-0.5,-5.25) {$a_2$};
                \node[hasse] (t) at (0,0) {};
                \node[hasse] (m1) at (0,-1.5) {};
                \node[hasse] (m2l) at (-1,-3) {};
                \node[hasse] (m2r) at (1,-3) {};
                \node[hasse] (m3) at (0,-4.5) {};
                \node[hasse] (b) at (0,-6) {};
                \draw (t)--(m1)--(m2l)--(m3)--(b) (m2r)--(m3);
                \draw[transform canvas={xshift=-2pt}](m1)--(m2r);
                \draw[transform canvas={xshift=2pt}](m1)--(m2r);
            %quiver1
            \node[gauge,label=below:{$2$}] (0) at (0.7,1) {};
            \node[gauge,label=below:{$1$}] (1) at (-0.3,1) {};
            \node[gauge,label=right:{$1$}] (2) at (1.7,1.7) {};            \node[gauge,label=right:{$1$}] (3) at (1.7,1) {};            \node[gauge,label=right:{$1$}] (4) at (1.7,0.3) {};
            \draw (0)--(1);
            \draw[transform canvas={xshift=-0.6pt,yshift=0.8pt}] (0)--(2);
            \draw[transform canvas={xshift=0.6pt,yshift=-0.8pt}] (0)--(2);
            \draw[transform canvas={xshift=0.6pt,yshift=0.8pt}] (0)--(4);
            \draw[transform canvas={xshift=-0.6pt,yshift=-0.8pt}] (0)--(4);
            \draw[transform canvas={yshift=1pt}] (0)--(3);
            \draw[transform canvas={yshift=-1pt}] (0)--(3);
            \draw (1.25,1.15)--(1.1,1)--(1.25,0.85);
            \draw (1.35,1.25)--(1.15,1.3)--(1.2,1.5);
            \draw (1.35,0.75)--(1.15,0.7)--(1.2,0.5);
            \draw[purple] (2) circle (0.25cm);
            \draw[purple] (3) circle (0.25cm);
            \draw[purple] (4) circle (0.25cm);
            %quiver2
            \node[gauge,label=below:{$2$}] (0) at (2.2,-1.5) {};
            \node[gauge,label=below:{$1$}] (1) at (1.2,-1.5) {};
            \node[gauge,label=left:{$1$}] (2) at (2.2,-.5) {};
            \node[gauge,label=below:{$1$}] (3) at (3.2,-1.5) {};
            \draw (0)--(1);
            \draw[transform canvas={xshift=-1pt}] (0)--(2);
            \draw[transform canvas={xshift=1pt}] (0)--(2);
            \draw (2.2+0.15,-1.5+0.65)--(2.2,-1.5+0.5)--(2.2-0.15,-1.5+0.65);
            \draw (2.2+0.15,-1.5+0.25)--(2.2,-1.5+0.4)--(2.2-0.15,-1.5+0.25);
            \draw[transform canvas={yshift=-1pt}] (0)--(3);
            \draw[transform canvas={yshift=1pt}] (0)--(3);
            \draw (2.2+0.55,-1.5+0.15)--(2.2+0.4,-1.5)--(2.2+0.55,-1.5-0.15);
            \draw[purple] (2) circle (0.25cm);
            \draw[purple] (3) circle (0.25cm);
            %quiver3
            \node[gauge,label=below:{$1$}] (0) at (2.5,-3.5) {};
            \node[gauge,label=below:{$1$}] (1) at (1.5,-3.5) {};
            \node[gauge,label=left:{$1$}] (2) at (2.5,-2.5) {};
            \node[gauge,label=below:{$1$}] (3) at (3.5,-3.5) {};
            \draw (0)--(1) (1)--(2);
            \draw[transform canvas={yshift=-1pt}] (0)--(3);
            \draw[transform canvas={yshift=1pt}] (0)--(3);
            \draw (2.5+0.55,-3.5+0.15)--(2.5+0.4,-3.5)--(2.5+0.55,-3.5-0.15);
            \draw[transform canvas={xshift=0.7pt,yshift=0.7pt}] (2)--(3);
            \draw[transform canvas={xshift=-0.7pt,yshift=-0.7pt}] (2)--(3);
            \draw (3.5-0.3,-3.5+0.5)--(3.5-0.6,-3.5+0.6)--(3.5-0.5,-3.5+0.3);
            \draw[purple] (3) circle (0.25cm);
            %quiver4
            \node[gauge,label=below:{$2$}] (0) at (-2.5,-3) {};
            \node[gauge,label=below:{$1$}] (1) at (-1.5,-3) {};
            \node[gauge,label=below:{$1$}] (2) at (-3.5,-3) {};
            \node[] (3) at (-2.13,-3) {};
            \node[] (4) at (-1.87,-3) {};
            \draw[transform canvas={yshift=-1pt}] (1)--(3);
            \draw[transform canvas={yshift=1pt}] (1)--(3);
            \draw (-2.5+0.75,-3+0.15)--(-2.5+0.6,-3)--(-2.5+0.75,-3-0.15);
            \draw[transform canvas={yshift=-2pt}] (0)--(4);
            \draw[transform canvas={yshift=2pt}] (0)--(4);
            \draw[transform canvas={yshift=0pt}] (0)--(4);
            \draw (-2.5+0.25,-3+0.15)--(-2.5+0.4,-3)--(-2.5+0.25,-3-0.15);
            \draw (0)--(2);
            \draw[purple] (1) circle (0.25cm);
            %quiver5
            \node[gauge,label=below:{$1$}] (0) at (1.2,-5.5) {};
            \node[gauge,label=below:{$1$}] (1) at (2.8,-5.5) {};
            \node[gauge,label=left:{$1$}] (2) at (2,-4.6) {};
            \draw (0)--(1) (1)--(2) (2)--(0);
            \end{tikzpicture}
            }
            \caption{Hasse diagram and decorated quiver of $d_4/\Sfrak_4(2)$.}
\label{hassed4s4type2}
\end{figure}

\subsection{\texorpdfstring{$\CG_{4,2}$ quiver with $\Sfrak_4$ gauging}{(A3,A7) gauge theory}}
As a next example, consider the $\CG_{4,2}$ quiver, for which one can quotient $\mathfrak{S}_2$, $\mathfrak{S}_2\times\mathfrak{S}_2$, $\mathfrak{S}_3$, or $\mathfrak{S}_4$. The resulting quivers are collected in Table \ref{tab:cog4}.

\input{Tablecog4}
Again, one verifies straightforwardly that the computed (inverse) volume ratios agree with the orders of the quotient groups: $\frac{\mathrm{vol}^{-1}\left(\CG_{4,2}/\Sfrak_2\right)}{\mathrm{vol}^{-1}\left(\CG_{4,2}\right)}=2$, $\frac{\mathrm{vol}^{-1}\left(\CG_{4,2}/\Sfrak_2\times\Sfrak_2\right)}{\mathrm{vol}^{-1}\left(\CG_{4,2}\right)}=4$, $\frac{\mathrm{vol}^{-1}\left(\CG_{4,2}/\Sfrak_3\right)}{\mathrm{vol}^{-1}\left(\CG_{4,2}\right)}=6$, and $\frac{\mathrm{vol}^{-1}\left(\CG_{4,2}/\Sfrak_4\right)}{\mathrm{vol}^{-1}\left(\CG_{4,2}\right)}=24$. 

\subsection{Orbifolds and covers of a hypersurface}
Recalling the hypersurface \Quiver{fig:quiver21loop} and its cover \Quiver{fig:quiver211tri}, one can study all possible discrete quotients of the latter.
Specifically, this is a complete graph with only two nodes and edge multiplicity $2g-2$, and connected to a background $\urm(1)$ gauge node. The $\Z_2$ quotient may be taken by either changing the edge multiplicity and changing the connection into non-simply laced, or quotient $\mathfrak{S}_2$ by merging the two edges. See Table \ref{tab:hypersurface} for a summary.

\input{Tablehsq}

The most interesting case is the moduli space $\Ccal(\text{\Quiver{fig:quiver211tri}})/\mathbb{Z}_q\!\rtimes\!\Sfrak_2$. The Hilbert series indicates that it is a complete intersection: it has $3$ generators in adjoint representation of $\surm(2)$ at degree $2$, and $3$ generators in adjoint representation of $\surm(2)$ at degree $4g-2$. Also there is one relation in the trivial representation at degree $4g$, and another relation in the trivial representation at degree $8g-4$.

\section{Generalisations}
\label{sec:general}
It is straightforward to generalise the $\mathfrak{S}_n$ quotient to quivers with other types of gauge groups with multiple adjoint hypermultiplets\footnote{This is the natural generalisation of the bouquets with different gauge nodes of \cite{Hanany:2018cgo}.}; see Table \ref{tab:generalisation}. For instance, Figures \ref{fig:generalise_Sp}, \ref{fig:generalise_SO}, and \ref{fig:generalise_O} illustrate such cases. The proof proceeds analogous to that Section~\ref{sec:Sn}. The following results are found:
\begin{align}
\Ccal(\text{\Quiver{fig:USp2nGadj}})&=\Ccal(\text{\Quiver{fig:mutatedsprose}})/\mathfrak{S}_n
    \\ \Ccal(\text{\Quiver{fig:so2n1Gadj}})&=\Ccal(\text{\Quiver{fig:mutatedso3rose}})/\mathfrak{S}_n
    \\ \Ccal(\text{\Quiver{fig:so2nGadj}})&=\Ccal(\text{\Quiver{fig:mutatedso2rose}})/\mathfrak{S}_n.
\end{align}

\begin{figure}[H]
\centering
\begin{subfigure}[H]{0.45\textwidth}
    \centering
    \begin{tikzpicture}[main/.style={draw,circle}]
    \node[main, label=left:$\usprm(2)$,color=orange!100] (top) at (0,2.5) {};
    \draw (top) to [out=135, in=45,looseness=8] node[pos=0.5,above]{$g$} (top);
    \node[main, label=right:$\usprm(2)$,color=orange!100] (topright) at ({2.5*cos(18)},{2.5*sin(18)}){};
    \draw (topright) to [out=135, in=45,looseness=8] node[pos=0.5,above]{$g$} (topright);
    \node[main, label=left:$\usprm(2)$,color=orange!100] (topleft) at ({-2.5*cos(18)},{2.5*sin(18)}){};
    \draw (topleft) to [out=135, in=45,looseness=8] node[pos=0.5,above]{$g$} (topleft);
    \node[main, label=below:$\usprm(2)$,color=orange!100] (bottomright) at ({2.5*cos(54)},{-2.5*sin(54)}){};
    \draw (bottomright) to [out=45, in=-45,looseness=8] node[pos=0.5,right]{$g$} (bottomright);
    \node[main, label=below:$\usprm(2)$,color=orange!100] (bottomleft) at ({-2.5*cos(54)},{-2.5*sin(54)}){};
    \draw (bottomleft) to [out=225, in=135,looseness=8] node[pos=0.5,left]{$g$} (bottomleft);
    \path (topright) -- (bottomright) node[midway,sloped] (dotsr) {$\cdots$};
    \path (topleft) -- (bottomleft) node[midway,sloped] (dotsl) {$\cdots$};

    \draw[-] (top)--(topright) node[pos=0.5,above,sloped]{$2g-2$}--(dotsr)--(bottomright)--(bottomleft)node[pos=0.5,below,sloped]{$2g-2$}--(dotsl)--(topleft)--(topright)node[pos=0.5,below,sloped]{$2g-2$}--(bottomleft)--(top)--(bottomright)--(topleft)--(top)node[pos=0.5,above,sloped]{$2g-2$};
        
    \end{tikzpicture}
    \caption{}
    \label{fig:mutatedsprose}
\end{subfigure}
\begin{subfigure}[H]{0.45\textwidth}
    \centering
    \begin{tikzpicture}[main/.style={draw,circle}]
    \node[main, label=below:$\usprm(2n)$,color=orange!100] (n) []{};
    \draw (n) to [out=135, in=45,looseness=8] node[pos=0.5,above]{$g$} (n);
    \end{tikzpicture}
    \caption{}
    \label{fig:USp2nGadj}
\end{subfigure}
\caption{\subref{fig:mutatedsprose}: Complete graph quiver with $n$ nodes of $\usprm(2)$ with $g$ adjoints and edges of multiplicity $2g-2$ connecting all pairs. \subref{fig:USp2nGadj}: Quiver for $\usprm(2n)$ gauge group with $g$ adjoints.}
\label{fig:generalise_Sp}
\end{figure}

\paragraph{$\usprm(4)$ Example.}
$\usprm(4)$ is used as an example to demonstrate the generalisation. A $\mathfrak{S}_2$ quotient on the Coulomb branch can be performed as follows:
\begin{equation}
\Ccal\left(\mathcal{Q}_{\usprm}(2,g)\right)=\Ccal\left(
        \vcenter{\hbox{
        \begin{tikzpicture}
            \node[gauge,label=below:{$\usprm(2)$},color=orange!100] (0) at (0,0) {};
            \node[gauge,label=below:{$\usprm(2)$},color=orange!100] (1) at (2,0) {};
            \node[label=above:{$2g-2$}] at (1,-0.1) {};
            \draw (0)--(1);
            \draw (0) to [out=135, in=45,looseness=8] node[pos=0.5,above]{$g$} (0);
            \draw (1) to [out=135, in=45,looseness=8] node[pos=0.5,above]{$g$} (1);
        \end{tikzpicture}
        }}\right) \quad \overset{\Sfrak_2}{\longrightarrow} \quad
\Ccal\left(\mathcal{Q}_{\usprm}(2,g)/\Sfrak_2 \right)=\Ccal\left(
    \vcenter{\hbox{
        \begin{tikzpicture}
           [main/.style={draw,circle}]
        \node[main, label=below:$\usprm(4)$,color=orange!100] (2) []{};
        \draw (2) to [out=135, in=45,looseness=8] node[pos=0.5,above]{$g$} (2);
        \end{tikzpicture}
    }}\right).
\end{equation}
The Hilbert series for the two Coulomb branches are calculated by an $\mathfrak{S}_2$ Molien sum
\begin{subequations}
\begin{align}
    \HS\left(\Ccal\left(\mathcal{Q}_{\usprm}(2,g)\right)\right)&=\frac{1}{(1-t^4)^2} + \left(\frac{1}{(1-t^2)(1-t^4)}\sum_{m_1=0,m_2>0} +\frac{1}{(1-t^2)(1-t^4)}\sum_{m_1>0,m2=0}\right.
    \notag \\
    &\quad \left.+\frac{1}{(1-t^2)^2}\sum_{m_1,m_2>0}\right) t^{(4g-4)(\vert m_1\vert+\vert m_2\vert)+(2g-2)(\vert m_1-m_2\vert+\vert m_1+m_2\vert)}
    \notag \\
    &=\frac{1+t^{8g-8}+2t^{8g-6}+2t^{12g-10}+t^{12g-8}+t^{20g-16}}{(1-t^4)^2(1-t^{8g-8})(1-t^{12g-12})}.
    \\
    \HS\left(\Ccal\left(\mathcal{Q}_{\usprm}(2,g)/\Sfrak_2\right)\right)&=\frac{1}{2}\left(\HS\left(\mathcal{Q}_{\usprm}(2,g)\right)+\frac{1}{1-t^8}+\frac{1}{1-t^4}\sum_{m_1=m_2>0} t^{(4g-4)(\vert m_1\vert+\vert m_2\vert)+(2g-2)(\vert m_1-m_2\vert+\vert m_1+m_2\vert)}\right)
    \notag \\
    &=\frac{1 + t^{8g-6} + t^{8g-4} + t^{8g-2} + t^{12g-10} + t^{12g-8} + t^{12g-6} + t^{20g-12}}{(1-t^4)(1-t^8)(1-t^{8g-8})(1-t^{12g-12})}.
\end{align}
\end{subequations}
and the last line agrees with the direct monopole formula evaluation of the $\usprm(4)$ theory. Thus, validating the discrete quotient proposal.
It then also follows that the (inverse) volume ratio agrees with the order quotient group: $\frac{\mathrm{vol}^{-1}\left(\mathcal{Q}_{\usprm}(2,g)/\Sfrak_2\right)}{\mathrm{vol}^{-1}\left(\mathcal{Q}_{\usprm}(2,g)\right)}=2$.

\begin{figure}[H]
\centering
\begin{subfigure}[H]{0.45\textwidth}
    \centering
    \begin{tikzpicture}[main/.style={draw,circle}]
    \node[main, label=left:$\sorm(3)$,color=orange!100] (top) at (0,2.5) {};
    \node[main, label=right:$\sorm(3)$,color=orange!100] (topright) at ({2.5*cos(18)},{2.5*sin(18)}){};
    \node[main, label=left:$\sorm(3)$,color=orange!100] (topleft) at ({-2.5*cos(18)},{2.5*sin(18)}){};
    \node[main, label=below:$\sorm(3)$,color=orange!100] (bottomright) at ({2.5*cos(54)},{-2.5*sin(54)}){};
    \node[main, label=below:$\sorm(3)$,color=orange!100] (bottomleft) at ({-2.5*cos(54)},{-2.5*sin(54)}){};
    \draw (top) to [out=135, in=45,looseness=8] node[pos=0.5,above]{$g$} (top);
    \draw (topright) to [out=135, in=45,looseness=8] node[pos=0.5,above]{$g$} (topright);
    \draw (topleft) to [out=135, in=45,looseness=8] node[pos=0.5,above]{$g$} (topleft);
    \draw (bottomright) to [out=45, in=-45,looseness=8] node[pos=0.5,right]{$g$} (bottomright);
    \draw (bottomleft) to [out=225, in=135,looseness=8] node[pos=0.5,left]{$g$} (bottomleft);
    \path (topright) -- (bottomright) node[midway,sloped] (dotsr) {$\cdots$};
    \path (topleft) -- (bottomleft) node[midway,sloped] (dotsl) {$\cdots$};

    \draw[-] (top)--(topright) node[pos=0.5,above,sloped]{$2g-2$}--(dotsr)--(bottomright)--(bottomleft)node[pos=0.5,below,sloped]{$2g-2$}--(dotsl)--(topleft)--(topright)node[pos=0.5,below,sloped]{$2g-2$}--(bottomleft)--(top)--(bottomright)--(topleft)--(top)node[pos=0.5,above,sloped]{$2g-2$};
        
    \end{tikzpicture}
    \caption{}
    \label{fig:mutatedso3rose}
\end{subfigure}
\begin{subfigure}[H]{0.45\textwidth}
    \centering
    \begin{tikzpicture}[main/.style={draw,circle}]
    \node[main, label=below:$\sorm(2n+1)$,color=orange!100] (n) []{};
    \draw (n) to [out=135, in=45,looseness=8] node[pos=0.5,above]{$g$} (n);
    \end{tikzpicture}
    \caption{}
    \label{fig:so2n1Gadj}
\end{subfigure}
\caption{\subref{fig:mutatedso3rose}: Complete graph quiver with $n$ nodes of $\sorm(3)$ with $g$ adjoints and edges of multiplicity $2g-2$ connecting all pairs. \subref{fig:so2n1Gadj}: Quiver for $\sorm(2n+1)$ gauge group with $g$ adjoints.}
\label{fig:generalise_SO}
\end{figure}

\begin{figure}[H]
\centering
\begin{subfigure}[H]{0.45\textwidth}
    \centering
    \begin{tikzpicture}[main/.style={draw,circle}]
    \node[main, label=left:$\orm(2)$,color=orange!100] (top) at (0,2.5) {};

    \node[main, label=right:$\orm(2)$,color=orange!100] (topright) at ({2.5*cos(18)},{2.5*sin(18)}){};

    \node[main, label=left:$\orm(2)$,color=orange!100] (topleft) at ({-2.5*cos(18)},{2.5*sin(18)}){};

    \node[main, label=below:$\orm(2)$,color=orange!100] (bottomright) at ({2.5*cos(54)},{-2.5*sin(54)}){};

    \node[main, label=below:$\orm(2)$,color=orange!100] (bottomleft) at ({-2.5*cos(54)},{-2.5*sin(54)}){};
    \draw (top) to [out=135, in=45,looseness=8] node[pos=0.5,above]{$g$} (top);
    \draw (topright) to [out=135, in=45,looseness=8] node[pos=0.5,above]{$g$} (topright);
    \draw (topleft) to [out=135, in=45,looseness=8] node[pos=0.5,above]{$g$} (topleft);
    \draw (bottomright) to [out=45, in=-45,looseness=8] node[pos=0.5,right]{$g$} (bottomright);
    \draw (bottomleft) to [out=225, in=135,looseness=8] node[pos=0.5,left]{$g$} (bottomleft);

    \path (topright) -- (bottomright) node[midway,sloped] (dotsr) {$\cdots$};
    \path (topleft) -- (bottomleft) node[midway,sloped] (dotsl) {$\cdots$};

    \draw[-] (top)--(topright) node[pos=0.5,above,sloped]{$2g-2$}--(dotsr)--(bottomright)--(bottomleft)node[pos=0.5,below,sloped]{$2g-2$}--(dotsl)--(topleft)--(topright)node[pos=0.5,below,sloped]{$2g-2$}--(bottomleft)--(top)--(bottomright)--(topleft)--(top)node[pos=0.5,above,sloped]{$2g-2$};
        
    \end{tikzpicture}
    \caption{}
    \label{fig:mutatedso2rose}
\end{subfigure}
\begin{subfigure}[H]{0.45\textwidth}
    \centering
    \begin{tikzpicture}[main/.style={draw,circle}]
    \node[main, label=below:$\orm(2n)$,color=orange!100] (n) []{};
    \draw (n) to [out=135, in=45,looseness=8] node[pos=0.5,above]{$g$} (n);
    \end{tikzpicture}
    \caption{}
    \label{fig:so2nGadj}
\end{subfigure}
\caption{\subref{fig:mutatedso3rose}: Complete graph quiver with $n$ nodes of $\orm(2)$ with edges of multiplicity $2g-2$ connecting all pairs. \subref{fig:so2n1Gadj}: Quiver for $\orm(2n)$ gauge group with $g$ adjoints.}
\label{fig:generalise_O}
\end{figure}

As in \cite{Hanany:2018cgo}, this statement can be further generalised to symplectic gauge groups with $g$ hypermultiplets in the traceless second rank anti-symmetric product $\Lambda^2$, and special orthogonal groups with $g$ hypermultiplets in traceless second rank symmetric product $S^2$; see Table \ref{tab:generalisation}. Again, the proof proceeds analogous to that in Section~\ref{sec:Sn}. The following results are found:
\begin{align}
    \Ccal(\text{\Quiver{fig:USp2nGadj2}})&=\Ccal(\text{\Quiver{fig:mutatedsprose2}})/\mathfrak{S}_n
    \\ \Ccal(\text{\Quiver{fig:so2n1Gadj2}})&=\Ccal(\text{\Quiver{fig:mutatedso3rose2}})/\mathfrak{S}_n
    \\ \Ccal(\text{\Quiver{fig:so2nGadj2}})&=\Ccal(\text{\Quiver{fig:mutatedso2rose2}})/\mathfrak{S}_n.
\end{align}

\begin{figure}[H]
\centering
\begin{subfigure}[H]{0.45\textwidth}
    \centering
    \begin{tikzpicture}[main/.style={draw,circle}]
    \node[main, label=left:$\usprm(2)$,color=orange!100] (top) at (0,2.5) {};
    \draw (top) to [out=135, in=45,looseness=8] node[pos=0.5,above]{$g\cdot\Lambda^2$} (top);
    \node[main, label=right:$\usprm(2)$,color=orange!100] (topright) at ({2.5*cos(18)},{2.5*sin(18)}){};
    \draw (topright) to [out=135, in=45,looseness=8] node[pos=0.5,above]{$g\cdot\Lambda^2$} (topright);
    \node[main, label=left:$\usprm(2)$,color=orange!100] (topleft) at ({-2.5*cos(18)},{2.5*sin(18)}){};
    \draw (topleft) to [out=135, in=45,looseness=8] node[pos=0.5,above]{$g\cdot\Lambda^2$} (topleft);
    \node[main, label=below:$\usprm(2)$,color=orange!100] (bottomright) at ({2.5*cos(54)},{-2.5*sin(54)}){};
    \draw (bottomright) to [out=45, in=-45,looseness=8] node[pos=0.5,right]{$g\cdot\Lambda^2$} (bottomright);
    \node[main, label=below:$\usprm(2)$,color=orange!100] (bottomleft) at ({-2.5*cos(54)},{-2.5*sin(54)}){};
    \draw (bottomleft) to [out=225, in=135,looseness=8] node[pos=0.5,left]{$g\cdot\Lambda^2$} (bottomleft);
    \path (topright) -- (bottomright) node[midway,sloped] (dotsr) {$\cdots$};
    \path (topleft) -- (bottomleft) node[midway,sloped] (dotsl) {$\cdots$};

    \draw[-] (top)--(topright) node[pos=0.5,above,sloped]{$2g-2$}--(dotsr)--(bottomright)--(bottomleft)node[pos=0.5,below,sloped]{$2g-2$}--(dotsl)--(topleft)--(topright)node[pos=0.5,below,sloped]{$2g-2$}--(bottomleft)--(top)--(bottomright)--(topleft)--(top)node[pos=0.5,above,sloped]{$2g-2$};
        
    \end{tikzpicture}
    \caption{}
    \label{fig:mutatedsprose2}
\end{subfigure}
\begin{subfigure}[H]{0.45\textwidth}
    \centering
    \begin{tikzpicture}[main/.style={draw,circle}]
    \node[main, label=below:$\usprm(2n)$,color=orange!100] (n) []{};
    \draw (n) to [out=135, in=45,looseness=8] node[pos=0.5,above]{$g\cdot\Lambda^2$} (n);
    \end{tikzpicture}
    \caption{}
    \label{fig:USp2nGadj2}
\end{subfigure}
\caption{\subref{fig:mutatedsprose}: Complete graph quiver with $n$ nodes of $\usprm(2)$ and edges of multiplicity $2g-2$ connecting all pairs. \subref{fig:USp2nGadj}: Quiver for $\usprm(2n)$ gauge group with $g\cdot\Lambda^2$ hypermultiplets.}
\label{fig:generalise_Sp_2}
\end{figure}

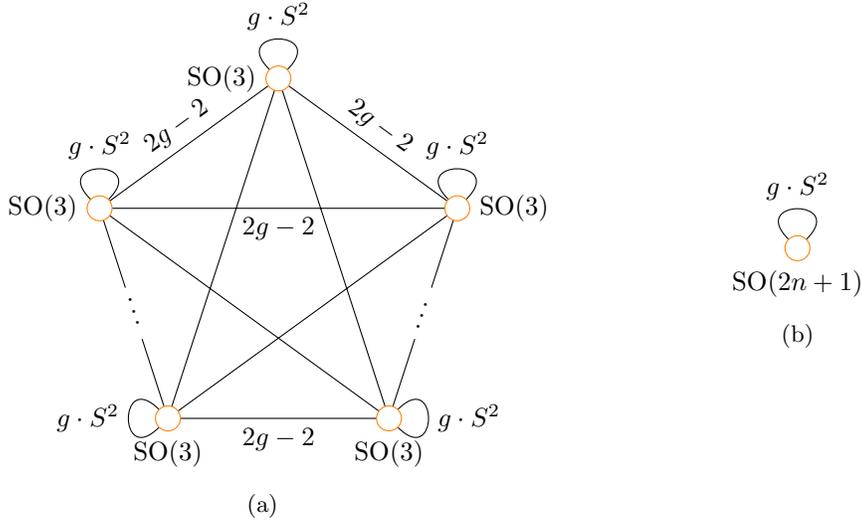
\begin{figure}[H]
\centering
\begin{subfigure}[H]{0.45\textwidth}
    \centering
    \begin{tikzpicture}[main/.style={draw,circle}]
    \node[main, label=left:$\sorm(3)$,color=orange!100] (top) at (0,2.5) {};
    \draw (top) to [out=135, in=45,looseness=8] node[pos=0.5,above]{$g\cdot S^2$} (top);
    \node[main, label=right:$\sorm(3)$,color=orange!100] (topright) at ({2.5*cos(18)},{2.5*sin(18)}){};
    \draw (topright) to [out=135, in=45,looseness=8] node[pos=0.5,above]{$g\cdot S^2$} (topright);
    \node[main, label=left:$\sorm(3)$,color=orange!100] (topleft) at ({-2.5*cos(18)},{2.5*sin(18)}){};
    \draw (topleft) to [out=135, in=45,looseness=8] node[pos=0.5,above]{$g\cdot S^2$} (topleft);
    \node[main, label=below:$\sorm(3)$,color=orange!100] (bottomright) at ({2.5*cos(54)},{-2.5*sin(54)}){};
    \draw (bottomright) to [out=45, in=-45,looseness=8] node[pos=0.5,right]{$g\cdot S^2$} (bottomright);
    \node[main, label=below:$\sorm(3)$,color=orange!100] (bottomleft) at ({-2.5*cos(54)},{-2.5*sin(54)}){};
    \draw (bottomleft) to [out=225, in=135,looseness=8] node[pos=0.5,left]{$g\cdot S^2$} (bottomleft);
    \path (topright) -- (bottomright) node[midway,sloped] (dotsr) {$\cdots$};
    \path (topleft) -- (bottomleft) node[midway,sloped] (dotsl) {$\cdots$};

    \draw[-] (top)--(topright) node[pos=0.5,above,sloped]{$2g-2$}--(dotsr)--(bottomright)--(bottomleft)node[pos=0.5,below,sloped]{$2g-2$}--(dotsl)--(topleft)--(topright)node[pos=0.5,below,sloped]{$2g-2$}--(bottomleft)--(top)--(bottomright)--(topleft)--(top)node[pos=0.5,above,sloped]{$2g-2$};
        
    \end{tikzpicture}
    \caption{}
    \label{fig:mutatedso3rose2}
\end{subfigure}
\begin{subfigure}[H]{0.45\textwidth}
    \centering
    \begin{tikzpicture}[main/.style={draw,circle}]
    \node[main, label=below:$\sorm(2n+1)$,color=orange!100] (n) []{};
    \draw (n) to [out=135, in=45,looseness=8] node[pos=0.5,above]{$g\cdot S^2$} (n);
    \end{tikzpicture}
    \caption{}
    \label{fig:so2n1Gadj2}
\end{subfigure}
\caption{\subref{fig:mutatedso3rose}: Complete graph quiver with $n$ nodes of $\sorm(3)$ with $g\cdot S^2$ hypermultiplets and edges of multiplicity $2g-2$ connecting all pairs. \subref{fig:so2n1Gadj}: Quiver for $\sorm(2n+1)$ gauge group with $g\cdot S^2$ hypermultiplets.}
\label{fig:generalise_SO_2}
\end{figure}

\begin{figure}[H]
\centering
\begin{subfigure}[H]{0.45\textwidth}
    \centering
    \begin{tikzpicture}[main/.style={draw,circle}]
    \node[main, label=left:$\orm(2)$,color=orange!100] (top) at (0,2.5) {};
    \draw (top) to [out=135, in=45,looseness=8] node[pos=0.5,above]{$g\cdot S^2$} (top);
    \node[main, label=right:$\orm(2)$,color=orange!100] (topright) at ({2.5*cos(18)},{2.5*sin(18)}){};
    \draw (topright) to [out=135, in=45,looseness=8] node[pos=0.5,above]{$g\cdot S^2$} (topright);
    \node[main, label=left:$\orm(2)$,color=orange!100] (topleft) at ({-2.5*cos(18)},{2.5*sin(18)}){};
    \draw (topleft) to [out=135, in=45,looseness=8] node[pos=0.5,above]{$g\cdot S^2$} (topleft);
   \node[main, label=below:$\orm(2)$,color=orange!100] (bottomright) at ({2.5*cos(54)},{-2.5*sin(54)}){};
    \draw (bottomright) to [out=45, in=-45,looseness=8] node[pos=0.5,right]{$g\cdot S^2$} (bottomright);
    \node[main, label=below:$\orm(2)$,color=orange!100] (bottomleft) at ({-2.5*cos(54)},{-2.5*sin(54)}){};
    \draw (bottomleft) to [out=225, in=135,looseness=8] node[pos=0.5,left]{$g\cdot S^2$} (bottomleft);
    \path (topright) -- (bottomright) node[midway,sloped] (dotsr) {$\cdots$};
    \path (topleft) -- (bottomleft) node[midway,sloped] (dotsl) {$\cdots$};

    \draw[-] (top)--(topright) node[pos=0.5,above,sloped]{$2g-2$}--(dotsr)--(bottomright)--(bottomleft)node[pos=0.5,below,sloped]{$2g-2$}--(dotsl)--(topleft)--(topright)node[pos=0.5,below,sloped]{$2g-2$}--(bottomleft)--(top)--(bottomright)--(topleft)--(top)node[pos=0.5,above,sloped]{$2g-2$};
        
    \end{tikzpicture}
    \caption{}
    \label{fig:mutatedso2rose2}
\end{subfigure}
\begin{subfigure}[H]{0.45\textwidth}
    \centering
    \begin{tikzpicture}[main/.style={draw,circle}]
    \node[main, label=below:$\orm(2n)$,color=orange!100] (n) []{};
    \draw (n) to [out=135, in=45,looseness=8] node[pos=0.5,above]{$g\cdot S^2$} (n);
    \end{tikzpicture}
    \caption{}
    \label{fig:so2nGadj2}
\end{subfigure}
\caption{\subref{fig:mutatedso3rose}: Complete graph quiver with $n$ nodes of $\orm(2)$ with edges of multiplicity $2g-2$ connecting all pairs. \subref{fig:so2n1Gadj}: Quiver for $\orm(2n)$ gauge group with $g$ $S^2$ hypermultiplets.}
\label{fig:generalise_O2_2}
\end{figure}

Those generalizations of $\mathfrak{S}_n$ quotient to orthosymplectic quivers are summarised in Table \ref{tab:generalisation}.

\begin{table}[H]
    \centering
    \begin{tabular}{c|c|c}
    \toprule
        $G$ & $G'$ & $V$ \\ \midrule
         $\urm(1)$ & $\urm(n)$ & $adj$\\
         $\usprm(2)$ & $\usprm(2n)$ & $adj,\ \Lambda^2$\\
         $\sorm(3)$ & $\sorm(2n+1)$ & $adj,\ S^2$\\
         $\orm(2)$ & $\orm(2n)$ & $adj,\ S^2$ \\ \bottomrule
    \end{tabular}
    \caption{The rank $1$ gauge nodes $G$ of the complete graph and the rank $n$ gauge node $G'$ after $\Sfrak_n$ quotient. $V$ is the allowed representation of the loop attached to $G$ and $G'$.}
    \label{tab:generalisation}
\end{table}

\paragraph{Weyl group quotient.}
In this section, only connected gauge groups are considered. Note that
\begin{equation}
\Ccal\left(\CG_{2,4g-4}\right)=\Ccal\left(
        \vcenter{\hbox{
        \begin{tikzpicture}
            \node[gauge,label=below:{$1$},color=orange!100] (0) at (0,0) {};
            \node[gauge,label=below:{$1$},color=orange!100] (1) at (1,0) {};
            \node[label=above:{$4g-4$}] at (0.5,-0.1) {};
            \draw (0)--(1);
        \end{tikzpicture}
        }}\right) \quad \overset{\Sfrak_2}{\longrightarrow} \quad
\Ccal\left(\ML_{\usprm(2),g}\right)=\Ccal\left(
    \vcenter{\hbox{
        \begin{tikzpicture}
           [main/.style={draw,circle}]
        \node[main, label=below:$\usprm(2)$,color=orange!100] (2) []{};
        \draw (2) to [out=135, in=45,looseness=8] node[pos=0.5,above]{$g$} (2);
        \end{tikzpicture}
    }}\right).
\end{equation}
This means that every $\usprm(2)$ node can be acquired from a $\Sfrak_2$ quotient of $\CG_{2,4g-4}$. Together with $\Sfrak_n$, one may expect an Abelian theory $\mathcal{T}_w$ (not necessarily a quiver), such that the $\Sfrak_2^n\rtimes\Sfrak_n=\mathcal{W}_{\usprm(2n)}$ quotient of the Coulomb branch of $\mathcal{T}_w$ is the Coulomb branch of $\ML_{\usprm(2n),g}$.

Now this result is generalised as the following: Suppose that the non-zero weights of the matter representation $\mathcal{R}$ contains all the non-zero roots of the gauge group $G_i$, then the Coulomb branch can be written as the Coulomb branch of an Abelian theory quotient the Weyl group of $G_i$ \cite{Teleman:2018wac}. The Abelian theory $\mathcal{T}_w$ is given by $\mathrm{rank}(G_i)$ copies of $\urm(1)$ gauge groups, and the hypermultiplet charges are given by non-zero weights in the original theory, minus one copy of the non-zero weights in the adjoint representation, i.e.
\begin{equation}
    \Ccal\left(\mathcal{Q}_G\right)=\Ccal\left(\mathcal{T}_w/\mathcal{W}_{G_i}\right).
    \label{eq:weyl}
\end{equation}
This generalises \eqref{eq2} in two ways. Firstly, in \eqref{eq:weyl} any representation \q{larger} than the adjoint can be included. Secondly, the gauge group $G$ can be any type.

In the orthogonal or symplectic case, from the Abelian theory the $\Z_2$ factors in the Weyl group of $G_i$ can be understood as the $\Z_2$ automorphisms between $\mathcal{N}_i$ and $\Bar{\mathcal{N}_i}$ of the hypermultiplet representations $\mathcal{R}_i=\mathcal{N}_i\oplus\Bar{\mathcal{N}_i}$. Thus, after the quotient half-hypermultiplets appear in the orthosymplectic quivers.

It is straightforward to see that in quivers \Figref{fig:UnGadj} and \Figref{fig:q3} the requirement for the hypermultiplet representations is satisfied. The requirement for a \q{larger} representation can be understood in the following way. For an Abelian theory, there is no negative contribution to the conformal dimension. In the corresponding non-Abelian theory, the negative contribution form the vector multiplet needs to be fully compensated. In general, for quivers with hypermultiplets representation \q{smaller} than the adjoint, one can only expect the Coulomb branch is birational to the Abelianized Coulomb branch \cite{Bullimore:2015lsa}.

The simplest example for orthosymplectic quivers are the following quivers:

\begin{figure}[H]
    \centering
    \begin{subfigure}[t]{0.3\textwidth}
    \centering
    \begin{tikzpicture}[main/.style={draw,circle}]
    \node[main, label=left:$\sorm(2k)$] (Dk) at (0,0) {};
    \node (Cn) at (0,-1) [flavour,label=below:$\usprm(2n)$]{};
    \draw (Dk) to [out=135, in=45,looseness=8] node[pos=0.5,above]{$S^2$} (Dk);
    \draw (Dk)--(Cn);
    \end{tikzpicture}
    \caption{}
    \label{fig:adhm1}
    \end{subfigure}
   \begin{subfigure}[t]{0.3\textwidth}
   \centering
    \begin{tikzpicture}[main/.style={draw,circle}]
    \node[main, label=left:$\sorm(2k+1)$] (Bk) at (0,0) {};
    \node (Cn) at (0,-1) [flavour,label=below:$\usprm(2n)$]{};
    \draw (Bk) to [out=135, in=45,looseness=8] node[pos=0.5,above]{$S^2$} (Bk);
    \draw (Bk)--(Cn);
    \end{tikzpicture}
    \caption{}
    \label{fig:adhm2}
    \end{subfigure}
   \begin{subfigure}[t]{0.3\textwidth}
   \centering
    \begin{tikzpicture}[main/.style={draw,circle}]
    \node[main, label=left:$\usprm(2k)$] (Ck) at (0,0) {};
    \node (Dn) at (0,-1) [flavour,label=below:$\orm(2n)$]{};
    \draw (Ck) to [out=135, in=45,looseness=8] node[pos=0.5,above]{$\text{Adj}$} (Ck);
    \draw (Ck)--(Dn);
    \end{tikzpicture}
    \caption{}
    \label{fig:adhm3}
    \end{subfigure}
   \begin{subfigure}[t]{0.3\textwidth}
   \centering
    \begin{tikzpicture}[main/.style={draw,circle}]
    \node[main, label=left:$\sorm(2k)$] (Dk) at (0,0) {};
    \node (Cn) at (0,-1) [flavour,label=below:$\usprm(2n)$]{};
    \draw (Dk) to [out=135, in=45,looseness=8] node[pos=0.5,above]{$\text{Adj}$} (Dk);
    \draw (Dk)--(Cn);
    \end{tikzpicture}
    \caption{}
    \label{fig:adhm4}
    \end{subfigure}
   \begin{subfigure}[t]{0.3\textwidth}
   \centering
    \begin{tikzpicture}[main/.style={draw,circle}]
    \node[main, label=left:$\sorm(2k+1)$] (Bk) at (0,0) {};
    \node (Cn) at (0,-1) [flavour,label=below:$\usprm(2n)$]{};
    \draw (Bk) to [out=135, in=45,looseness=8] node[pos=0.5,above]{$\text{Adj}$} (Bk);
    \draw (Bk)--(Cn);
    \end{tikzpicture}
    \caption{}
    \label{fig:adhm5}
    \end{subfigure}
    \caption{ADHM-like quivers.}
    \label{fig:adhm}
\end{figure}
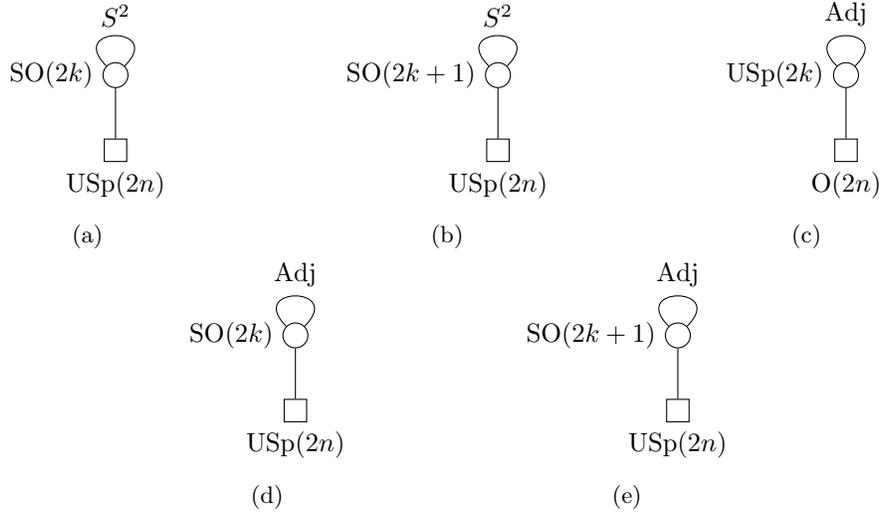
 With \eqref{eq:weyl}, the Coulomb branches of those quivers are summarised in Table \ref{tab:CB}. For completeness, the Weyl groups are listed in Table \ref{tab:weyl}.
 
\begin{table}[H]
 \centering
 \begin{subtable}{0.485\textwidth}
    \centering
    \begin{tabular}{c|c}
    \toprule
        Quiver & Coulomb branch  \\ \midrule
         \Quiver{fig:adhm1} & $(A_{2n+3})^k/\mathcal{W}_{D_k}$ \\
         \Quiver{fig:adhm2} & $(A_{2n+3})^k/\mathcal{W}_{B_k}$\\
         \Quiver{fig:adhm3} & $(A_{2n-1})^k/\mathcal{W}_{C_k}$\\
         \Quiver{fig:adhm4} & $(A_{2n-1})^k/\mathcal{W}_{D_k}$\\
         \Quiver{fig:adhm5} & $(A_{2n-1})^k/\mathcal{W}_{B_k}$\\
        \bottomrule
    \end{tabular}
    \caption{}
    \label{tab:CB}
\end{subtable}
\begin{subtable}{0.485\textwidth}
    \centering
    \begin{tabular}{c|c|c}
    \toprule
        G & $\mathcal{W}_G$ & $|\mathcal{W}_G|$  \\ \midrule
         $A_k$ & $\Sfrak_k$ & $k!$\\
         $B_k$ & $\mathbb{Z}_2^k\rtimes\Sfrak_k$ & $2^k\cdot k!$\\
         $C_k$ & $\mathbb{Z}_2^k\rtimes\Sfrak_k$ & $2^k\cdot k!$\\
         $D_k$ & $\mathbb{Z}_2^{k-1}\rtimes\Sfrak_k$ & $2^{k-1}\cdot k!$\\
        \bottomrule
    \end{tabular}
    \caption{The Weyl group of classical Lie algebras.}
    \label{tab:weyl}
\end{subtable}
\caption{\subref{tab:CB}: The Coulomb branches of the quivers in \Figref{fig:adhm}. \subref{tab:weyl}: The Weyl group of classical Lie algebras.}
\label{}
\end{table}

\section{Conclusions and outlook}
\label{sec:conclusions}

In this work, the $\Sfrak_n$ discrete quotient on the Coulomb branch of $3d$ $\mathcal{N}=4$ quiver with a bouquet of $n$ $\urm(1)$ is generalised to any quiver with a complete graph of $n$ $\urm(1)$. This result is proven on the level of the monopole formula by using the generalised Molien sum. This $\mathfrak{S}_n$ plays the role of both outer-automorphism of the quiver and Weyl group of gauge group. This result is also generalised to complete graphs of any rank $1$ nodes with loops in certain representations.

Another discrete action considered is $\Z_q^{n-1}$. This can be achieved by changing the edge multiplicity to $q$-fold or the charge to $q$-times. These two operations on quivers cannot be distinguished on Coulomb branch level, but can be distinguished by their Higgs branches. The idea behind both actions is to reflect the discrete action on the magnetic lattices into an action on the Coulomb branch geometry. Further, $\Z_q^{n-1}$ and $\Sfrak_n$ can be combined into $\Z_q^{n-1}\!\rtimes\!\Sfrak_n$ action on the Coulomb branch.

These results are expanding our knowledge of relations between different quivers. It allows us to determine whether a Coulomb branch can be obtained as an orbifold of another, and construct new quivers following the rules. For example, a quiver construction of $a_2/\Sfrak_4$ and a new quiver construction of $d_4/\Sfrak_4$ were obtained in this paper.

There are plenty of directions to explore in the future. First of all, the Hasse diagrams of Coulomb branches of quivers with $g$ adjoint hypermultiplets is not fully understood yet. Secondly, for the theories which cannot be Abelianised into a well-defined theory in the sense of \cite{Teleman:2018wac}, it is an open challenge to derive the $\Sfrak_n$ action on the chiral ring generators. Thirdly, it is interesting to classify all quivers whose Coulomb branches are orbifolds of flat space. For example, the orbifold quivers for the complex reflection groups are analysed in an upcoming work. Finally, it remains to be explored what those discrete actions on quivers imply on $4d$ theories. That is to say, the discrete action is well-defined on the magnetic quiver, but to what extend can this be transferred to the 4d $\mathcal{N}=2$ theory (Argyres-Douglas or class $\mathcal{S}$).

\paragraph{Acknowledgements.}
The authors would like to thank Mathew Bullimore, Michael Finkelberg, Lorenzo Foscolo, Paul Levy, Constantin Teleman for insightful discussions and interesting comments. The authors would like to thank the organisers of the \emph{Symplectic Singularities and Supersymmetric QFT} workshop in Amiens where many inspiring discussions took place.
The work of AH, GK, CL, and DL is partially supported by STFC Consolidated Grants ST/T000791/1 and ST/X000575/1. The work of GK is supported by STFC DTP research studentship grant ST/X508433/1.
The work of MS is supported by Austrian Science Fund (FWF), START project STA 73-N. MS also acknowledges support from the Faculty of Physics, University of Vienna.

\appendix

\section{Monopole formula}
\label{app:monopole}

The Coulomb branch chiral ring operators are monopole operators dressed by complex scalars in the vector multiplet. The Hilbert series counts those dressed monopole operators, graded by their conformal dimension. A bare monopole operator is labelled by its magnetic charge $m\in\text{Hom}(\urm(1),G)=\Lambda_\omega^{G\spcheck}$, which is the dual lattice of the weight lattice of the gauge group $G$. For gauge group $G$ and hypermultiplet representation $\mathcal{R}$, the conformal dimension is given by \cite{Borokhov:2002cg,Gaiotto:2008ak,Benna:2009xd,Bashkirov:2010kz}
\begin{equation}
    \Delta(m)=-\sum_{\alpha\in\Phi^+}|\langle\alpha,m\rangle|+\frac{1}{4}\sum_{\omega_i\in\mathcal{R}}|\langle\omega_i,m\rangle| .
    \label{cd1}
\end{equation}
 The positive roots of $\text{Lie}(G)$ are denoted by $\Phi^+$, and the $\omega_i$ are the weights for the matter representation $\mathcal{R}$ for the hypermultiplets. If the representation $\mathcal{R}$ is polarizable, $\mathcal{R}=\mathcal{N}\oplus\Bar{\mathcal{N}}$, then the conformal dimension simplifies to
\begin{equation}
    \Delta(m)=-\sum_{\alpha\in\Phi^+}|\langle\alpha,m\rangle|+\frac{1}{2}\sum_{\omega_i\in\mathcal{N}}|\langle\omega_i,m\rangle| .
    \label{cd2}
\end{equation}

Then the refined Hilbert series is given by:
\begin{equation}
\HS(z,t)=\sum_{m\in\Lambda_\omega^{G\spcheck}/\mathcal{W}_G}z^{J(m)}t^{2\Delta(m)}P_G[G_m](t) \,.
    \label{hs}
\end{equation}
Here $t$ is a fugacity for R-charge, and $z$ is a fugacity for counting topological quantum numbers $J(m)$. There is a dressing factor, $P_G[G_m](t)$, accounting for the dressing of the monopole operators by the complex scalar in the vector multiplet, and it depends on the residual gauge group $G_m$ that is unbroken by the magnetic charges. The dressing factor is counting Casimir invariants of the unbroken gauge group $G_m$. 

\paragraph{Weyl chamber and dressing factor.}
Here detailed explanation of the role of Weyl chamber and the dressing factor $P_G[G_m](t)$ is given, since this is where the $\Sfrak_n$ is acting in the monopole formula. 

The principal Weyl chamber for $\urm(n)$ in \Quiver{fig:q3} can be taken as $\{m_1\geq\cdots\geq m_n\}$ in $\mathbb{R}^n$. This is $\mathfrak{S}_n$ quotient on $\mathbb{R}^n$, the principal Weyl chamber for $\urm(1)^n$ in \Quiver{fig:q4}. The magnetic lattice that is summed over is the intersection of principal Weyl chamber with $\Z^n$.

The principal Weyl chamber is viewed as a polyhedral cone, as in \cite{Hanany:2016ezz,Hanany:2016pfm}. One can decompose the principal Weyl chamber $\sigma$ into a union of loci,
\begin{equation}
    \sigma=\bigcup_i{\sigma_i}
    \label{decomposition}
\end{equation}

 The summation over magnetic lattice in \eqref{hs} is a summation over all the loci $\sigma_i\cap\Z^n$. Each locus is fixed by a subgroup of the Weyl group $\mathcal{W}_G$ and gives a pattern of breaking of the gauge group. 

For a given magnetic charge $m$, the unbroken gauge group $G_m$ is a subgroup of $G$ which commute with the magnetic charges $m$. The whole Weyl group of $G$ is also broken to the Weyl group of $G_m$, denoted by $\mathcal{W}_m$. The Weyl group of $G_m$, $\mathcal{W}_m$, is just the subgroup of $\mathcal{W}_G$ that fixes the locus $\sigma_i$ that $m$ lives in.

The complex scalars of the unbroken gauge group $G_m$ combines into Casimirs, giving the ring $\mathbb{C}[\mathfrak{g}_m]^{G_m}=\mathbb{C}[\mathfrak{h}_m]^{\mathcal{W}_m} $, whose Hilbert series gives the dressing factor and the calculation is explained in next appendix. This $\mathcal{W}_m$ action on the $\mathbb{C}[\mathfrak{h}_m]$ justifies the action on the dressing factor that were used in the main context.

Notice that the dressing factors $P_G[G_m](t)$ have the same order of pole at $t=1$. The residue at $t=1$ is related by a factor of $1/|\mathcal{W}_m|$, since it only counts $\mathcal{W}_m$ invariant polynomials. It immediately follows that the volumes of the Coulomb branches of quivers \Quiver{fig:q3} and \Quiver{fig:q4} are related by an $|\Sfrak_n|$ quotient, as a consistentency check of \eqref{eq2}. 

\section{Molien sum}
\label{app:molien}

The Molien sum is used to compute the Hilbert series of an orbifold $V/\G$, where $V$ is a $n$-dim module of discrete group $\G$, i.e. a $n$-dim complex vector space with $\G$ acting on with certain representation:
\begin{equation}
    \HS=\frac{1}{\vert \G\vert}\sum_{\g\in \G}\frac{1}{\text{det}(1-\g\cdot t)}.
\end{equation}
To explain this formula, recall the orthogonality of characters, i.e. the sum of characters of any non-trivial irrep of $\G$ is zero:
\begin{equation}
\frac{1}{\vert \G\vert}\sum_{\g\in \G}\chi(\g)= \begin{cases}
1 &\text{if $\g$ in trivial irrep}\\
0 &\text{if $\g$ in non-trivial irrep}
\end{cases}
\end{equation}
For any $\g\in \G$, there exists a basis $(x_1,\dots,x_n)$ of $V$ diagonalise $\g$, such that $\g=\text{diag}(\g_1,\dots,\g_n)$ under this basis. The Molien sum is independent of basis. The basis of the space of polynomials $\prod_{d=0}^\infty Sym^d(V)$ can be encoded into a formal series:
\begin{equation}
    \frac{1}{(1-x_1)\cdots(1-x_n)}
\end{equation}

The element $\g$ acts on this series as:
\begin{equation}
    \frac{1}{(1-x_1)\cdots(1-x_n)} \quad \overset{\g}{\longrightarrow} \quad \frac{1}{(1-\g_1 x_1)\cdots(1-\g_n x_n)}.
\end{equation}
If $x_i$ is set to $x_i=t$, the coefficient of $t^d$ is the character of $\g$ on $Sym^d(V)$:
\begin{equation}
    \frac{1}{(1-\g_1 t)\cdots(1-\g_n t)}=\sum_{d=0}^\infty \chi_d(\g)t^d.
\end{equation}
Now the sum over the group $\G$ is performed, only the trivial rep will remain and each copy of trivial rep will contribute one to the coefficients. One can see from Molien sum, the result is exactly the Hilbert series:
\begin{align}
\frac{1}{\vert \G\vert}\sum_{\g\in \G}\frac{1}{\text{det}(1-\g\cdot t)}&=
\frac{1}{\vert \G\vert}\sum_{\g\in \G}\frac{1}{(1-\g_1t)\dots(1-\g_nt)}
\notag \\
&=\frac{1}{\vert \G\vert}\sum_{\g\in \G}\sum_{d=0}^\infty \chi_d(\g)t^d
\notag \\
&=\sum_{d=0}^\infty \text{dim}\left((Sym^d(V))^\G\right) t^d
\notag \\
&=\HS.
\end{align}

The Molien sum can be generalised for any variety with $\G$ acting on it:
\begin{align}
    \HS&=\sum_{\g\in \G}\HS_0^\g \notag\\
    &=\frac{1}{\vert \G\vert}\sum_{\g\in \G}\sum_{d=0}^\infty \chi_d(\g)t^d,
\end{align}
where $\HS_0$ is the Hilbert series of the original variety and $\chi_d(\g)$ is the character of $\g$ in the vector space spanned by polynomials of degree $d$. See \cite{Bourget:2020bxh} for more examples.
The key of Molien sum is using the orthogonality property of characters to throw away any terms transforming non-trivially under $\G$, so that the remaining terms are invariant under the action of $\G$. Note that the Molien sum is representation dependent. The same logic can be extend to continuous group, for example, the Weyl integral for Lie groups.

\section{\texorpdfstring{Outer $\Sfrak_4$ action on $a_2$}{Outer S4 action on a2}}
\label{app:s4}

In this section the $\Sfrak_4$ action on the ring of $a_2$ following \cite{2023arXiv230807398F} is discussed. The ring can be expressed as a $3\!\times\!3$ complex matrix $M\! \in \! \mathrm{SL}(3,\C)$ with conditions traceless and the rank no greater than $1$: $\C[M^i_j]/\langle \Tr(M)=0,\ \rk(M)\leq 1 \rangle$, with $i,j=1,2,3$. We can also use an alternative choice of coordinates by set $M^i_j=x_iy_j$: $\C[x_iy_j]/\langle x_1y_1+x_2y_2+x_3y_3=0 \rangle$, with $i,j=1,2,3$. This set of coordinates automatically impose the condition of $\rk(M)\leq 1$. In the matrix form:
\begin{equation}
\begin{pmatrix}
    x_1y_1& x_1y_2& x_1y_3\\
    x_2y_1& x_2y_2& x_2y_3\\
    x_3y_1& x_3y_2& x_3y_3
\end{pmatrix}.
\end{equation}

The action of $\Sfrak_4=\Sfrak_3\rtimes\Z_2^2$ are generated by the action of $\Sfrak_3$ together with the action of $\mathrm{Klein}_4=\Z_2^2$. Start with the action of $\mathrm{Klein}_4$, which is a subgroup of $\mathrm{SL}(3,\C)$:
\begin{equation}
    (x_1,y_1)\to(-x_1,-y_1),\ (x_2,y_2)\to(-x_2,-y_2),\ (x_3,y_3)\to(-x_3,-y_3). \label{k4}
\end{equation}
It is easy to see that there are only two independent $\Z_2$ actions on the ring of $a_2$, for example, the $(x_1,y_1)\to(-x_1,-y_1)$ together with $(x_2,y_2)\to(-x_2,-y_2)$ can be composed to $(x_3,y_3)\to(-x_3,-y_3)$. The fundamental invariants under this action (after eliminating with the relation) are $x_iy_i$ and $x_i^2y_j^2$ where $i\neq j$. We have the induced relations by requiring the following matrix has rank no greater than $1$ (any $2\times2$ sub-matrix has zero determinant):
\begin{equation}
\begin{pmatrix}
    x_1^2y_1^2& x_1^2y_2^2& x_1^2y_3^2\\
    x_2^2y_1^2& x_2^2y_2^2& x_2^2y_3^2\\
    x_3^2y_1^2& x_3^2y_2^2& x_3^2y_3^2
\end{pmatrix}. \label{Ma2k4}
\end{equation}

Now let us look at the $\Sfrak_3$ action, which is not a subgroup of $\mathrm{SL}(3,\C)$. Here the invariants under $\mathrm{Klein}_4$ are labelled as $\alpha_i=x_{i+1}y_{i+1}-x_{i-1}y_{i-1}$, $\beta_i=x_i^2y_{i-1}^2$ and $\gamma_i=x_{i-1}^2y_i^2$. The action of $\Sfrak_3$ is:
\begin{equation}
    \alpha_i\to\alpha_{\sigma(i)},\ \beta_i\to\beta_{\sigma(i)},\ \gamma_i\to\gamma_{\sigma(i)}. \label{outers3}
\end{equation}
Combine these two actions \eqref{k4} and \eqref{outers3}, there is the action of $\Sfrak_4$ on $a_2$. We call it outer $\Sfrak_4$ since it is not a subgroup of $\mathrm{SL}(3,\C)$ but belongs to its outer-automorphism. One can examine it and find consistency with the action in Section \ref{sec:a2s3}.

\bibliographystyle{JHEP}
\bibliography{bibli.bib}

\end{document}

%% file: Tablea2.tex
\begin{table}[H]
\centering
\scalebox{.9}{
\begin{tabular}{|c|c|c|c|c|}
\hline 
Label & Quiver & Discrete Quotient & HS & Volume \\ \hline 
$\Ccal(\CG_{3,1})=a_2$ & 
\raisebox{-.5 \height}{\begin{tikzpicture}
\node[gauge,label=below:{$1$}] (0) at (0,0) {};
            \node[gauge,label=below:{$1$}] (1) at (1,0) {};
            \node[gauge,label=above:{$1$}] (2) at (0.5,0.7) {};
            \draw[] (0)--(1);
            \draw[] (0)--(2);
            \draw[] (1)--(2);
        \end{tikzpicture}} & & $\frac{1+4t^2+t^4}{(1-t^2)^4}$ & $\frac{3}{8}$ \\  \hline 
        $\Ccal(\CG_{3,2})=a_2/\mathbb{Z}_2^2$ & 
        \raisebox{-.5 \height}{\begin{tikzpicture}
        \node[gauge,label=below:{$1$}] (0) at (0,0) {};
            \node[gauge,label=below:{$1$}] (1) at (1,0) {};
            \node[gauge,label=above:{$1$}] (2) at (0.5,0.7) {};
            \draw[transform canvas={yshift=-1pt}] (0)--(1);
            \draw[transform canvas={yshift=1pt}] (0)--(1);
            \draw[transform canvas={yshift=-0.8pt,
             xshift=0.6pt}] (0)--(2);
            \draw[transform canvas={yshift=0.8pt,
             xshift=-0.6pt}] (0)--(2);
            \draw[transform canvas={yshift=0.8pt,
             xshift=0.6pt}] (1)--(2);
            \draw[transform canvas={yshift=-0.8pt,
             xshift=-0.6pt}] (1)--(2);
        \end{tikzpicture}} & $\mathbb{Z}_2^2\cong Klein_4$ & $\frac{1+4t^4+t^8}{(1-t^2)^2(1-t^4)^2}$ & $\frac{3}{32}$ \\  \hline
$a_2/\mathbb{Z}_2^2\!\rtimes\!\Sfrak_2$ & 
        \raisebox{-.5 \height}{\begin{tikzpicture}
        \node[gauge, label=below:{$2$}] (2) []{};
        \draw (2) to [out=135, in=45,looseness=8] node[pos=0.5,above]{$2$} (2);
        \node[gauge,label=below:{$1$}] (1) at (1,0) {};
        \draw[transform canvas={yshift=-1pt}] (2)--(1);
        \draw[transform canvas={yshift=1pt}] (2)--(1);
        \end{tikzpicture}} & $\mathbb{Z}_2^2\!\rtimes\!\Sfrak_2\cong Dih_4$ & $\frac{1 + t^2 + 2 t^4 + 4 t^6 + 2 t^8 + t^{10} + t^{12}}{(1-t^4)^4}$ & $\frac{3}{64}$\\  \hline
$a_2/\mathbb{Z}_2^2\rtimes\Sfrak_3\cong a_2/\Sfrak_4$ & 
        \raisebox{-.5 \height}{\begin{tikzpicture}
        \node[gauge, label=below:{$3$}] (3) []{};
        \draw (3) to [out=135, in=45,looseness=8] node[pos=0.5,above]{$2$} (3);
        \end{tikzpicture}} & $\mathbb{Z}_2^2\rtimes\Sfrak_3\cong\Sfrak_4$ & $\frac{1+2t^6+2t^8+t^{14}}{(1-t^4)^3(1-t^6)}$ & $\frac{1}{64}$ \\  \hline
\end{tabular}}
\caption{Quivers from discrete quotient of $a_2$.}
\label{tab:a_2}
\end{table} 
%\end{center}
%\caption{}
%\label{fig:a_2}
%\end{figure}

%% file: Tablea3.tex
\begin{table}[H]
\centering
\scalebox{.9}{
\begin{tabular}{|c|c|c|c|c|}
\hline 
Label & Quiver & Discrete Quotient & HS & Volume \\ \hline 
$a_3$ & 
\raisebox{-.5 \height}{\begin{tikzpicture}
        \node[gauge,label=below:{$1$}] (0) at (0,0) {};
            \node[gauge,label=below:{$1$}] (1) at (1,0) {};
            \node[gauge,label=above:{$1$}] (2) at (0,1) {};
            \node[gauge,label=above:{$1$}] (3) at (1,1) {};
            \draw[] (0)--(1) (2)--(3) (0)--(2) (1)--(3);
        \end{tikzpicture}} & & $\frac{(1 + t^2) (1 + 8 t^2 + t^4)}{(1-t^2)^6}$ & $\frac{5}{8}$ \\  \hline 
        $a_3/\Sfrak_2\cong \overline{\text{n.min }B_2}$ & 
        \raisebox{-.5 \height}{\begin{tikzpicture}
        \node[gauge,label=below:{$1$}] (0) at (0,0) {};
            \node[gauge,label=below:{$2$}] (1) at (1,0) {};
            \node[gauge,label=below:{$1$}] (2) at (2,0) {};
            \draw (1) to [out=135, in=45,looseness=8] node[pos=0.5,above]{} (1);
            \draw[] (0)--(1) (1)--(2);
        \end{tikzpicture}} & $\Sfrak_2$ & $\frac{(1 + t^2) (1 + 3 t^2 + t^4)}{(1-t^2)^6}$ & $\frac{5}{16}$ \\  \hline
$a_3/\Sfrak_2\!\times\!\Sfrak_2$ & 
        \raisebox{-.5 \height}{\begin{tikzpicture}
        \node[gauge,label=below:$2$] (2) []{};
        \draw (2) to [out=135, in=45,looseness=8] node[pos=0.5,above]{} (2);
        \node[gauge,label=below:{$2$}] (1) at (1,0) {};
        \draw (1) to [out=135, in=45,looseness=8] node[pos=0.5,above]{} (1);
        \draw[] (2)--(1);
        \end{tikzpicture}} & $\Sfrak_2\!\times\!\Sfrak_2\cong \mathrm{Klein}_4$ & $\frac{1 + 3 t^2 + 11 t^4 + 10 t^6 + 11 t^8 + 3 t^{10} + t^{12}}{(1-t^2)^3(1-t^4)^3}$ & $\frac{5}{32}$\\  \hline
$a_3/\mathbb{Z}_2^3$ & 
        \raisebox{-.5 \height}{\begin{tikzpicture}
\node[gauge,label=below:{$1$}] (0) at (0,0) {};
            \node[gauge,label=below:{$1$}] (1) at (1,0) {};
            \node[gauge,label=above:{$1$}] (2) at (0,1) {};
            \node[gauge,label=above:{$1$}] (3) at (1,1) {};
            \draw[transform canvas={yshift=-1pt}] (0)--(1) (2)--(3);
            \draw[transform canvas={yshift=1pt}] (0)--(1) (2)--(3);
            \draw[transform canvas={xshift=-1pt}] (0)--(2) (1)--(3);
            \draw[transform canvas={xshift=1pt}] (0)--(2) (1)--(3);
        \end{tikzpicture}} & $\mathbb{Z}_2^3$ & $\frac{(1 + t^4) (1 + 8 t^4 + t^8)}{(1-t^2)^3(1-t^4)^3}$ & $\frac{5}{64}$ \\  \hline
$a_3/\mathbb{Z}_2^3\!\rtimes\!\Sfrak_2$ & 
        \raisebox{-.5 \height}{\begin{tikzpicture}
\node[gauge,label=below:{$1$}] (0) at (0,0) {};
            \node[gauge,label=below:{$2$}] (1) at (1,0) {};
            \node[gauge,label=below:{$1$}] (2) at (2,0) {};
            \draw (1) to [out=135, in=45,looseness=8] node[pos=0.5,above]{} (1);
            \draw[transform canvas={yshift=1pt}] (0)--(1) (1)--(2);
            \draw[transform canvas={yshift=-1pt}] (0)--(1) (1)--(2);
        \end{tikzpicture}} & $\mathbb{Z}_2^3\!\rtimes\!\Sfrak_2\cong \Z_2\!\times\!Dih_4$ & $\frac{(1 + t^4) (1 - t^2 + 5 t^4 - t^6 + t^8)}{(1-t^2)^3(1-t^4)^3}$ & $\frac{5}{128}$ \\  \hline
$a_3/\mathbb{Z}_2^3\!\rtimes\!\Sfrak_2\!\times\!\Sfrak_2$ & 
        \raisebox{-.5 \height}{\begin{tikzpicture}
        \node[gauge,label=below:$2$] (2) []{};
        \draw (2) to [out=135, in=45,looseness=8] node[pos=0.5,above]{} (2);
        \node[gauge,label=below:{$2$}] (1) at (1,0) {};
        \draw (1) to [out=135, in=45,looseness=8] node[pos=0.5,above]{} (1);
        \draw[transform canvas={yshift=1pt}] (2)--(1);
        \draw[transform canvas={yshift=-1pt}] (2)--(1);
        \end{tikzpicture}} & $\mathbb{Z}_2^3\!\rtimes\!\Sfrak_2\!\times\!\Sfrak_2$ & $\frac{1 - 2 t^2 + 6 t^4 - 4 t^6 + 8 t^8 - 4 t^{10} + 6 t^{12} - 2 t^{14} + t^{16}}{(1-t^2)^3(1-t^4)^2(1-t^8)}$ & $\frac{5}{256}$ \\  \hline
\end{tabular}}
\caption{Quivers from discrete quotient of $a_3$.}
\label{tab:a_3}
\end{table} 

%% file: Tabled4.tex
\begin{landscape}
\begin{table}[H]
\centering
\scalebox{.8}{
\begin{tabular}{|c|c|c|c|c|}
\hline 
Label & Quiver & Discrete Quotient & HS & Volume \\ \hline 
$d_4$ & 
\raisebox{-.5 \height}{\begin{tikzpicture}
\node[gauge,label=left:{$2$}] (0) at (0.5,0.5) {};
            \node[gauge,label=below:{$1$}] (1) at (0,0) {};
            \node[gauge,label=below:{$1$}] (2) at (1,0) {};
            \node[gauge,label=above:{$1$}] (3) at (0,1) {};
            \node[gauge,label=above:{$1$}] (4) at (1,1) {};
            \draw[] (0)--(1) (0)--(2) (0)--(3) (0)--(4);
        \end{tikzpicture}} &  & $\frac{(1 + t^2) (1 + 17 t^2 + 48 t^4 + 17 t^6 + t^8)}{(1 - t^2)^{10}}$ & $\frac{21}{128}$ \\  \hline
$d_4/\Sfrak_2 (1)$ & 
        \raisebox{-.5 \height}{\begin{tikzpicture}
        \node[gauge,label=below:{$2$}] (0) at (0.5,0.5) {};
        \node[gauge,label=below:{$1$}] (1) at (0,0) {};
        \node[gauge,label=below:{$2$}] (2) at (1.2,0.5) {};
        \node[gauge,label=above:{$1$}] (3) at (0,1) {};
        \draw (2) to [out=135, in=45,looseness=8] node[pos=0.5,above]{} (2);
        \draw[] (0)--(1) (0)--(2) (0)--(3);
        \end{tikzpicture}} & $\Sfrak_2$ & $\frac{(1 + t^2) (1 + 10 t^2 + 20 t^4 + 10 t^6 + t^8)}{(1 - t^2)^{10}}$ & $\frac{21}{256}$\\  \hline
$d_4/\Sfrak_2 (2)$ & 
        \raisebox{-.5 \height}{\begin{tikzpicture}
        \node[gauge,label=left:{$2$}] (0) at (0.5,0.5) {};
        \node[gauge,label=below:{$1$}] (1) at (0,0) {};
        \node[gauge,label=below:{$1$}] (2) at (1,0) {};
        \node[gauge,label=above:{$1$}] (3) at (0,1) {};
        \node[gauge,label=above:{$1$}] (4) at (1,1) {};
        \draw[transform canvas={xshift=0.7pt,yshift=-0.7pt}] (0)--(4);
        \draw[transform canvas={xshift=-0.7pt,yshift=0.7pt}] (0)--(4);
        \draw[transform canvas={xshift=0.7pt,yshift=0.7pt}] (0)--(2);
        \draw[transform canvas={xshift=-0.7pt,yshift=-0.7pt}] (0)--(2);
        \draw[] (0)--(1) (0)--(3);
        \draw (0.9-0.15,0.9-0.05)--(0.7,0.7)--(0.9-0.05,0.9-0.15);
        \draw (0.9-0.15,0.1+0.05)--(0.7,0.3)--(0.9-0.05,0.1+0.15);
        \end{tikzpicture}} & $\Z_2\cong\Sfrak_2$ & $\frac{(1 + 3 t^2 + 6 t^4 + 3 t^6 + t^8) (1 + 8 t^2 + 55 t^4 + 64 t^6 + 55 t^8 + 8 t^{10} + t^{12})}{(1 - t^2)^5(1 - t^4)^5}$ & $\frac{21}{256}$\\  \hline
$d_4/\Sfrak_2\!\times\!\Sfrak_2 (1)$ & 
        \raisebox{-.5 \height}{\begin{tikzpicture}
        \node[gauge,label=below:{$2$}] (0) at (0.6,0.5) {};
        \node[gauge,label=below:{$2$}] (1) at (0,0.5) {};
        \node[gauge,label=below:{$2$}] (2) at (1.2,0.5) {};
        \draw (1) to [out=135, in=45,looseness=8] node[pos=0.5,above]{} (1);
        \draw (2) to [out=135, in=45,looseness=8] node[pos=0.5,above]{} (2);
        \draw[] (0)--(1) (0)--(2);
        \end{tikzpicture}} & $\Sfrak_2\!\times\!\Sfrak_2$ & $\frac{1 + 10 t^2 + 55 t^4 + 150 t^6 + 288 t^8 + 336 t^{10} + 288 t^{12} + 150 t^{14} + 55 t^{16} + 10 t^{18} + t^{20}}{(1 - t^2)^5(1 - t^4)^5}$ & $\frac{21}{512}$\\  \hline
$d_4/\Sfrak_2\!\times\!\Sfrak_2 (2)$ & 
        \raisebox{-.5 \height}{\begin{tikzpicture}
        \node[gauge,label=left:{$2$}] (0) at (0.5,0.5) {};
        \node[gauge,label=below:{$1$}] (1) at (0,0) {};
        \node[gauge,label=below:{$1$}] (2) at (1,0) {};
        \node[gauge,label=above:{$1$}] (3) at (0,1) {};
        \node[gauge,label=above:{$1$}] (4) at (1,1) {};
        \draw[transform canvas={xshift=0.7pt,yshift=-0.7pt}] (0)--(4);
        \draw[transform canvas={xshift=-0.7pt,yshift=0.7pt}] (0)--(4);
        \draw[transform canvas={xshift=0.7pt,yshift=0.7pt}] (0)--(2) (0)--(3);
        \draw[transform canvas={xshift=-0.7pt,yshift=-0.7pt}] (0)--(2) (0)--(3);
        \draw[] (0)--(1);
        \draw (0.9-0.15,0.9-0.05)--(0.7,0.7)--(0.9-0.05,0.9-0.15);
        \draw (0.9-0.15,0.1+0.05)--(0.7,0.3)--(0.9-0.05,0.1+0.15);
        \draw (0.1+0.15,0.9-0.05)--(0.3,0.7)--(0.1+0.05,0.9-0.15);
        \end{tikzpicture}} & $\Z_2^2\cong\Sfrak_2\!\times\!\Sfrak_2$ & $\frac{1 + 5 t^2 + 45 t^4 + 130 t^6 + 314 t^8 + 354 t^{10} + 314 t^{12} + 130 t^{14} + 45 t^{16} + 5 t^{18} + t^{20}}{(1 - t^2)^5(1 - t^4)^5}$ & $\frac{21}{512}$\\  \hline
$d_4/\Sfrak_3 (1)$ & 
        \raisebox{-.5 \height}{\begin{tikzpicture}
        \node[gauge,label=below:{$2$}] (0) at (0.6,0.5) {};
        \node[gauge,label=below:{$1$}] (1) at (0,0.5) {};
        \node[gauge,label=below:{$3$}] (2) at (1.2,0.5) {};
        \draw (2) to [out=135, in=45,looseness=8] node[pos=0.5,above]{} (2);
        \draw[] (0)--(1) (0)--(2);
        \end{tikzpicture}} & $\Sfrak_3$ & $\frac{(1 + t^2) (1 + 3 t^2 + 6 t^4 + 3 t^6 + t^8)}{(1 - t^2)^{10}}$ & $\frac{7}{256}$\\  \hline
$d_4/\Sfrak_3 (2)$ & 
        \raisebox{-.5 \height}{\begin{tikzpicture}
        \node[gauge,label=below:{$2$}] (0) at (0.5,0.5) {};
        \node[gauge,label=below:{$1$}] (1) at (0,0) {};
        \node[gauge,label=below:{$2$}] (2) at (1.3,0.5) {};
        \node[gauge,label=above:{$1$}] (3) at (0,1) {};
        \draw (2) to [out=135, in=45,looseness=8] node[pos=0.5,above]{} (2);
        \draw[] (0)--(1) (0)--(2) (0)--(3);
        \draw[transform canvas={yshift=1.5pt}] (0)--(2);
        \draw[transform canvas={yshift=-1.5pt}] (0)--(2);
        \draw (1,0.6)--(0.85,0.5)--(1,0.4);
        \end{tikzpicture}} & $\Z_3\!\rtimes\!\Sfrak_2\cong\Sfrak_3$ & $\frac{(1 + t^2) (1 + 9 t^2 + 31 t^4 + 119 t^6 + 234 t^8 + 392 t^{10} + 645 t^{12} + 540 t^{14} + 645 t^{16} + 392 t^{18} + 234 t^{20} + 119 t^{22} + 31 t^{24} + 9 t^{26} + t^{28})}{(1 - t^2)^5(1 - t^6)^5}$ & $\frac{7}{256}$\\  \hline
$d_4/\Sfrak_4 (1)$ & 
        \raisebox{-.5 \height}{\begin{tikzpicture}
        \node[gauge,label=below:{$2$}] (0) at (0,0.5) {};
        \node[gauge,label=below:{$4$}] (1) at (0.75,0.5) {};
        \draw (1) to [out=135, in=45,looseness=8] node[pos=0.5,above]{} (1);
        \draw[] (0)--(1);
        \end{tikzpicture}} & $\Sfrak_4$ & $\frac{1 + 3 t^2 + 13 t^4 + 25 t^6 + 46 t^8 + 48 t^{10} + 46 t^{12} + 25 t^{14} + 13 t^{16} + 3 t^{18} + t^{20}}{(1 - t^2)^5(1 - t^4)^5}$ & $\frac{7}{1024}$\\  \hline
$d_4/\Sfrak_4 (2)$ & 
        \raisebox{-.5 \height}{\begin{tikzpicture}
        \node[gauge,label=below:{$2$}] (0) at (0.8,0.5) {};
        \node[gauge,label=below:{$1$}] (1) at (0,0.5) {};
        \node[gauge,label=below:{$3$}] (2) at (1.6,0.5) {};
        \draw (2) to [out=135, in=45,looseness=8] node[pos=0.5,above]{} (2);
        \draw[] (0)--(1);
        \draw[transform canvas={yshift=1pt}] (0)--(2);
        \draw[transform canvas={yshift=-1pt}] (0)--(2);
        \draw (1.3,0.6)--(1.15,0.5)--(1.3,0.4);
        \end{tikzpicture}} & $\Sfrak_4$ & $\frac{1 + 3 t^2 + 13 t^4 + 25 t^6 + 46 t^8 + 48 t^{10} + 46 t^{12} + 25 t^{14} + 13 t^{16} + 3 t^{18} + t^{20}}{(1 - t^2)^5(1 - t^4)^5}$ & $\frac{7}{1024}$\\  \hline
\end{tabular}}
\caption{Quivers from discrete quotient of $d_4$.}
\label{tab:d_4}
\end{table} 
\end{landscape}

%% file: Tablecog4.tex
\begin{table}[H]
\centering
\scalebox{.9}{
\begin{tabular}{|c|c|c|c|c|}
\hline 
Label & Quiver & Discrete Quotient & HS & Volume \\ \hline 
$\CG_{4,2}$ & 
\raisebox{-.5 \height}{\begin{tikzpicture}
\node[gauge,label=below:{$1$}] (0) at (0,0) {};
            \node[gauge,label=below:{$1$}] (1) at (1,0) {};
            \node[gauge,label=above:{$1$}] (2) at (0,1) {};
            \node[gauge,label=above:{$1$}] (3) at (1,1) {};
            \draw[transform canvas={yshift=-1pt}] (0)--(1) (2)--(3);
            \draw[transform canvas={yshift=1pt}] (0)--(1) (2)--(3);
            \draw[transform canvas={xshift=-1pt}] (0)--(2) (1)--(3);
            \draw[transform canvas={xshift=1pt}] (0)--(2) (1)--(3);
            \draw[transform canvas={yshift=0.7pt,
             xshift=-0.7pt}] (0)--(3);
            \draw[transform canvas={yshift=-0.7pt,
             xshift=0.7pt}] (0)--(3);
             \draw[transform canvas={yshift=-0.7pt,
             xshift=-0.7pt}] (1)--(2);
            \draw[transform canvas={yshift=0.7pt,
             xshift=0.7pt}] (1)--(2);
        \end{tikzpicture}} & & $\frac{1 + 6t^6 + 5t^8 + 5t^{12} + 6t^{14} + t^{20}}{(1-t^2)^3(1-t^6)^2(1-t^8)}$ & $\frac{1}{96}$ \\  \hline 
$\CG_{4,2}/\Sfrak_2$ & 
        \raisebox{-.5 \height}{\begin{tikzpicture}
        \node[gauge,label=below:{$1$}] (0) at (0,0) {};
            \node[gauge,label=below:{$1$}] (1) at (1,0) {};
            \node[gauge,label=left:{$2$}] (2) at (0.5,0.7) {};
            \draw (2) to [out=135, in=45,looseness=8] node[pos=0.5,above]{$2$} (2);
            \draw[transform canvas={yshift=-1pt}] (0)--(1);
            \draw[transform canvas={yshift=1pt}] (0)--(1);
            \draw[transform canvas={yshift=-0.8pt,
             xshift=0.6pt}] (0)--(2);
            \draw[transform canvas={yshift=0.8pt,
             xshift=-0.6pt}] (0)--(2);
            \draw[transform canvas={yshift=0.8pt,
             xshift=0.6pt}] (1)--(2);
            \draw[transform canvas={yshift=-0.8pt,
             xshift=-0.6pt}] (1)--(2);
        \end{tikzpicture}} & $\Sfrak_2$ & $\frac{1 + 4 t^6 + 5 t^8 + 2 t^{10} + 2 t^{12} + 5 t^{14} + 4 t^{16} + t^{22}}{(1-t^2)^2(1-t^4)(1-t^6)^2(1-t^8)}$ & $\frac{1}{192}$ \\  \hline
$\CG_{4,2}/\Sfrak_2 \!\times\! \Sfrak_2$ & 
        \raisebox{-.5 \height}{\begin{tikzpicture}
        \node[gauge, label=below:{$2$}] (2) []{};
        \draw (2) to [out=135, in=45,looseness=8] node[pos=0.5,above]{$2$} (2);
        \node[gauge,label=below:{$2$}] (1) at (1,0) {};
        \draw (1) to [out=135, in=45,looseness=8] node[pos=0.5,above]{$2$} (1);
        \draw[transform canvas={yshift=-1pt}] (2)--(1);
        \draw[transform canvas={yshift=1pt}] (2)--(1);
        \end{tikzpicture}} & $\Sfrak_2\! \times\! \Sfrak_2$ & $\frac{1 + 2 t^6 + 6 t^8 + 2 t^{10} + 2 t^{12} + 2 t^{14} + 6 t^{16} + 2 t^{18} + t^{24}}{(1-t^2)(1-t^4)^2(1-t^6)^2(1-t^8)}$ & $\frac{1}{384}$\\  \hline
$\CG_{4,2}/\Sfrak_3$ & 
        \raisebox{-.5 \height}{\begin{tikzpicture}
        \node[gauge, label=below:{$3$}] (2) []{};
        \draw (2) to [out=135, in=45,looseness=8] node[pos=0.5,above]{$2$} (2);
        \node[gauge,label=below:{$1$}] (1) at (1,0) {};
        \draw[transform canvas={yshift=-1pt}] (2)--(1);
        \draw[transform canvas={yshift=1pt}] (2)--(1);
        \end{tikzpicture}} & $\Sfrak_3$ & $\frac{1 + 2 t^6 + 3 t^8 + 4 t^{10} + 2 t^{12} + 2 t^{14} + 4 t^{16} + 3 t^{18} + 2 t^{20} + t^{26}}{(1-t^2)(1-t^4)(1-t^6)^3(1-t^8)}$ & $\frac{1}{576}$ \\  \hline
$\CG_{4,2}/\Sfrak_4$ & 
        \raisebox{-.5 \height}{\begin{tikzpicture}
                \node[gauge, label=below:{$4$}] (3) []{};
        \draw (3) to [out=135, in=45,looseness=8] node[pos=0.5,above]{$2$} (3);
        \end{tikzpicture}} & $\Sfrak_4$ & $\frac{1 + 2 t^8 + 3 t^{10} + 4 t^{12} + t^{14} + 2 t^{16} + t^{18} + 4 t^{20} + 
 3 t^{22} + 2 t^{24} + t^{32}}{(1-t^4)(1-t^6)^3(1-t^8)^2}$ & $\frac{1}{2304}$ \\  \hline
\end{tabular}}
\caption{Quivers from discrete quotient of $\CG_{4,2}$.}
\label{tab:cog4}
\end{table} 

%% file: Tablehsq.tex
%\begin{landscape}
\begin{table}[H]
\hspace*{-30pt}
\scalebox{.9}{
\begin{tabular}{|c|c|c|c|c|}
\hline 
Label & Quiver & Discrete Quotient & HS & Volume \\ \hline 
$\Ccal(\text{\Quiver{fig:quiver211tri}})$ & 
\raisebox{-.5 \height}{\begin{tikzpicture}
\node[gauge,label=left:{$1$}] (0) at (0,0) {};
            \node[gauge,label=right:{$1$}] (1) at (1,0) {};
            \node[gauge,label=above:{$1$}] (2) at (0.5,0.7) {};
            \node[label=above:{$2g-2$}] (3) at (0.5,-0.8) {};
            \draw (0)--(1);
            \draw (0)--(2);
            \draw (1)--(2);
        \end{tikzpicture}} & & $\frac{1+t^2+2t^{2g-1}-2t^{2g+1}-t^{4g-2}-t^{4g}}{(1-t^2)^3(1-t^{2g-1})^2}$ & $\frac{g}{(2g-1)^2}$ \\  \hline 
$\Ccal(\text{\Quiver{fig:quiver21loop}})=\Ccal(\text{\Quiver{fig:quiver211tri}})/\Sfrak_2$ & 
        \raisebox{-.5 \height}{\begin{tikzpicture}
        \node[gauge, label=below:{$2$}] (2) []{};
        \draw (2) to [out=135, in=45,looseness=8] node[pos=0.5,above]{$g$} (2);
        \node[gauge,label=below:{$1$}] (1) at (1,0) {};
        \draw (2)--(1);
        \end{tikzpicture}} & $\Sfrak_2$ & $\frac{1-t^{4g}}{(1-t^2)^3(1-t^{2g-1})^2}$ & $\frac{g}{2(2g-1)^2}$ \\  \hline
$\Ccal(\text{\Quiver{fig:quiver211tri}})/\mathbb{Z}_2$ & 
        \raisebox{-.5 \height}{\begin{tikzpicture}
            \node[gauge,label=left:{$1$}] (0) at (0,0) {};
            \node[gauge,label=right:{$1$}] (1) at (1,0) {};
            \node[gauge,label=above:{$1$}] (2) at (0.5,0.7) {};
            \node[label=above:{$4g-4$}] (3) at (0.5,-0.8) {};
            %\node[label=center:{$2$}] (4) at (0.05,0.5) {};
            %\node[label=center:{$2$}] (5) at (0.9,0.5) {};
            \draw (0)--(1);
            \draw[transform canvas={xshift=-0.5pt, yshift=0.7pt}] (0)--(2);
            \draw[transform canvas={xshift=0.5pt, yshift=-0.7pt}] (0)--(2);
            \draw[transform canvas={xshift=0.5pt, yshift=0.7pt}] (1)--(2);
            \draw[transform canvas={xshift=-0.5pt, yshift=-0.7pt}] (1)--(2);
            \draw (0.08,0.28)--(0.25,0.35)--(0.23,0.16);
            \draw (0.92,0.28)--(0.75,0.35)--(0.77,0.16);
        \end{tikzpicture}} & $\mathbb{Z}_2$ & $\frac{1 + t^2 + 4 t^{4g-2} - 4 t^{4g}  - t^{8g-4} - t^{8g-2}}{(1-t^2)^3(1-t^{4g-2})^2}$ & $\frac{g}{2(2g-1)^2}$\\  \hline
$\Ccal(\text{\Quiver{fig:quiver211tri}})/\mathbb{Z}_2\!\rtimes\!\Sfrak_2$ & 
        \raisebox{-.5 \height}{\begin{tikzpicture}
        \node[gauge, label=below:{$2$}] (2) []{};
        \draw (2) to [out=135, in=45,looseness=8] node[pos=0.5,above]{$2g-1$} (2);
        \node[gauge,label=below:{$1$}] (1) at (1,0) {};
        %\node[label=above:{$2$}] (3) at (0.52,-0.1) {};
        \draw (0.45,0.1)--(0.6,0)--(0.45,-0.1);
        \draw[transform canvas={yshift=-1pt}] (2)--(1);
        \draw[transform canvas={yshift=1pt}] (2)--(1);
        \end{tikzpicture}} & $\mathbb{Z}_2\!\rtimes\!\Sfrak_2$ & $\frac{(1-t^{4g})(1-t^{8g-4})}{(1-t^2)^3(1-t^{4g-2})^3}$ & $\frac{g}{4(2g-1)^2}$ \\  \hline
$\Ccal(\text{\Quiver{fig:quiver211tri}})/\mathbb{Z}_q$ & 
        \raisebox{-.5 \height}{\begin{tikzpicture}
            \node[gauge,label=left:{$1$}] (0) at (0,0) {};
            \node[gauge,label=right:{$1$}] (1) at (1,0) {};
            \node[gauge,label=above:{$1$}] (2) at (0.5,0.7) {};
            \node[label=above:{$4g-4$}] (3) at (0.5,-0.8) {};
            %\node[label=center:{$2$}] (4) at (0.05,0.5) {};
            %\node[label=center:{$2$}] (5) at (0.9,0.5) {};
            \draw (0)--(1);
            \draw[] (0)--(2);
            \draw[] (1)--(2);
            \draw (0.08,0.28)--(0.25,0.35)--(0.23,0.16);
            \draw (0.92,0.28)--(0.75,0.35)--(0.77,0.16);
            \node[label=above:{$q$}] (4) at (0.05,0.1) {};
            \node[label=above:{$q$}] (5) at (0.95,0.1) {};
        \end{tikzpicture}} & $\mathbb{Z}_q$ & $\frac{1 + t^2 + 2q t^{2qg-q} - 2q t^{2qg-q+2}  - t^{4qg-2q} - t^{4qg-2q+2}}{(1-t^2)^3(1-t^{2qg-q})^2}$ & $\frac{g}{q(2g-1)^2}$ \\  \hline
$\Ccal(\text{\Quiver{fig:quiver211tri}})/\mathbb{Z}_q\!\rtimes\!\Sfrak_2$ & 
        \raisebox{-.5 \height}{\begin{tikzpicture}
        \node[gauge, label=below:{$2$}] (2) []{};
        \draw (2) to [out=135, in=45,looseness=8] node[pos=0.5,above]{$2g-1$} (2);
        \node[gauge,label=below:{$1$}] (1) at (1,0) {};
        %\node[label=above:{$2$}] (3) at (0.52,-0.1) {};
        \draw (0.45,0.1)--(0.6,0)--(0.45,-0.1);
        \draw[] (2)--(1);
        \node[label=above:{$q$}] (3) at (0.5,0) {};
        \end{tikzpicture}} & $\mathbb{Z}_q\!\rtimes\!\Sfrak_2$ & $\frac{1+(q-1)t^{2qg-q}-qt^{2qg-q+2}-qt^{4qg-2q}-(q-1)t^{4qg-2q+2}+t^{6qg-3q+2}}{(1-t^2)^3(1-t^{2qg-q})^3}$ & $\frac{g}{2q(2g-1)^2}$ \\  \hline
\end{tabular}}
\caption{Quivers from discrete quotients of $a_2$.}
\label{tab:hypersurface}
\end{table}
%\end{landscape}